\documentclass[aps,prd,twocolumn,amsmath,amssymb,showpacs,footnote,superscriptaddress,reprint]{revtex4-1}

\usepackage{graphicx}
\usepackage{dcolumn}
\usepackage{bm}

\usepackage[dvipdfm]{hyperref} % Si on compile en dvipdf

\hypersetup
    {
        colorlinks,%
        citecolor=blue,%
        linkcolor=blue,%
        urlcolor=blue,%
    }

\begin{document}

\title{Resonant excitation of black holes by massive bosonic fields and giant ringings}

\author{Yves D\'ecanini}
\email{decanini@univ-corse.fr}

\affiliation{Equipe Physique
Th\'eorique - Projet COMPA, \\ SPE, UMR 6134 du CNRS
et de l'Universit\'e de Corse,\\
Universit\'e de Corse, BP 52, F-20250 Corte,
France}

\author{Antoine Folacci}
\email{folacci@univ-corse.fr}

\affiliation{Equipe Physique
Th\'eorique - Projet COMPA, \\ SPE, UMR 6134 du CNRS
et de l'Universit\'e de Corse,\\
Universit\'e de Corse, BP 52, F-20250 Corte,
France}

\author{Mohamed Ould El Hadj}
\email{ould-el-hadj@univ-corse.fr}

\affiliation{Equipe Physique
Th\'eorique - Projet COMPA, \\ SPE, UMR 6134 du CNRS
et de l'Universit\'e de Corse,\\
Universit\'e de Corse, BP 52, F-20250 Corte,
France}

\date{\today}

\begin{abstract}

We consider the massive scalar field, the Proca field and the Fierz-Pauli field in the Schwarzschild spacetime and we focus more particularly on their long-lived quasinormal modes. We show numerically that the associated excitation factors have a strong resonant behavior and we confirm this result analytically from semiclassical considerations based on the properties of the unstable circular geodesics on which a massive particle can orbit the black hole. The conspiracy of (i) the long-lived behavior of the quasinormal modes and (ii) the resonant behavior of their excitation factors induces intrinsic giant ringings, i.e., ringings of huge amplitude. Such ringings, which are moreover slowly decaying, are directly constructed from the retarded Green function. If we describe the source of the black hole perturbation by an initial value problem with Gaussian initial data, i.e., if we consider the excitation of the black hole from an extrinsic point of view, we can show that these extraordinary ringings are still present. This suggests that physically realistic sources of perturbations should generate giant and slowly decaying ringings and that their existence could be used to constrain ultralight bosonic field theory interacting with black holes.

\end{abstract}


\maketitle

\section{Introduction}

Ultralight bosonic fields are important ingredients of the fundamental theories beyond the standard model of elementary particles and the standard model of cosmology based on Einstein's general relativity. They are predicted by string theories as well as by higher-dimensional field theories and they could contribute to the dark matter content of the Universe and explain, without dark energy, its accelerated expansion. However, the particles associated with these fields are so light and are so weakly coupled with the visible sector particles that they have escaped detection so far. Thus, it is rather exciting to realize that ultralight bosonic fields interacting with black holes (BHs) could lead to ``macroscopic" effects (see, e.g., Refs.~\cite{Arvanitaki:2009fg,Rosa:2009ei,Arvanitaki:2010sy,Burt:2011pv,Cardoso:2011xi,
Rosa:2011my,Barranco:2012qs,Pani:2012vp,Dolan:2012yt,Brito:2013wya,
Brito:2013yxa,Babichev:2013una,Brito:2013xaa,Hod:2013dka,Barranco:2013rua,Decanini:2014kha} for recent works on this subject) which could be used to provide strong evidence for the existence of these new particles and to constrain the parameters defining the associated field theories.

In this context, we have highlighted in our recent work dealing with the massive spin-2 field in the Schwarzschild spacetime a new and unexpected effect in BH physics \cite{Decanini:2014kha} : around particular values of the graviton mass, the excitation factors of the long-lived quasinormal modes (QNMs) have a strong resonant behavior. This effect has an immediate fascinating consequence : it induces giant and slowly decaying ringings when the Schwarzschild BH is excited by an external perturbation. So, if massive gravity is relevant to physics, such extraordinary BH ringings could be observed by the future gravitational wave detectors and allow us to test the various massive gravity theories. Otherwise, their absence could be used to impose constraints on the graviton mass and to further support Einstein's general relativity.

In Ref.~\cite{Decanini:2014kha}, we focused only on the Fierz-Pauli theory \cite{Fierz:1939zz,Fierz:1939ix} in the Schwarzschild spacetime, a field theory which has been developed in great details by Brito, Cardoso and Pani in Ref.~\cite{Brito:2013wya} and which can be obtained, e.g., by linearization of the pathology-free bimetric theory of Hassan, Schmidt-May and von Strauss \cite{Hassan:2012wr}, an extension, in curved spacetime, of the fundamental work of de Rham, Gabadadze and Tolley \cite{deRham:2010ik,deRham:2010kj}. In our opinion, the effects we have described in this rather limited context are a general feature of all massive bosonic field theories in arbitrary BH spacetimes and, in this article, we intend to discuss more particularly, in the Schwarzschild spacetime, the cases of the real massive scalar field and of the massive vector field described by the usual Proca theory \cite{Konoplya:2005hr,Rosa:2011my}. We shall also discuss at greater length the case of the massive gravity theory already considered in Ref.~\cite{Decanini:2014kha}.

Before entering into the technical part of our work, it seems to us necessary to recall and to indicate some important points concerning the long-lived QNMs of BHs. First, we note that they should not be confused with the long-lived quasibound states (see, e.g., Refs.~\cite{Deruelle:1974zy,Damour:1976kh,Zouros:1979iw,Detweiler:1980uk} for important pioneering work on this particular topic) which have been the subject of recent work dealing with ultralight scalar fields as well as with massive spin-1 and spin-2 fields \cite{Arvanitaki:2009fg,Rosa:2009ei,Arvanitaki:2010sy,Burt:2011pv,Cardoso:2011xi,
Rosa:2011my,Barranco:2012qs,Pani:2012vp,Dolan:2012yt,Brito:2013wya,Barranco:2013rua}. The long-lived QNMs of BHs are resonant modes which can be encountered in various situations and, in particular, in the two important following contexts : (i) when we consider massless fields in the Kerr spacetime and that we increase the BH angular momentum up to the extremal limit (see Ref.~\cite{Detweiler:1980gk} for a pioneering work on this topic) and (ii) when we consider massive fields in the Schwarzschild spacetime and that we increase the mass field up to the QNM disappearance (see Ref.~\cite{Simone:1991wn} for a pioneering work on this topic). These weakly damped QNMs are really interesting. Indeed, BHs are usually considered as poor oscillators but, in the two contexts previously mentioned, they have very large quality factors and one might intuitively think that the QNMs are then easier to detect. However, it is in fact important to keep in mind that a long-lived QNM can be observed only if the corresponding quasinormal excitation factor is not too small because, in that case, it can be excited easily. We recall, in particular, that the long-lived QNMs of the rapidly rotating Kerr BH have quasinormal excitation factors which vanish in the extremal limit \cite{Ferrari:1984zz}. As a consequence, contrary to initial expectations, the rapidly rotating Kerr BH is not easier to detect. As far as the long-lived QNMs of the massive fields are concerned, the situation is very different as we have already noted in Ref.~\cite{Decanini:2014kha} and as we shall explain here in more detail. In general, the quasinormal excitation factors have a huge amplitude when the QNMs are weakly damped and the conspiracy of these two behaviors induces giant ringings so that the detectability of the BH is considerably improved.

Our paper is organized as follows. We only focus on the $(\ell,n)$ QNMs which are governed by a Regge-Wheeler-type equation (here, $\ell$ denotes the angular momentum index while $n$ is the overtone index). We are therefore concerned with all the QNMs of the scalar field (here $\ell \in \mathbb{N}$), all the odd-parity QNMs of the Proca field (here $\ell \in \mathbb{N}^\ast$), the even-parity $\ell=0$ QNMs of the Proca field and the odd-parity $\ell=1$ QNMs of the Fierz-Pauli field. In Sec.~II, we consider the excitation factors ${\cal {B}}_{\ell n}$ corresponding to the complex quasinormal frequencies $\omega_{\ell n}$ of the $(\ell,n)$ QNMs and we study their evolution as functions of the dimensionless coupling constant ${\tilde \alpha}=2M\mu /{m_\mathrm{P}}^2$ (here $M$, $\mu$ and ${m_\mathrm{P}}$ denote, respectively, the mass of the BH, the rest mass of the field and the Planck mass). We show numerically that, in general, they present a resonant behavior around a critical value ${\tilde \alpha}_{\ell n}$ with a maximum which increases rapidly with the angular momentum index $\ell$ and decreases with the overtone index $n$. We furthermore note that, in a large range around the critical value ${\tilde \alpha}_{\ell n}$, the imaginary part of the quasinormal frequency $\omega_{\ell n}$ is very small, i.e., that the corresponding $(\ell,n)$ QNM is weakly damped. So, because the quasinormal excitation factors and the quasinormal frequencies can be used to quantify intrinsically the amplitude and the decay of the BH ringing, we show that, when ${\tilde \alpha}$ is near, above and far above one of the critical values ${\tilde \alpha}_{\ell n}$, a slowly decaying giant ringing is generated. Moreover, we study numerically the role of the angular momentum index $\ell$  and of the overtone index $n$ by comparing some giant ringings constructed directly from the retarded Green function. In Sec.~III, we confirm analytically some of the previous results by using semiclassical considerations based on the properties of the unstable circular geodesics on which a massive particle can orbit the Schwarzschild BH. Of course, the point of view developed in Sec.~II is an intrinsic one, i.e., it depends only on the BH properties. With astrophysical and physical considerations in mind, it is necessary to examine the role of the source of the field perturbation and to check that the resonant effects previously discussed are not neutralized in the presence of a realistic external perturbation. In other words, it is necessary to develop an extrinsic point of view and to study ringings constructed from the quasinormal excitation coefficients. Indeed, they permit us to include the contribution of the source of the perturbation into the BH response (see Ref.~\cite{Berti:2006wq} for a clear analysis of the distinction between the quasinormal excitation factors and the quasinormal excitation coefficients). In Sec.~IV, we describe the source of the perturbation by an initial value problem (see, e.g., Refs.~\cite{Leaver:1986gd,Andersson:1996cm,Berti:2006wq} for previous works using such an approach) and we show that the use of quasinormal excitation coefficients still leads to giant ringings. In a conclusion, we discuss some limitations of our work as well as possible extensions.

Throughout this article, we adopt units such that $\hbar = c = G = 1$. We consider the exterior of the Schwarzschild BH of
mass $M$ defined by the metric
\begin{equation} \label{Schw_metric}
ds^2= -(1-2M/r)dt^2+ (1-2M/r)^{-1}dr^2+ r^2 d\sigma_2^2
\end{equation}
where $d\sigma_2^2$ denotes the metric on the unit $2$-sphere $S^2$ and with the Schwarzschild coordinates $(t,r)$ which satisfy $t \in ]-\infty, +\infty[$ and $r \in ]2M,+\infty[$. We also use the so-called tortoise coordinate $r_\ast \in ]-\infty,+\infty[$ defined from the
radial Schwarzschild coordinate $r$ by $dr/dr_\ast=(1-2M/r)$ and given by $r_\ast(r)=r+2M \ln[r/(2M)-1]$ and assume a harmonic time dependence $\exp(-i\omega t)$ for all fields.

\section{Resonant behavior of quasinormal excitation factors and intrinsic giant ringings.}

In the Schwarzschild spacetime, the time-dependent Regge-Wheeler equation
\begin{equation}\label{Phi_ell1}
\left[-\frac{\partial^2 }{\partial t^2}+\frac{\partial^2}{\partial r_\ast^2}-V_\ell(r)  \right] \phi_\ell (t,r)=0
\end{equation}
with the effective potential $V_\ell(r)$ given by
\begin{equation}\label{pot_RW_Schw}
V_\ell(r) = \left(1-\frac{2M}{r} \right) \left(\mu^2+
\frac{A(\ell)}{r^2} + \beta \frac{2M}{r^3}\right)
\end{equation}
governs the partial amplitudes $\phi_\ell (t,r)$ of (i) the modes of the massive scalar field (we then have $\ell \in \mathbb{N}$, $A(\ell)= \ell (\ell+1)$ and $\beta =1$), (ii) the odd-parity modes of the Proca field [we then have $\ell \in \mathbb{N}^\ast$, $A(\ell)= \ell (\ell+1)$ and $\beta =0$], (iii) the even-parity $\ell=0$ mode of the Proca field [we then have $A(\ell)= 2$ and $\beta =-3$] and (iv) the odd-parity $\ell =1$ mode of the Fierz-Pauli field [we then have $A(\ell)= 6$ and $\beta =-8$]. This is obvious for the scalar field; for the Proca field, see Refs.~\cite{Konoplya:2005hr,Rosa:2011my} and for the Fierz-Pauli field, see Ref.~\cite{Brito:2013wya}.

The retarded Green function $G^\mathrm{ret}_\ell(t;r,r')$ associated with the partial amplitude $\phi_\ell (t,r)$ is a solution of
\begin{equation}\label{Gret}
\left[-\frac{\partial^2 }{\partial t^2}+\frac{\partial^2}{\partial r_\ast^2}-V_\ell(r)  \right] G^\mathrm{ret}_\ell(t;r,r')=-\delta (t)\delta (r_\ast-r_\ast')
\end{equation}
\noindent satisfying the condition $G_\ell^\mathrm{ret}(t;r,r') = 0$ for $t \le 0$. It can be written as
\begin{equation}\label{Gret_om}
G_\ell^\mathrm{ret}(t;r,r')=-\int_{-\infty +ic}^{+\infty +ic}  \frac{d\omega}{2\pi}  \frac{\phi^\mathrm{in}_{\omega \ell}(r_<) \phi^\mathrm{up}_{\omega \ell}(r_>)}{W_\ell (\omega)} e^{-i\omega t}
\end{equation}
\noindent where $c>0$, $r_< =\mathrm{min} (r,r')$, $r_> =\mathrm{max} (r,r')$ and with $W_\ell (\omega)$ denoting the Wronskian of the functions $\phi^\mathrm{in}_{\omega \ell}$ and $\phi^\mathrm{up}_{\omega \ell}$. These two functions are linearly independent solutions of the Regge-Wheeler equation
\begin{equation}\label{RW}
\frac{d^2 \phi_{\omega \ell}}{dr_\ast^2} + \left[ \omega^2 -V_\ell(r)\right]  \phi_{\omega \ell}=0.
\end{equation}
When $\mathrm{Im} (\omega) > 0$, $\phi^\mathrm{in}_{\omega \ell}$ is uniquely defined by its ingoing behavior at the event horizon $r=2M$ (i.e., for $r_\ast \to -\infty$)
\begin{subequations}\label{bc_in}
\begin{equation}\label{bc_1_in}
\phi^\mathrm{in}_{\omega \ell} (r) \underset{r_\ast \to -\infty}{\sim} e^{-i\omega r_\ast}
\end{equation}
and, at spatial infinity $r \to +\infty$ (i.e., for $r_\ast \to +\infty$), it has an
asymptotic behavior of the form
\begin{eqnarray}\label{bc_2_in}
& & \phi^\mathrm{in}_{\omega  \ell}(r) \underset{r_\ast \to +\infty}{\sim}
 \left[ \frac{\omega}{p(\omega)}
\right]^{1/2}  \nonumber \\
& & \quad \times \left(A^{(-)}_\ell (\omega) e^{-i[p(\omega)
r_\ast + [M\mu^2/p(\omega)] \ln(r/M)]}\right. \nonumber \\
& & \quad \quad  \left. + A^{(+)}_\ell (\omega) e^{+i[p(\omega) r_\ast +
[M\mu^2/p(\omega)] \ln(r/M)]} \right).
\end{eqnarray}
\end{subequations}
Similarly, $\phi^\mathrm{up}_{\omega \ell }$ is uniquely defined by its outgoing behavior at spatial infinity
\begin{subequations}\label{bc_up}
\begin{equation}\label{bc_1_up}
\phi^\mathrm{up}_{\omega \ell} (r) \underset{r_\ast \to +\infty}{\sim}  \left[ \frac{\omega}{p(\omega)}
\right]^{1/2} e^{+i[p(\omega) r_\ast +
[M\mu^2/p(\omega)] \ln(r/M)]}
\end{equation}
and, at the horizon, it has an asymptotic behavior of the form
\begin{equation}\label{bc_2_up}
\phi^\mathrm{up}_{\omega \ell }(r) \underset{r_\ast \to -\infty}{\sim}
B^{(-)}_\ell (\omega) e^{-i\omega r_\ast}  + B^{(+)}_\ell (\omega) e^{+i\omega r_\ast}.
\end{equation}
\end{subequations}
In Eqs.~(\ref{bc_in}) and (\ref{bc_up}), $p(\omega)=\left( \omega^2 - \mu^2 \right)^{1/2}$ denotes the ``wave number"
while $A^{(-)}_\ell (\omega)$, $A^{(+)}_\ell (\omega)$, $B^{(-)}_\ell (\omega)$ and $B^{(+)}_\ell (\omega)$ are complex amplitudes which, like the $\mathrm{in}$ and $\mathrm{up}$ modes, can be defined by analytic continuation in the full complex $\omega$ plane (or, more precisely, in a well-chosen multisheeted Riemann surface). By evaluating the Wronskian $W_\ell (\omega)$ at $r_\ast \to -\infty$ and $r_\ast \to +\infty$, we obtain
\begin{equation}\label{Well}
W_\ell (\omega) =2i\omega A^{(-)}_\ell (\omega) = 2i\omega B^{(+)}_\ell (\omega).
\end{equation}

If the Wronskian $W_\ell (\omega)$ vanishes, the functions $\phi^\mathrm{in}_{\omega \ell}$ and $\phi^\mathrm{up}_{\omega \ell}$ are linearly dependent and propagate inward at the horizon and outward at spatial infinity, a behavior which defines the QNMs. The zeros of the Wronskian lying in the lower part of the complex $\omega$ plane are the frequencies of the $(\ell ,n)$ QNMs. The contour of integration in Eq.~(\ref{Gret_om}) may be deformed in order to capture them (see, e.g., Ref.~\cite{Leaver:1986gd}). By Cauchy's theorem and if we do not take into account the ``prompt" contribution (arising from the arcs at $|\omega|=\infty$) and the ``tail" contribution (associated with the various cuts), we can extract from the retarded Green function (\ref{Gret_om}) a residue series over the quasinormal frequencies $\omega_{\ell n}$ lying in the fourth quadrant of the complex $\omega$ plane. We then obtain the contribution describing the BH ringing. It is given by
\begin{equation}\label{Gret_ell_QNMsum}
G_\ell^{\mathrm{ret} \, \mathrm{QNM}}(t;r,r')= \sum_n G_{\ell n}^{\mathrm{ret} \, \mathrm{QNM}}(t;r,r')
\end{equation}
with
\begin{eqnarray}\label{Gret_ell_QNM}
&& G_{\ell n}^{\mathrm{ret} \, \mathrm{QNM}}(t;r,r')=2 \, \mathrm{Re} \left[ {\cal B}_{\ell n}
{\tilde \phi}_{\ell n}(r) {\tilde \phi}_{\ell n}(r') \right. \nonumber\\
&&  \left.  \times e^{-i[\omega_{\ell n} t - p(\omega_{\ell n})r_\ast - p(\omega_{\ell n})r'_\ast - [M\mu^2/p(\omega_{\ell n})] \ln(rr'/M^2)]} \right].\nonumber\\
&&
\end{eqnarray}
Here
\begin{equation}\label{Excitation F}
{\cal B}_{\ell n} = \left(\frac{1}{2 p(\omega)} \frac{A^{(+)}_\ell (\omega)}{\frac{dA^{(-)}_\ell (\omega)}{d\omega}}   \right)_{\omega=\omega_{\ell n}}
\end{equation}
denotes the excitation factor corresponding to the complex frequency $\omega_{\ell n}$. In Eq.~(\ref{Gret_ell_QNM}), the real part symbol $\mathrm{Re}$ has been introduced to take into account the symmetry of the quasinormal frequency spectrum with respect to the imaginary $\omega$ axis and the modes ${\tilde \phi}_{\ell n}(r)$ are defined by
\begin{eqnarray}\label{QNM norm}
&& {\tilde \phi}_{\ell n}(r) \equiv   \phi^\mathrm{in}_{\omega_{\ell n} \ell}(r)  {\Big/} \left[  \left[\omega_{\ell n}/p(\omega_{\ell n})
\right]^{1/2}A^{(+)}_\ell (\omega_{\ell n}) \right. \nonumber\\
&& \left. \quad\qquad\qquad\qquad \times e^{i[p(\omega_{\ell n})r_\ast +
[M\mu^2/p(\omega_{\ell n})] \ln(r/M)]} \right]
\end{eqnarray}
and are therefore normalized so that ${\tilde \phi}_{\ell n}(r) \sim 1$ as $r \to +\infty $. In the sum (\ref{Gret_ell_QNMsum}), $n=0$ corresponds to the fundamental QNM (i.e., the least damped one) and $n=1,2,\dots $ to the overtones.

It is important to recall that quasinormal retarded Green functions such as (\ref{Gret_ell_QNM}) do not provide physically relevant results at ``early times" due to their exponentially divergent behavior as $t$ decreases. In fact, it is necessary to determine, from physical considerations, the time beyond which they can be used and this time is the starting time $t_\mathrm{start}$ of the BH ringing. This is the so-called ``time-shift problem" (see, e.g., Ref.~\cite{Berti:2006wq} for a clear discussion of this problem and references therein for related works). It can be ``easily" solved for massless fields. We first note that the QNMs are semiclassically associated with the peak of the effective potential located close to $r_\ast \approx 0 $. Then, by assuming that the source at $r'_\ast$ and the observer at $r_\ast$ are far from the BH (i.e., that $r_\ast,r'_\ast \gg 2M$) we have $t_\mathrm{start} \approx r_\ast + r'_\ast$ which is approximatively the time taken for the signal to travel from the source to the peak of the potential and then to reach the observer. For massive fields, the previous considerations must be slightly modified : it is necessary to take into account the dispersive behavior of the QNMs and therefore to define $t_\mathrm{start}$ from group velocities. From the dispersion relation $p(\omega)=\left( \omega^2 - \mu^2 \right)^{1/2}$, we can show that the group velocity corresponding to the quasinormal frequency $\omega_{\ell n}$ is given by $v_\mathrm{g}=\mathrm{Re}[p(\omega_{\ell n})]/\mathrm{Re}[\omega_{\ell n}]$. This can be confirmed by noting that the factor $\exp [ -i\omega_{\ell n} t + ip(\omega_{\ell n})r_\ast ] $ appearing in (\ref{Gret_ell_QNM}) leads to the phase velocity $v_\mathrm{p}=\mathrm{Re}[\omega_{\ell n}]/\mathrm{Re}[p(\omega_{\ell n})]$ [here it is necessary to neglect the term $i[M\mu^2/p(\omega_{\ell n})] \ln(r/M)$, an assumption formally valid for large $r_\ast$]. Because the peak of the effective potential still remains located close to $r_\ast \approx 0 $, we then obtain $t_\mathrm{start} \approx (r_\ast + r'_\ast)\mathrm{Re}[\omega_{\ell n}]/\mathrm{Re}[p(\omega_{\ell n})]$, a result which depends on the angular momentum index $\ell$ and the overtone index $n$.

\begin{figure}
\centering
\includegraphics[height=3cm,width=8.5cm]{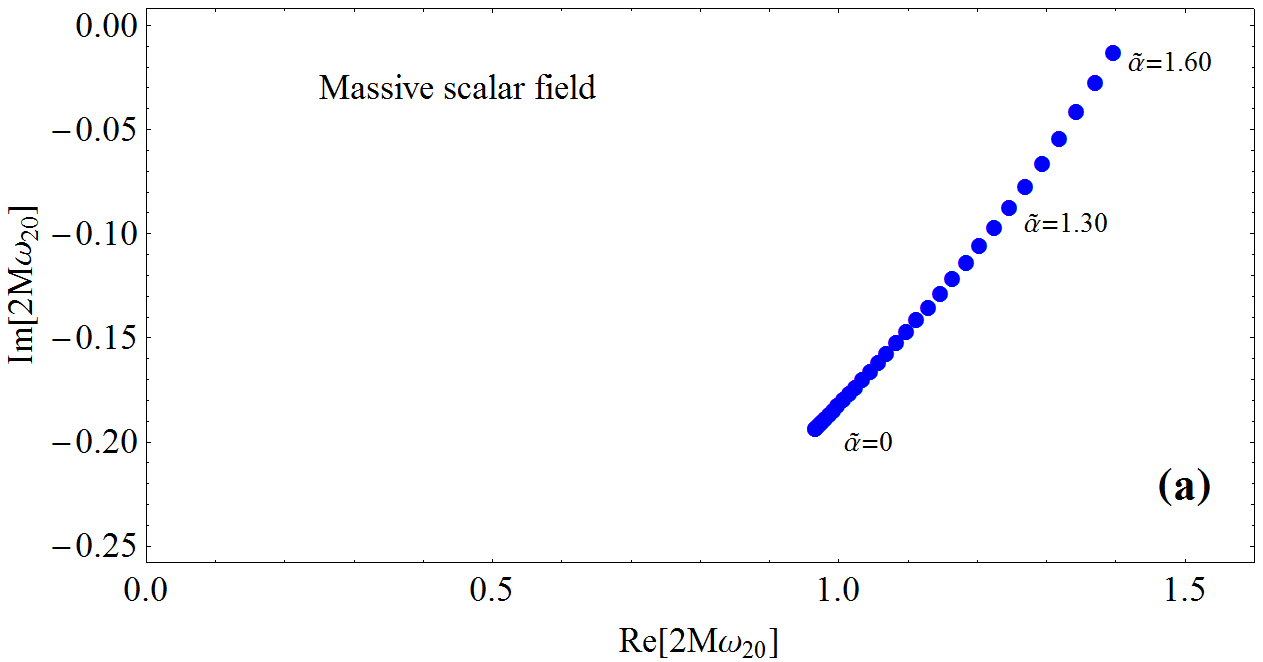}
\centering
\vspace{0.2cm}
\includegraphics[height=4cm,width=8.5cm]{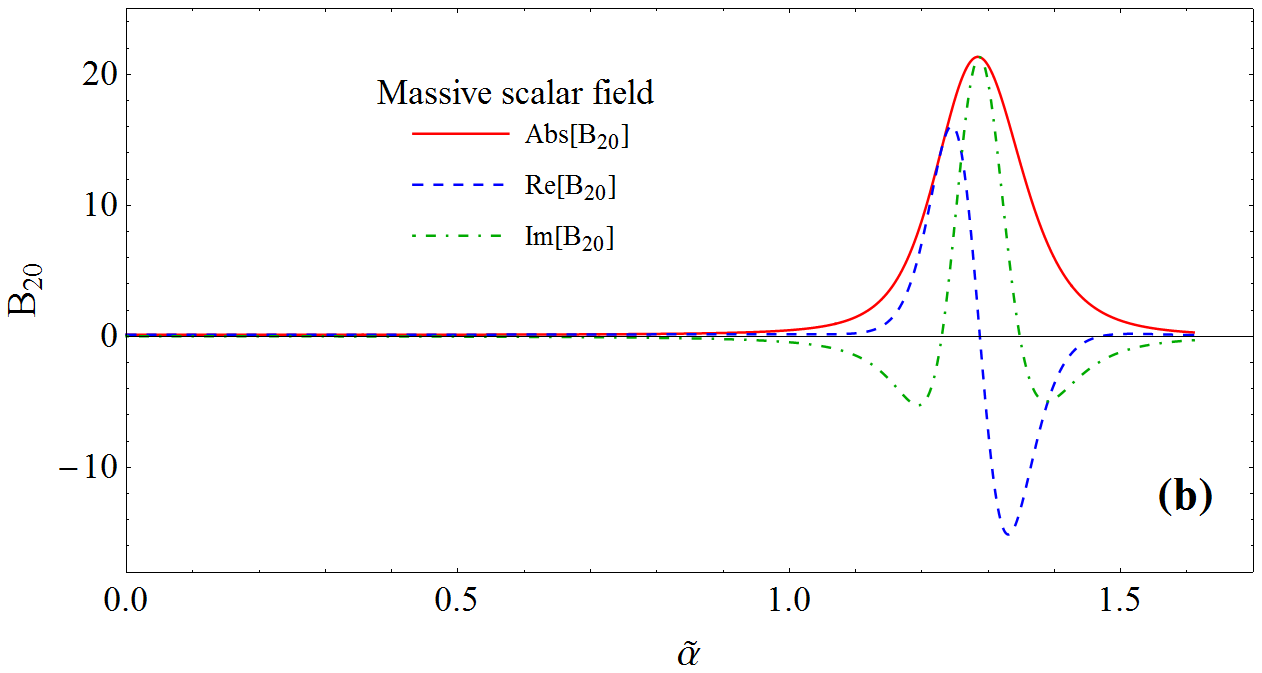}
\centering
\vspace{0.2cm}
\includegraphics[height=4cm,width=8.5cm ]{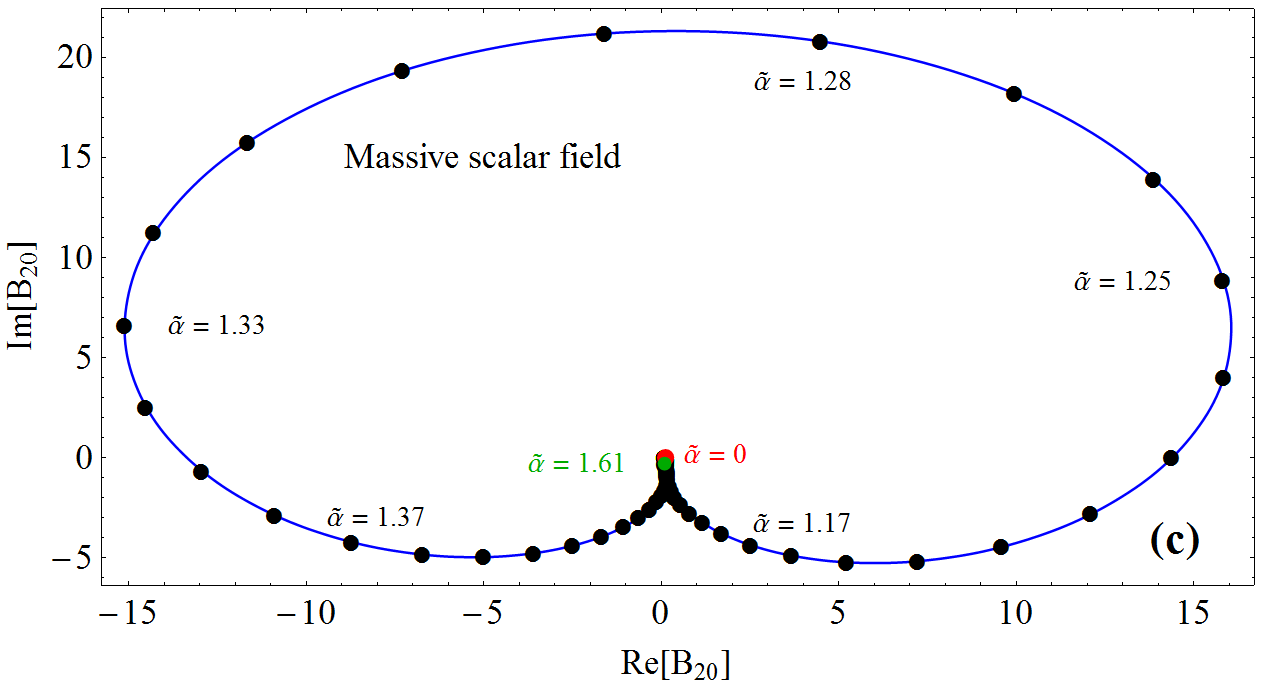}
\centering
\vspace{0.2cm}
\includegraphics[height=4cm,width=8.5cm ]{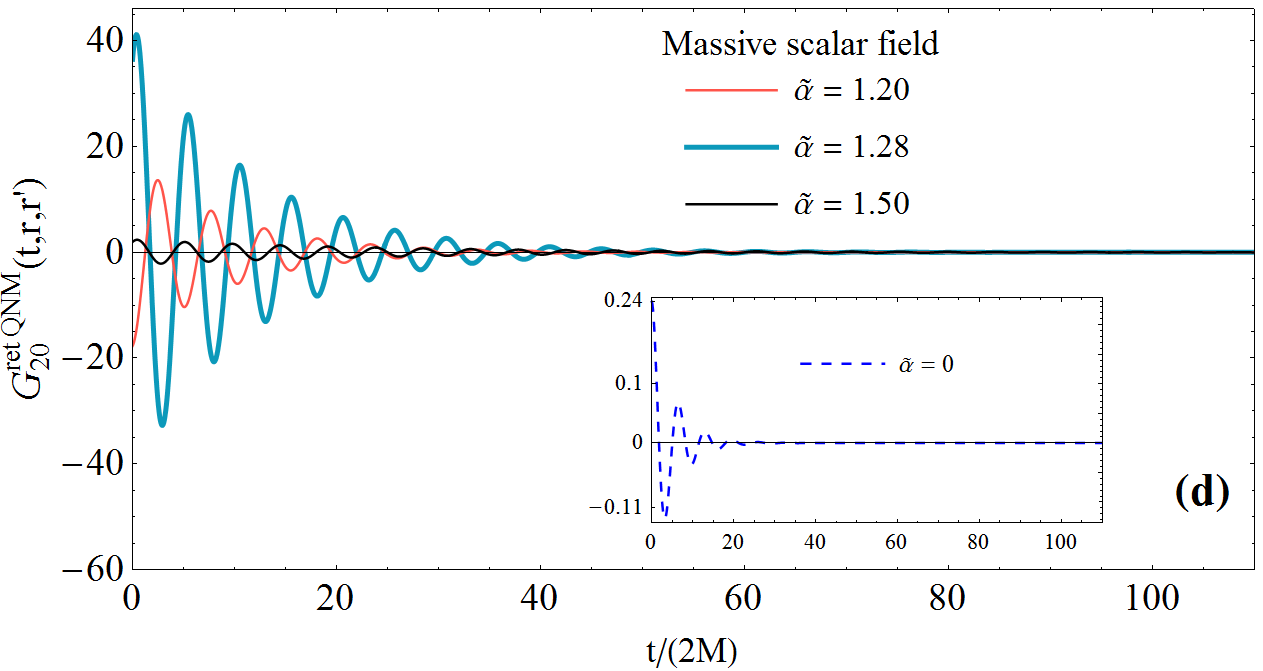}
\centering
\vspace{0.2cm}
\includegraphics[height=4cm,width=8.5cm ]{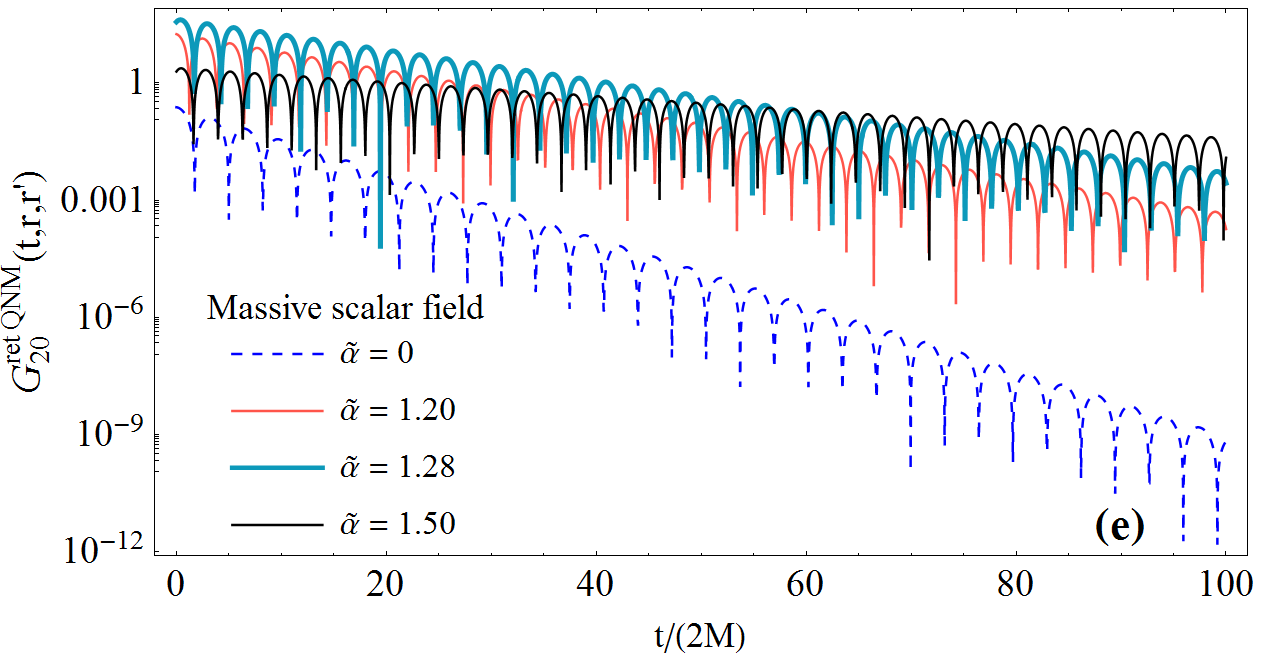}
\setlength\abovecaptionskip{0.25ex}
\vspace{-0.3cm}
\caption{\label{fig:QNM20_INT} The $(\ell = 2, n = 0)$ QNM of the massive scalar field. (a) The complex quasinormal frequency $2M\omega_{20}$ for $\tilde\alpha = 0, 0.05,\ldots, 1.55, 1.60$. (b) The resonant behavior of the excitation factor ${\cal B}_{20}$. (c) The excitation factor ${\cal B}_{20}$  for $\tilde\alpha = 0, 0.01,\ldots, 1.60, 1.61$. (d) and (e) Some intrinsic ringings corresponding to values of the mass near and above the critical value  $\tilde\alpha_{20}$. We compare them with the ringing corresponding to the massless scalar field. The results are obtained from $(\ref{Gret_ell_QNM})$ with $r = 50M$ and $r' = 10M$.}
\end{figure}

\begin{figure}
\centering
\includegraphics[height=3cm,width=8.5cm]{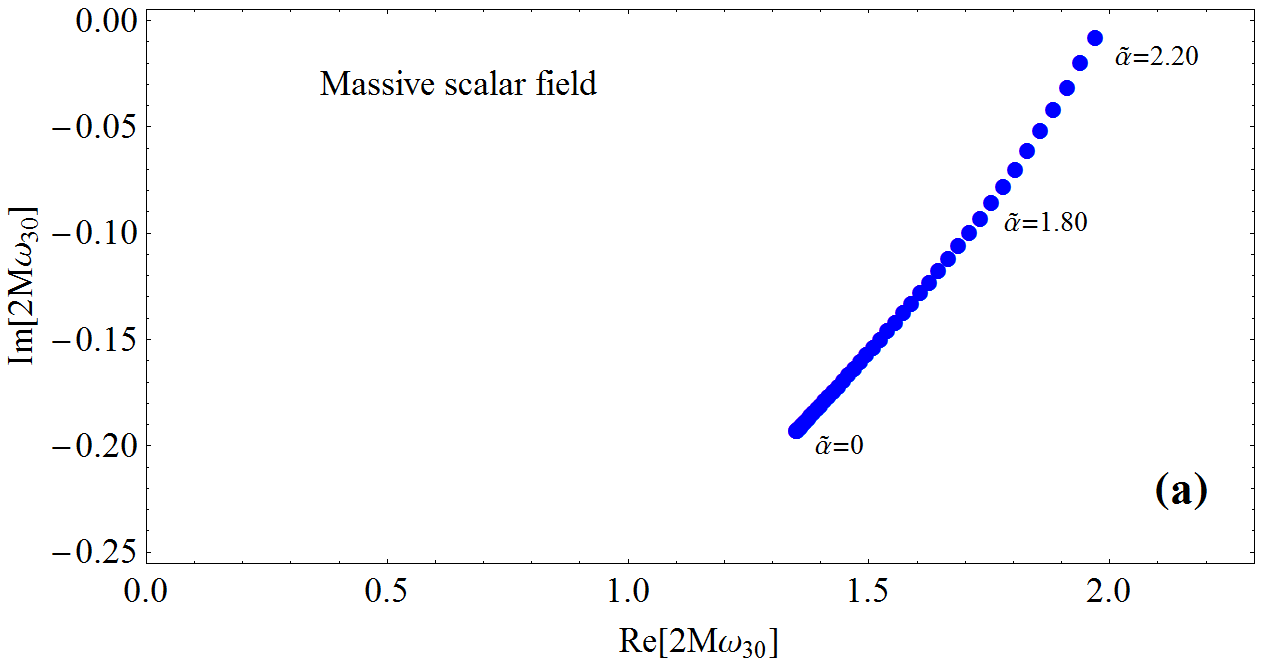}
\centering
\vspace{0.2cm}
\includegraphics[height=4cm,width=8.5cm]{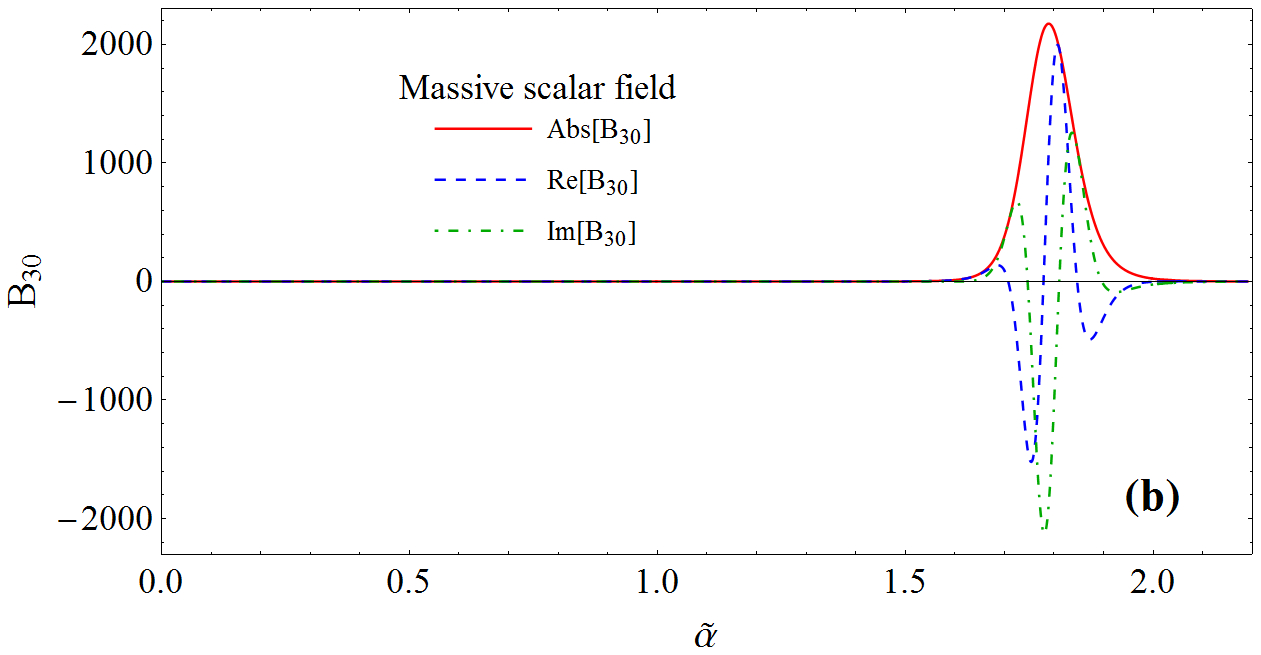}
\centering
\vspace{0.2cm}
\includegraphics[height=4cm,width=8.5cm ]{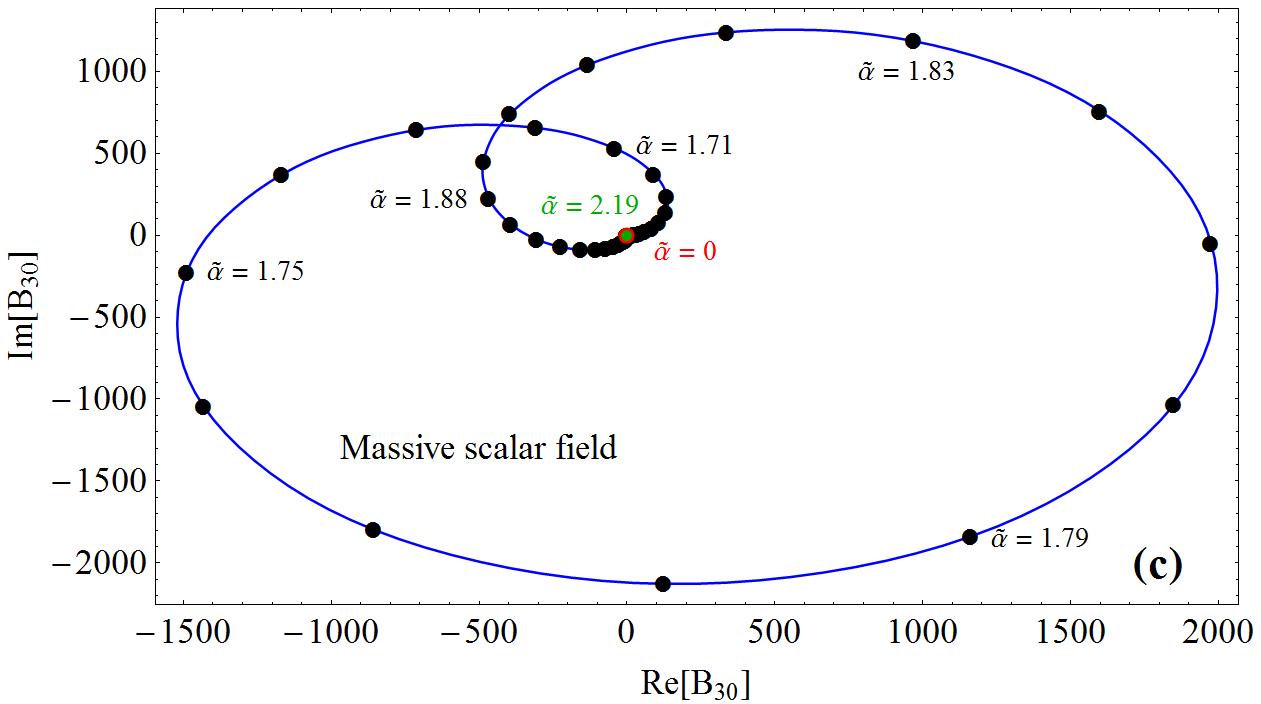}
\centering
\vspace{0.2cm}
\includegraphics[height=4cm,width=8.5cm ]{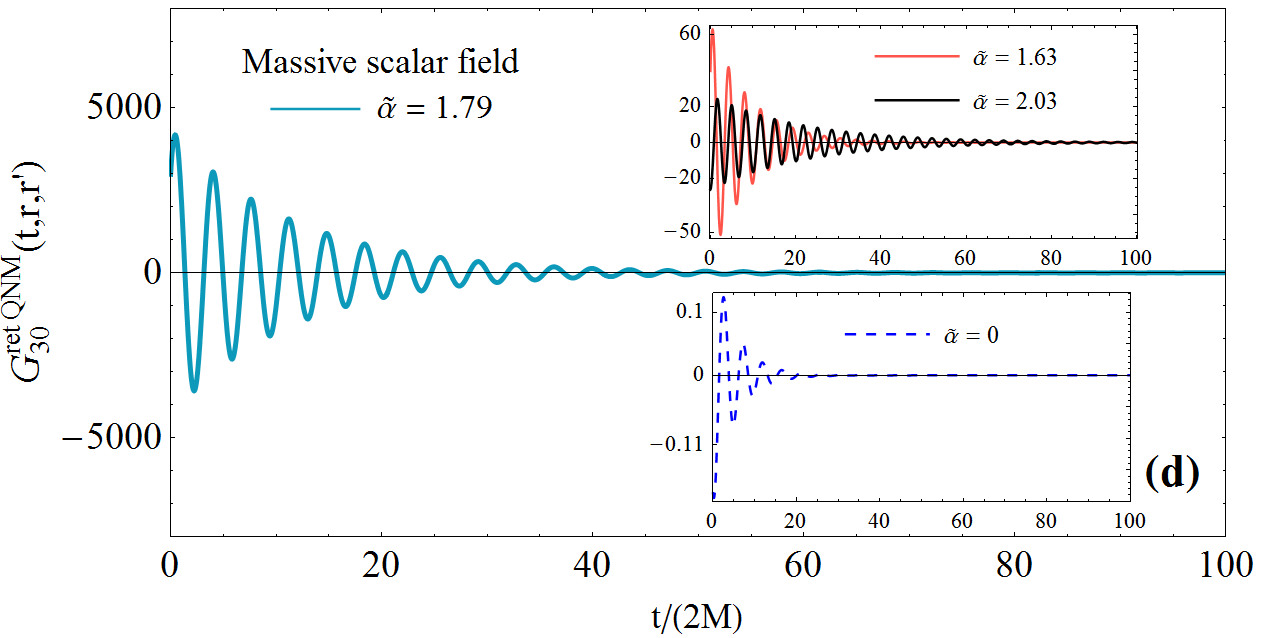}
\centering
\vspace{0.2cm}
\includegraphics[height=4cm,width=8.5cm ]{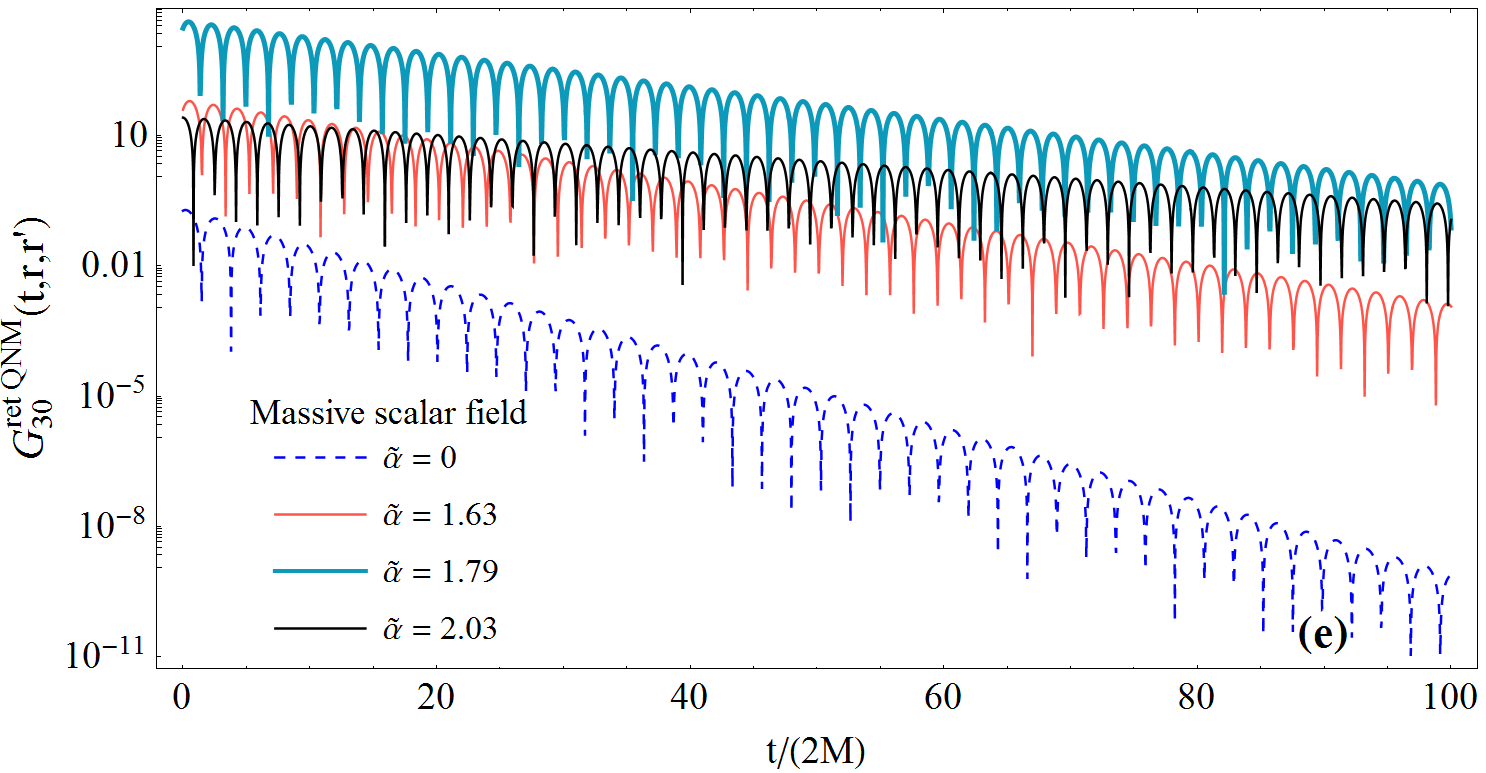}
\setlength\abovecaptionskip{0.25ex}
\vspace{-0.3cm}
\caption{\label{fig:QNM30_INT} The $(\ell = 3, n = 0)$ QNM of the massive scalar field. (a) The complex quasinormal frequency $2M\omega_{30}$ for $\tilde\alpha = 0, 0.05,\ldots, 2.15, 2.20$. (b) The resonant behavior of the excitation factor ${\cal B}_{30}$. (c) The excitation factor ${\cal B}_{30}$  for $\tilde\alpha = 0, 0.01,\ldots, 2.18, 2.19$. (d) and (e) Some intrinsic ringings corresponding to values of the mass near and above the critical value  $\tilde\alpha_{30}$. We compare them with the ringing corresponding to the massless scalar field. The results are obtained from $(\ref{Gret_ell_QNM})$ with $r = 50M$ and $r' = 10M$.}
\end{figure}

\begin{figure}
\centering
\includegraphics[height=3cm,width=8.5cm]{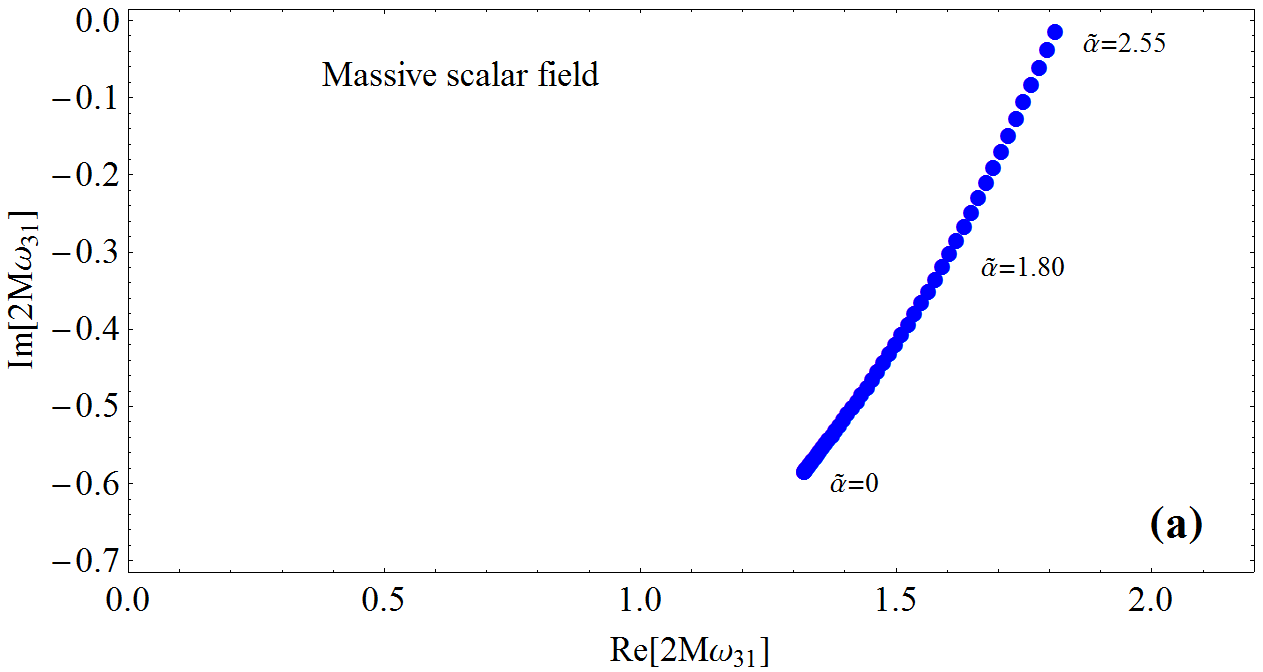}
\centering
\vspace{0.2cm}
\includegraphics[height=4cm,width=8.5cm]{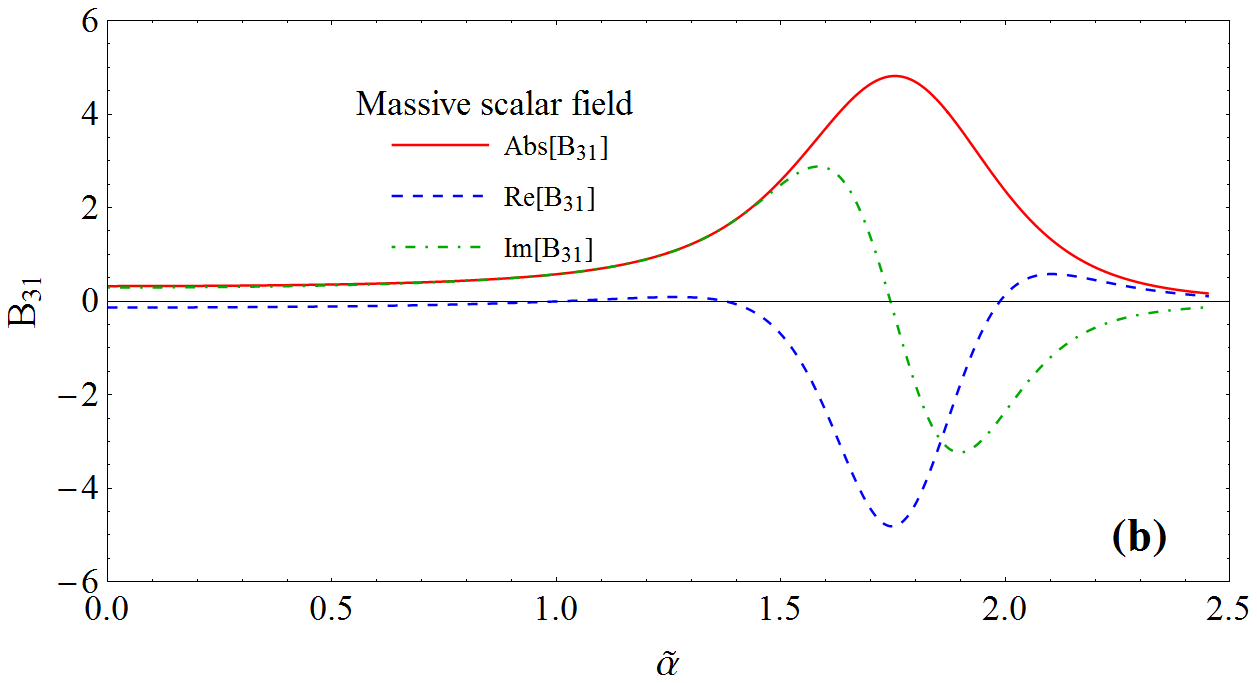}
\centering
\vspace{0.2cm}
\includegraphics[height=4cm,width=8.5cm ]{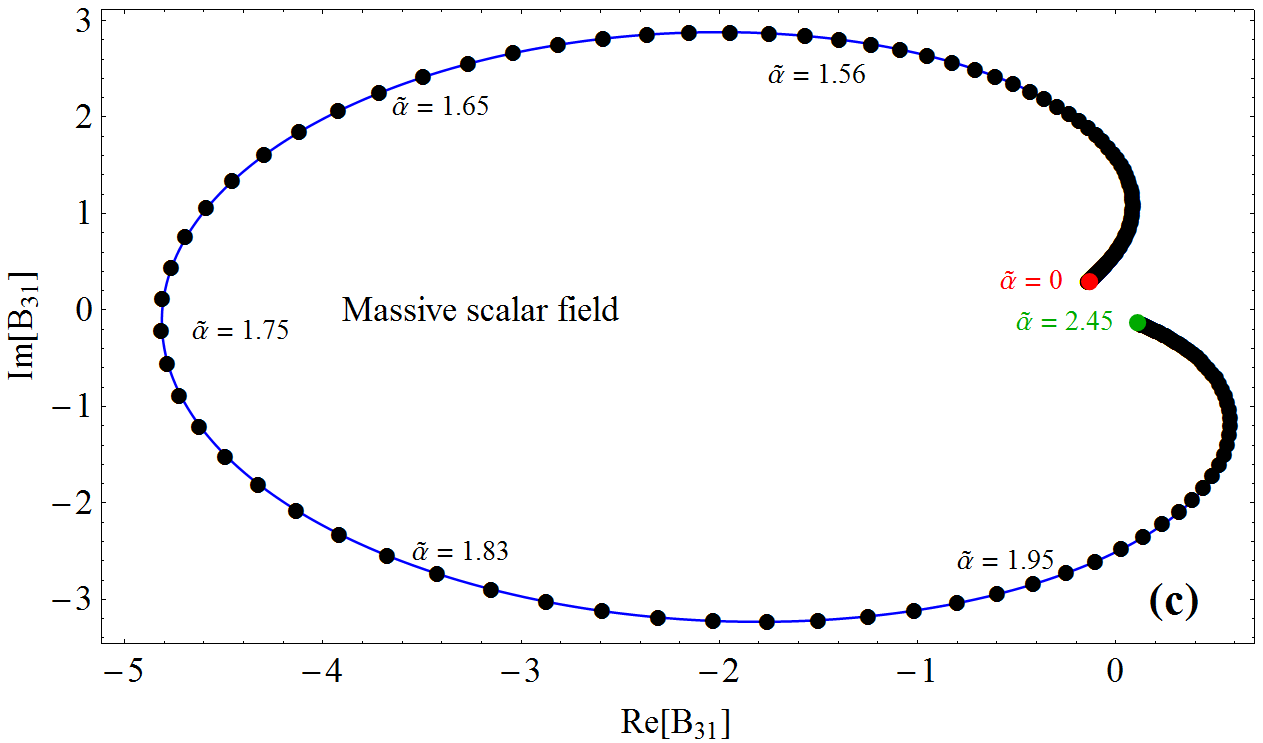}
\centering
\vspace{0.2cm}
\includegraphics[height=4cm,width=8.5cm ]{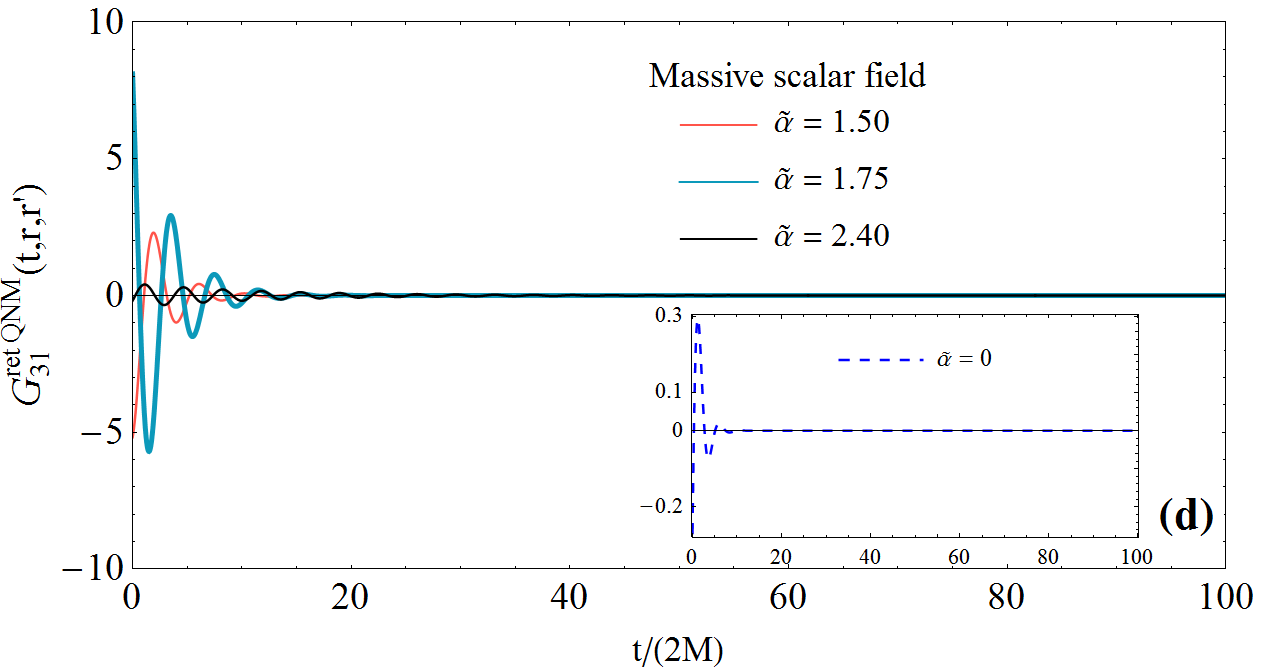}
\centering
\vspace{0.2cm}
\includegraphics[height=4cm,width=8.5cm ]{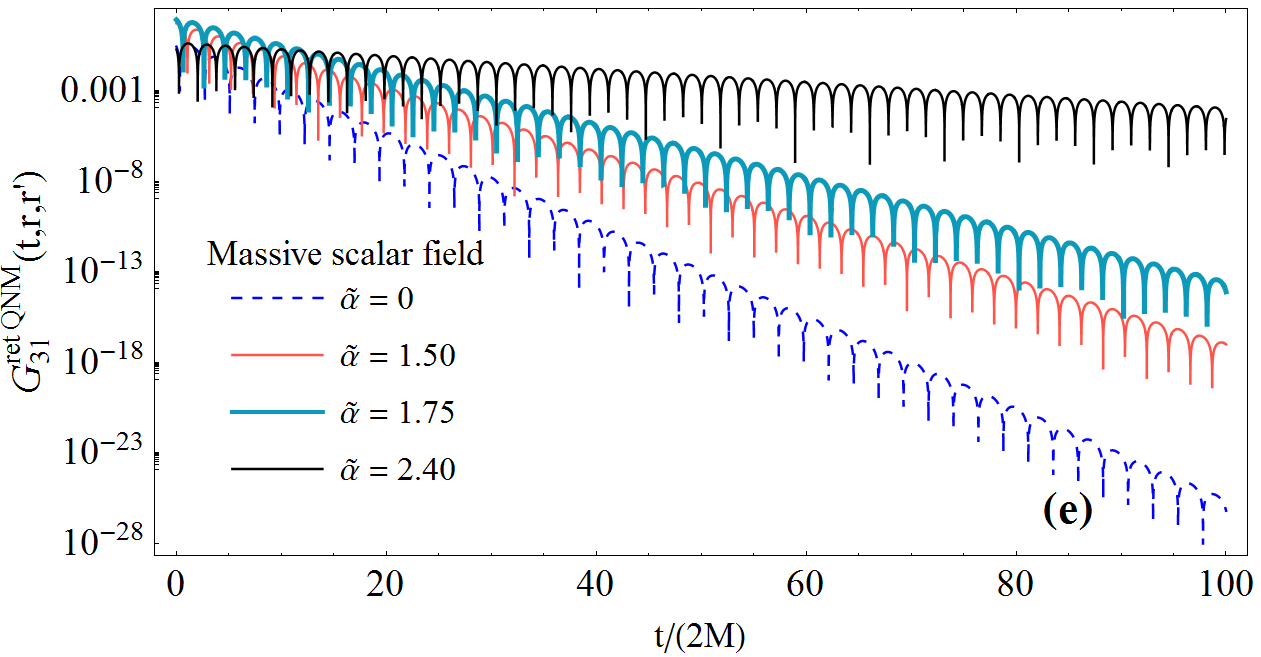}
\setlength\abovecaptionskip{0.25ex}
\vspace{-0.3cm}
\caption{\label{fig:QNM31_INT} The $(\ell = 3, n = 1)$ QNM of the massive scalar field. (a) The complex quasinormal frequency $2M\omega_{31}$ for $\tilde\alpha = 0, 0.05,\ldots, 2.50, 2.55$. (b) The resonant behavior of the excitation factor ${\cal B}_{31}$. (c) The excitation factor ${\cal B}_{31}$  for $\tilde\alpha = 0, 0.01,\ldots, 2.44, 2.45$. (d) and (e) Some intrinsic ringings corresponding to values of the mass near and above the critical value  $\tilde\alpha_{31}$. We compare them with the ringing corresponding to the massless scalar field. The results are obtained from $(\ref{Gret_ell_QNM})$ with $r = 50M$ and $r' = 10M$.}

\end{figure}

\begin{figure}
\centering
\includegraphics[height=3cm,width=8.5cm]{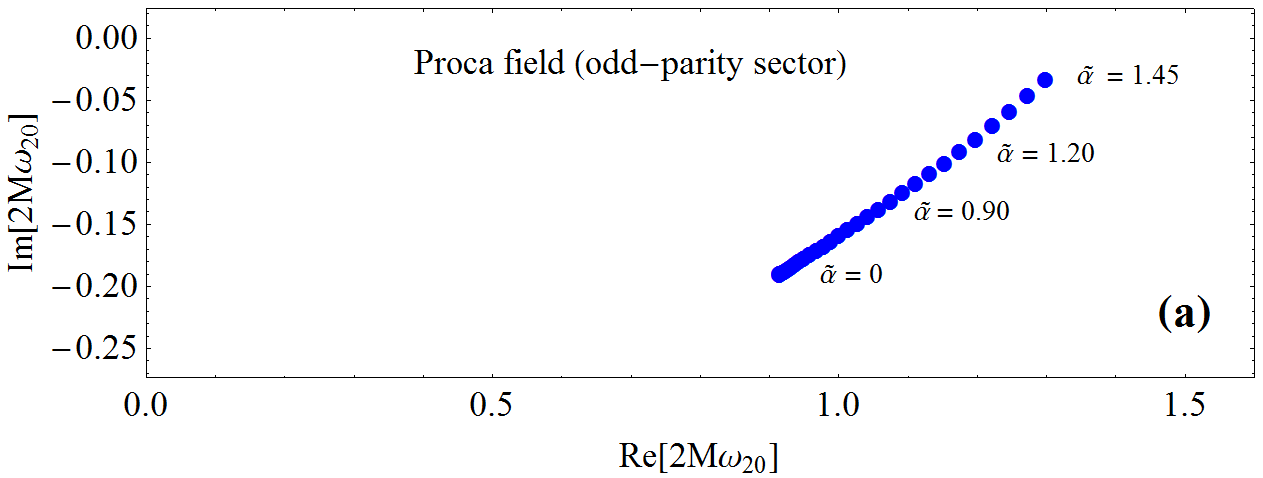}
\centering
\vspace{0.2cm}
\includegraphics[height=4cm,width=8.5cm]{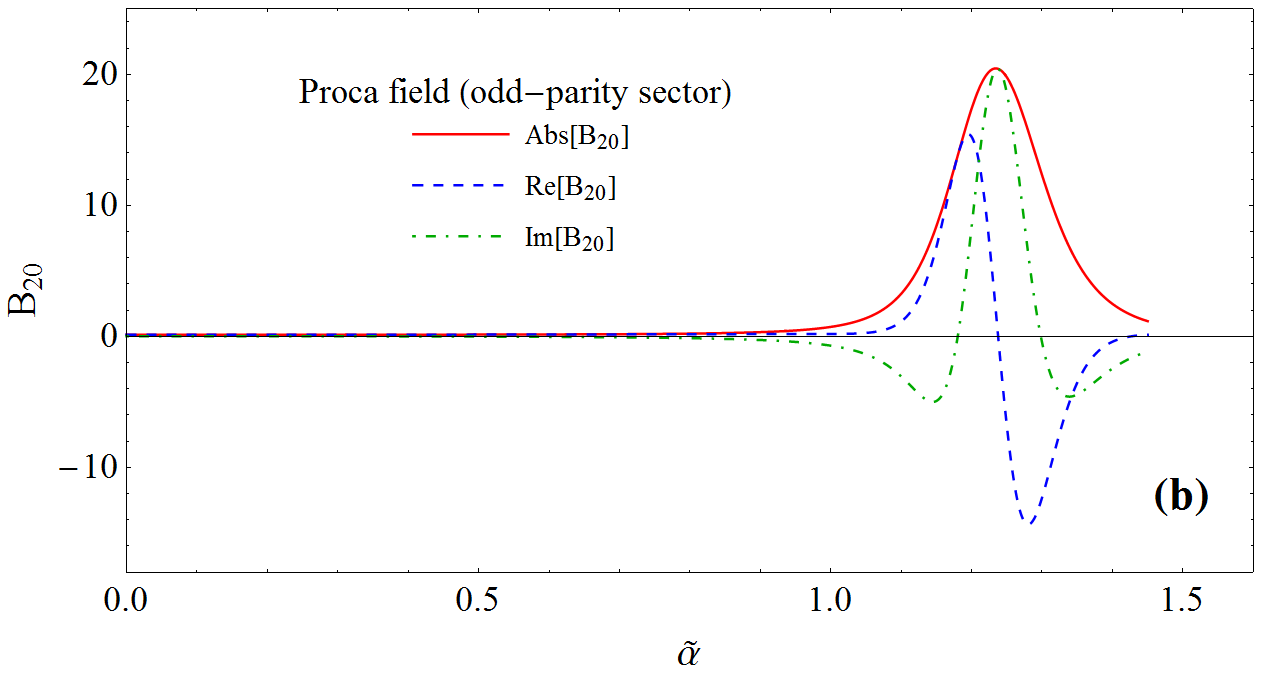}
\centering
\vspace{0.2cm}
\includegraphics[height=4cm,width=8.5cm ]{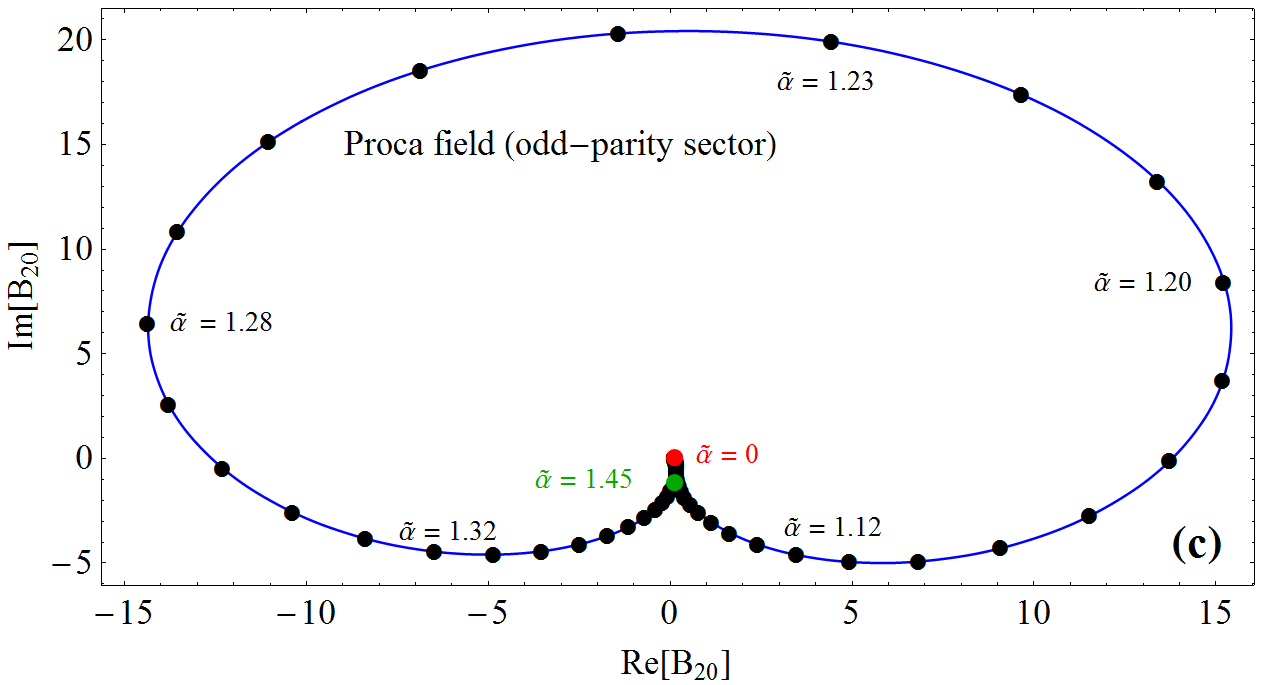}
\centering
\vspace{0.2cm}
\includegraphics[height=4cm,width=8.5cm ]{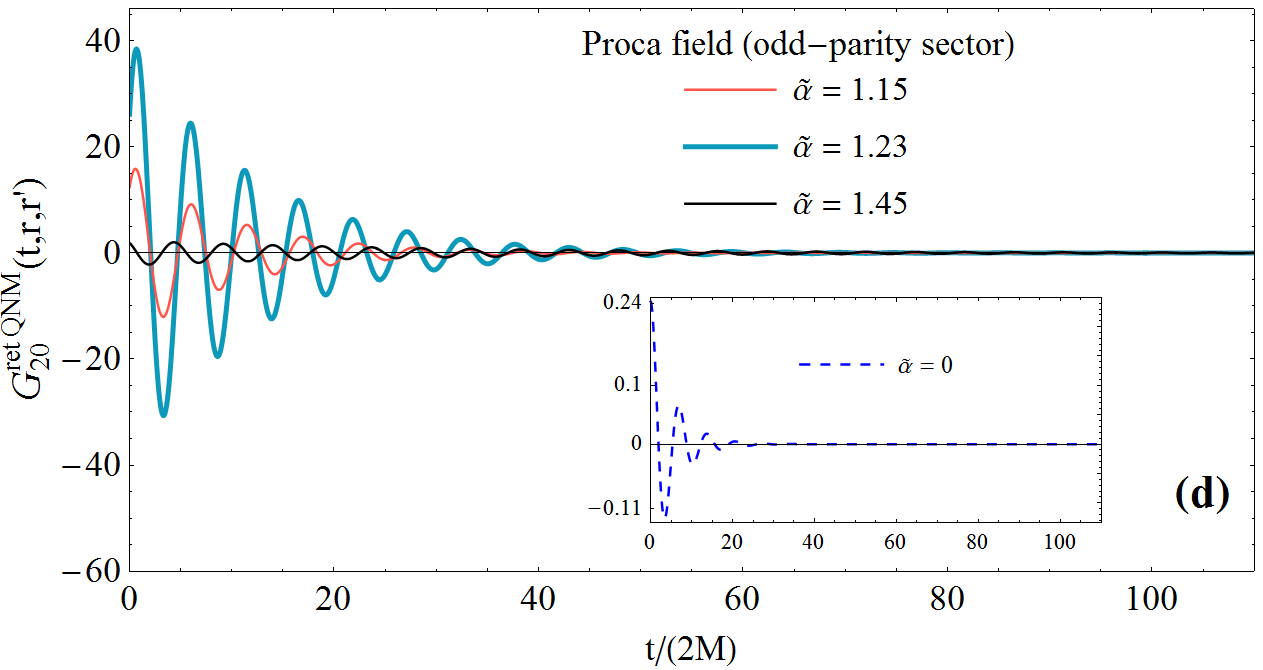}
\centering
\vspace{0.2cm}
\includegraphics[height=4cm,width=8.5cm ]{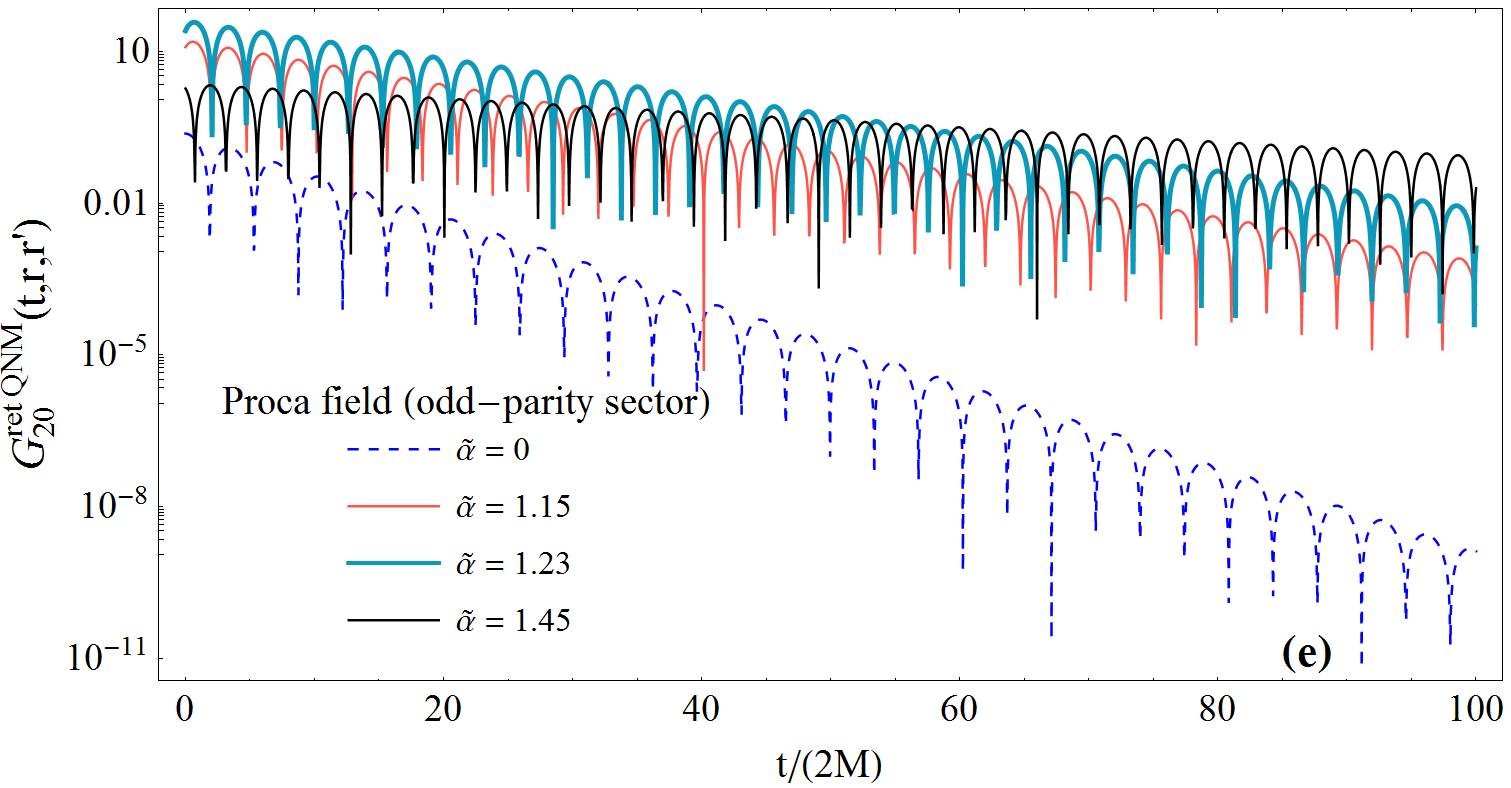}
\setlength\abovecaptionskip{0.25ex}
\vspace{-0.3cm}
\caption{\label{fig:QNM20_s1odd_INT} The odd-parity $(\ell = 2, n = 0)$ QNM of the Proca field. (a) The complex quasinormal frequency $2M\omega_{20}$ for $\tilde\alpha = 0, 0.05,\ldots, 1.40, 1.45$. (b) The resonant behavior of the excitation factor ${\cal B}_{20}$. (c) The excitation factor ${\cal B}_{20}$  for $\tilde\alpha = 0, 0.01,\ldots, 1.44, 1.45$. (d) and (e) Some intrinsic ringings corresponding to values of the mass near and above the critical value  $\tilde\alpha_{20}$. We compare them with the ringing corresponding to the massless vector field. The results are obtained from $(\ref{Gret_ell_QNM})$ with $r = 50M$ and $r' = 10M$.}
\end{figure}

\begin{figure}
\centering
\includegraphics[height=3cm,width=8.5cm]{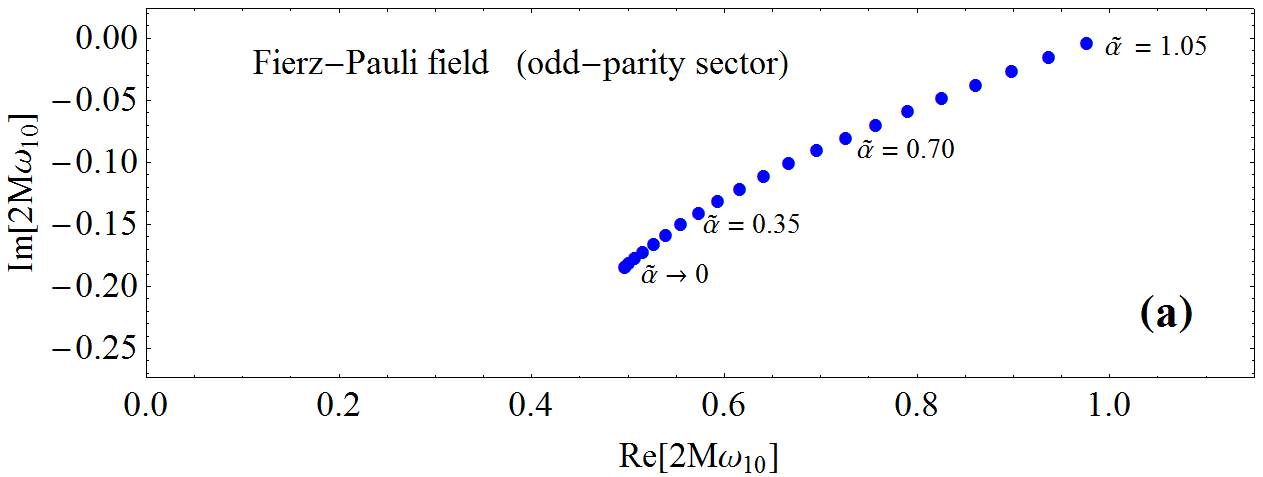}
\centering
\vspace{0.2cm}
\includegraphics[height=4cm,width=8.5cm]{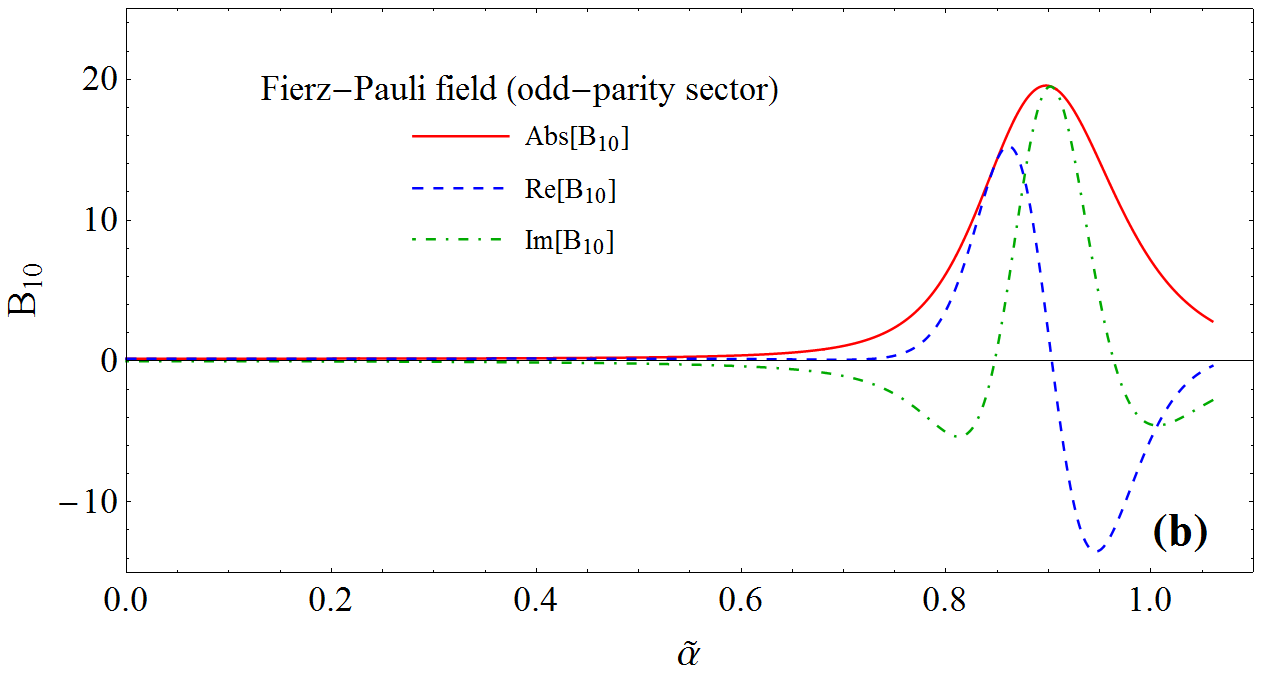}
\centering
\vspace{0.2cm}
\includegraphics[height=4cm,width=8.5cm ]{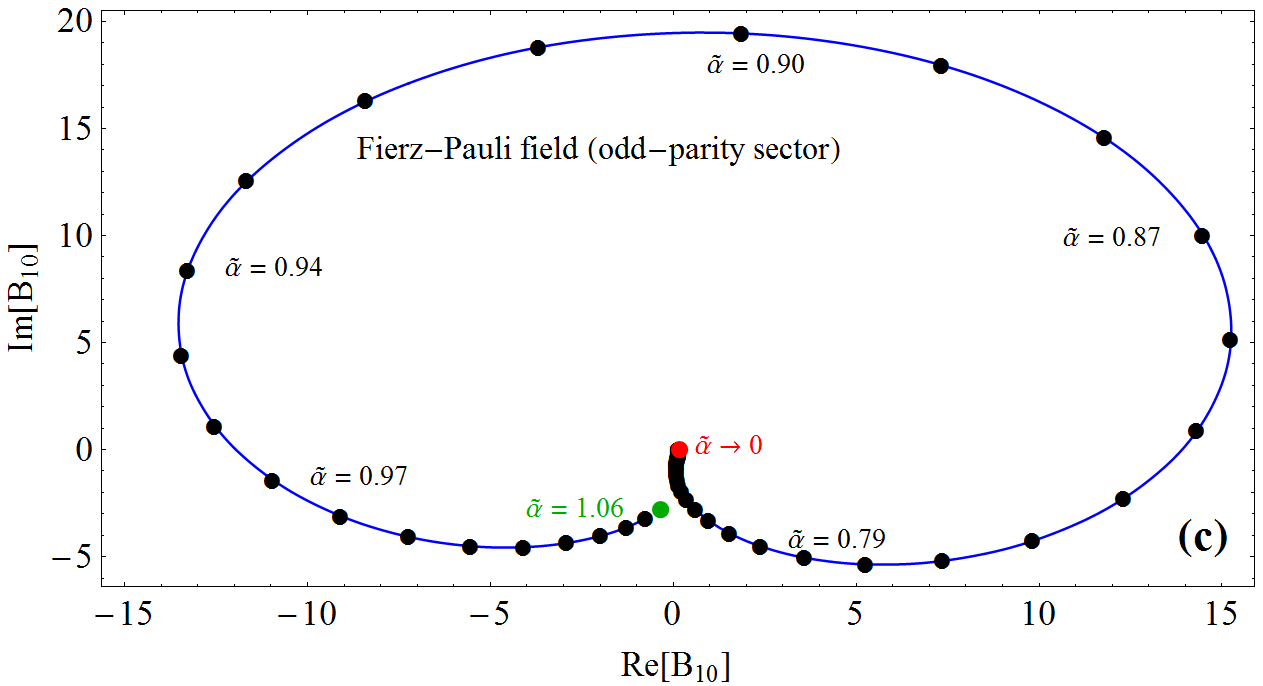}
\centering
\vspace{0.2cm}
\includegraphics[height=4cm,width=8.5cm ]{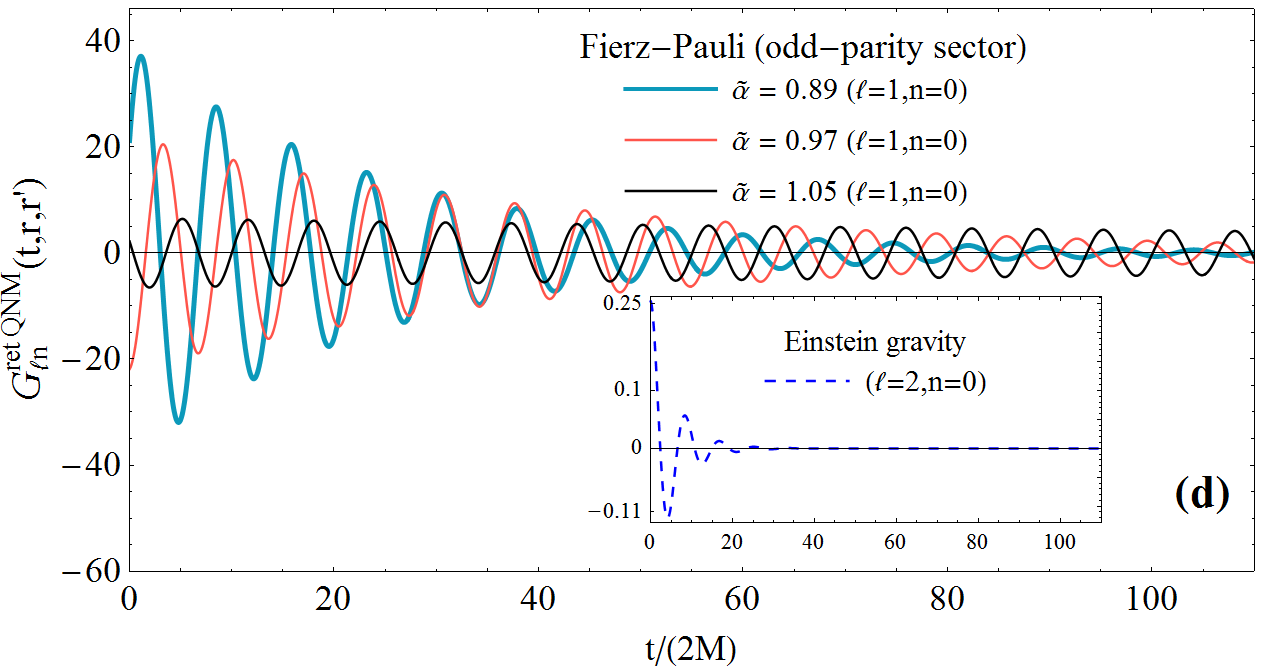}
\centering
\vspace{0.2cm}
\includegraphics[height=4cm,width=8.5cm ]{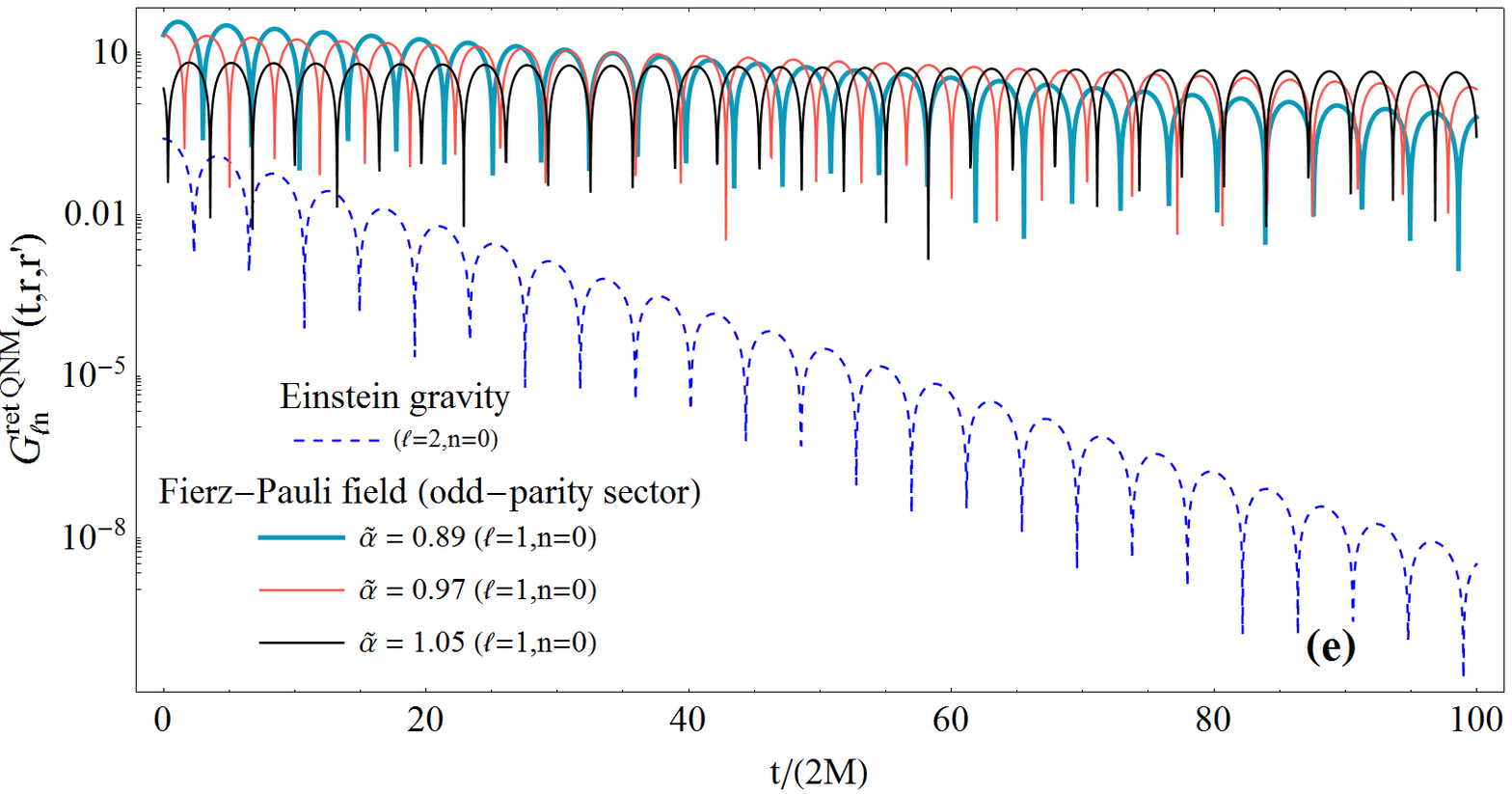}
\setlength\abovecaptionskip{0.25ex}
\vspace{-0.3cm}
\caption{\label{fig:QNM10_s2odd_INT} The odd-parity $(\ell = 1, n = 0)$ QNM of the Fierz-Pauli field. (a) The complex quasinormal frequency $2M\omega_{10}$ for $\tilde\alpha = 0, 0.05,\ldots, 1.00, 1.05$. (b) The resonant behavior of the excitation factor ${\cal B}_{10}$. (c) The excitation factor ${\cal B}_{10}$  for $\tilde\alpha = 0, 0.01,\ldots, 1.05, 1.06$. (d) and (e) Some intrinsic ringings corresponding to values of the mass near and above the critical value  $\tilde\alpha_{10}$. We compare them with the ringing corresponding to the odd-parity $(\ell = 2, n = 0)$ QNM of the massless spin-2 field. The results are obtained from $(\ref{Gret_ell_QNM})$ with $r = 50M$ and $r' = 10M$.}
\end{figure}

The quasinormal frequencies $\omega_{\ell n}$ can be determined by using the method developed by Leaver \cite{Leaver:1985ax} for massless theories and extended to massive fields by Konoplya and Zhidenko \cite{Konoplya:2004wg}. We have numerically implemented this method by modifying the Hill determinant approach of Majumdar and Panchapakesan \cite{mp}. The excitation factors ${\cal B}_{\ell n}$ can be obtained by integrating numerically the Regge-Wheeler equation (\ref{RW}) for $\omega=\omega_{\ell n}$ with the Runge-Kutta method and then by comparing its solution to asymptotic expansions with ingoing and outgoing behavior at spatial infinity. In order to obtain stable results for ``large" values of the mass parameter, it has been necessary to decode, by Pad\'e summation, the information hidden in the divergent part of the asymptotic expansions considered (see, e.g., Chap.~8 of Ref.~\cite{BenderOrszag1978} for information on Pad\'e summation).

In Fig.~\ref{fig:QNM20_INT}, we consider the $(\ell=2,n=0)$ QNM of the massive scalar field [it should be noted that the resonant effects we shall describe for this particular QNM also exists for the $(\ell=0,n=0)$ and $(\ell=1,n=0)$ QNMs but are much more attenuated]. We display the effect of the mass on its complex quasinormal frequency $\omega_{20}$ [see Fig.~\ref{fig:QNM20_INT}(a)] and on the associated quasinormal excitation factor ${\cal B}_{20}$ [see Figs.~\ref{fig:QNM20_INT}(b) and \ref{fig:QNM20_INT}(c)]. In Fig.~\ref{fig:QNM20_INT}(b), we can observe the strong resonant behavior of ${\cal B}_{20}$. It occurs around the critical value ${\tilde \alpha}_{20} \approx 1.28447$ and it is important to note that near, above and ``far above" ${\tilde \alpha}_{20}$ this QNM is weakly and even very weakly damped [see Fig.~\ref{fig:QNM20_INT}(a)]. As a consequence, for masses in a rather large range around ${\tilde \alpha}_{20}$, the BH ``intrinsic" ringings constructed from the quasinormal part (\ref{Gret_ell_QNM}) of the retarded Green function (\ref{Gret_om}) and which are generated by the scalar perturbation have a huge amplitude and are, moreover, slowly or even very slowly decaying [see Fig.~\ref{fig:QNM20_INT}(d) and Fig.~\ref{fig:QNM20_INT}(e)]. We compare them with the ringing generated by the massless scalar field which allows us to highlight the giant behavior of the ringings induced by the massive scalar field. It should be noted that we have plotted the ringings for $r = 50M$ and $r' = 10M$ but similar results can be obtained for various locations of the source and the observer. It is important to remark that, for ${\tilde \alpha} \to {\tilde \alpha}_d \approx 1.65...$, $\mathrm{Im} [\omega_{20}]$ and ${\cal B}_{20}$ vanish and that above ${\tilde \alpha} = {\tilde \alpha}_d$ the $(\ell=2,n=0)$ QNM disappears. As a consequence, as ${\tilde \alpha}$ increases from ${\tilde \alpha} =0$, the amplitude of the BH ringing increases, becomes huge in the large domain around ${\tilde \alpha}_{20} $ and then decreases when ${\tilde \alpha} \to {\tilde \alpha}_d$. However, it should be noted that when we explore the range ${\tilde \alpha} \to {\tilde \alpha}_d$, we encounter strong numerical instabilities and, in particular, it is very difficult to obtain numerically the vanishing of ${\cal B}_{20}$.

\begin{widetext}

\begin{table}

\caption{\label{tab:table1} Massive scalar field. A sample of the first quasinormal frequencies $\omega_{\ell n}$ and excitation factors ${\cal B}_{\ell n}$ for ${\tilde \alpha}=0$ (massless scalar field) and for ${\tilde \alpha}={\tilde \alpha}_{\ell n} $ (massive scalar field with the mass parameter corresponding to the maximum of the excitation factor). For a given angular momentum index $\ell$, only the excitation factors of the lowest overtones present a strong resonant behavior.}

\smallskip

\centering

\resizebox{\textwidth}{!}{%

\begin{tabular}{ccccrcccc}

\hline
\hline
& $(\ell,n)$  &  \multicolumn{1}{c}{$ {\tilde \alpha}_{\ell n}$} &

$2M \omega_{\ell n}$ &

\multicolumn{1}{c}{${\cal B}_{\ell n}$} & $|{\cal B}_{\ell n}|$ &  $2M \omega_{\ell n}$  & \multicolumn{1}{c} {$ {\cal B}_{\ell n}$ }& $  |{\cal B}_{\ell n}|$  \\

&   &   &  $  \mathrm{for} \, {\tilde \alpha}=0   $ &\multicolumn{1}{c}{ $  \mathrm{for} \, {\tilde \alpha}=0   $ }& $  \mathrm{for} \, {\tilde \alpha}=0   $

& $ \mathrm{for} \, {\tilde \alpha}={\tilde \alpha}_{\ell n}   $  &  $ \mathrm{for} \, {\tilde \alpha}={\tilde \alpha}_{\ell n}   $&$ \mathrm{for} \, {\tilde \alpha}={\tilde \alpha}_{\ell n}   $\\

\hline

& $(0,0)$  & $0.27999$ & $0.220910 - 0.209791i$ & $0.212349 - 0.059275i$ & $0.220467$  & $0.227442 - 0.179033i$    &\phantom{0} $0.263003 - 0.016875i$ & $0.263544$  \\

\hline

& $(1,0)$ & $0.77371$  & $0.585872 - 0.195320i$  & $-0.150670 + 0.022814i$ & $0.152388$ &  $0.721915 - 0.103844i$    &  $-0.7542 - 0.5041i $  & $0.907094$  \\

& $(1,1)$ & $0.74303$  & $0.528897 - 0.612515i$  & $0.028966 + 0.188821i$ & $0.191030$  &  $0.528356 - 0.524389i$    &   $-0.027649 + 0.261726i$  & $0.263182$ \\

\hline

& $(2,0)$ & $1.28447$  & $0.967288 - 0.193517i$  & $0.119355 + 0.013428i$ & $0.120108$  &  $1.240695 - 0.090415i$    & \phantom{00} $1.800 + 21.264i$ & $21.340064$   \\

& $(2,1)$ & $1.22223$  &  $0.927709 - 0.591196i$ & $0.035511 - 0.264290i$ & $0.266665$  &  $1.033952 - 0.403778 i$    & \phantom{0} $0.55839 - 0.63603i$ & $0.846367$  \\

& $(2,2)$ & $1.23193$  & $0.861088 - 1.017117i$  & $-0.286082 + 0.045925i$ & $0.289745$   &  $0.857086 - 0.868873i$    &  $-0.462994 - 0.063382i$ & $0.467312$  \\

\hline

& $(3,0)$ & $1.78928$  & $1.350732 - 0.192999i$  & $-0.093638 - 0.040471i$ & $0.102010$   &  $1.750318 - 0.087332i$    & \phantom{0}$1.094\times10^3 \,- 1.880\times 10^3i$ & $2.175612\times 10^3$   \\

& $(3,1)$ &  $1.75358$  & $1.321343 - 0.584570i$  & $-0.134112 + 0.294158i$ & $
0.323288$    &  $1.578868 - 0.333911i$   &  $-4.8046 - 0.3367i$ & $4.816410$  \\

& $(3,2)$  & $1.68072$  &  $1.267252 - 0.992016i$ & $0.487872 + 0.120390i$ & $0.608262$   & $1.355312 - 0.741016i$    &\phantom{0}  $1.08005 + 1.12461i$ & $1.559250$ \\

& $(3,3)$ & $1.72590$   & $1.197546 - 1.422442 i $ & $-0.092651 - 0.517332i $   & $0.525563$ & $1.190664 - 1.213516i$ &\phantom{0} $0.15209 - 0.98334i$ & $0.995033$  \\

\hline

& $(4,0)$ & $2.29161$  & $1.734831 - 0.192783i $  & $0.068044 + 0.059131i $ & $0.090147 $  & $2.255734 - 0.086470i $  & $ -5.54\times 10^5 + 5.02\times10^5$ & $7.476094\times10^5 $   \\

& $(4,1)$ & $2.28611$ & $1.711616 - 0.581782i$ & $0.247593 - 0.275444i$ & $0.370367$ & $2.122809 - 0.297760i$ &\phantom{00} $22.926 + 49.222i$ & $54.299074$    \\

& $(4,2)$  & $2.19818$ & $1.667384 - 0.980650i$ & $-0.621067 - 0.421239i$ & $0.750444$ & $1.890542 - 0.641944i$ &\phantom{00} $0.3739 - 6.7715i$ & $6.781840$    \\

& $(4,3)$ & $2.15102$ & $1.606576 - 1.394963i $ & $-0.289498 + 0.967789i $ & $1.010161$ & $1.684235 - 1.084401i $ &\phantom{0} $-2.58160 + 2.20255i $ & $3.393505$     \\

& $(4,4)$ & $2.22179$ & $1.535465 - 1.828039i $ & $ 1.020115 - 0.205096i$ & $1.040528 $ & $ 1.525887 - 1.558210i$ &\phantom{00} $2.28265\, + 0.38472i$ & $2.314843 $   \\

\hline
\hline

\end{tabular}%
}
\end{table}

\begin{table}

\caption{\label{tab:table2} Proca field (odd-parity sector). A sample of the first quasinormal frequencies $\omega_{\ell n}$ and excitation factors ${\cal B}_{\ell n}$ for ${\tilde \alpha}=0$ (massless vector field) and for ${\tilde \alpha}={\tilde \alpha}_{\ell n} $ (massive vector field with the mass parameter corresponding to the maximum of the excitation factor). For a given angular momentum index $\ell$, only the excitation factors of the lowest overtones present a strong resonant behavior.}

\smallskip

\centering

\resizebox{\textwidth}{!}{%

\begin{tabular}{ccccrcccc}

\hline
\hline
& $(\ell,n)$  &  \multicolumn{1}{c}{$ {\tilde \alpha}_{\ell n}$} &

$2M \omega_{\ell n}$ &

\multicolumn{1}{c}{${\cal B}_{\ell n}$} & $|{\cal B}_{\ell n}|$ &  $2M \omega_{\ell n}$  & \multicolumn{1}{c} {$ {\cal B}_{\ell n}$ }& $  |{\cal B}_{\ell n}|$  \\

&   &   &  $  \mathrm{for} \, {\tilde \alpha}=0   $ &\multicolumn{1}{c}{ $  \mathrm{for} \, {\tilde \alpha}=0   $ }& $  \mathrm{for} \, {\tilde \alpha}=0   $

& $ \mathrm{for} \, {\tilde \alpha}={\tilde \alpha}_{\ell n}   $  &  $ \mathrm{for} \, {\tilde \alpha}={\tilde \alpha}_{\ell n}   $&$ \mathrm{for} \, {\tilde \alpha}={\tilde \alpha}_{\ell n}   $\\

\hline

& $(1,0)$ & $0.68831$  & $0.496527\, -0.184975 i$  & $-0.1614\, +0.0119 i$ & $0.1618$ &  $0.634396\, -0.086904 i$    &  $-0.7650\, -0.4629 i $  & $0.8942$  \\

& $(1,1)$ & $0.70276$  & $0.429031\, -0.587335 i$  & $0.0118\, +0.1809 i$ & $0.1813$  &  $0.417568\, -0.493170 i$    &   $-0.0428\, +0.2496 i$  & $0.2533$ \\

\hline

& $(2,0)$ & $1.23422$  & $0.915191\, -0.190009 i$  & $0.1212\, +0.0186 i$ & $0.1226$  &  $1.190479\, -0.084580 i$    & \phantom{00} $1.9966\, +20.3582 i$ & $20.4559$   \\

& $(2,1)$ & $1.17146$  &  $0.873085\, -0.581420 i$ & $0.0466\, -0.2590 i$ & $0.2631$  &  $0.974408\, -0.390809 i$    & \phantom{0} $0.5426\, -0.6172 i$ & $0.8218$  \\

& $(2,2)$ & $1.21000$  & $0.802373\, -1.003175 i$  & $-0.2729\, +0.0325 i$ & $0.2749$   &  $0.791564\, -0.851199 i$    &  $-0.4409\, -0.0729 i$ & $0.4469$  \\

\hline

& $(3,0)$ & $1.75351$  & $1.313797\, -0.191232 i$  & $-0.0934\, -0.0435 i$ & $0.1031$   &  $1.714716\, -0.084437 i$    & \phantom{0}$1.0184\times10^3\, -1.8006\times10^3 i$ & $2.0686\times10^3$   \\

& $(3,1)$ &  $1.71623$  & $1.283475\, -0.579457 i$  & $-0.1419\, +0.2883 i$ & $
0.3214$    &  $1.539459\, -0.326015 i$   &  $-4.6621\, -0.2973 i$ & $4.6716$  \\

& $(3,2)$  & $1.64677$  &  $1.227664\, -0.984133 i$ & $0.4725\, +0.1314 i$ & $0.4905$   & $1.311423\, -0.731576 i$    &\phantom{0}  $1.0402\, +1.0982 i$ & $1.5127$ \\

\hline
\hline

\end{tabular}%
}
\end{table}

\begin{table}

\caption{\label{tab:table3} Proca field (even-parity $\ell=0$ QNMs). A sample of the first quasinormal frequencies $\omega_{\ell n}$ and excitation factors ${\cal B}_{\ell n}$ for ${\tilde \alpha} \to 0$ and for ${\tilde \alpha}={\tilde \alpha}_{\ell n} $ (Proca field with the mass parameter corresponding to the maximum of the excitation factor). For a given angular momentum index $\ell$, only the excitation factors of the lowest overtones present a strong resonant behavior.}

\smallskip

\centering

\resizebox{\textwidth}{!}{%

\begin{tabular}{ccccrcccccc}

\hline
\hline
& $(\ell,n)$  &  \multicolumn{1}{c}{$ {\tilde \alpha}_{\ell n}$} &

$2M \omega_{\ell n}$ &

\multicolumn{1}{c}{${\cal B}_{\ell n}$} & $|{\cal B}_{\ell n}|$ &  $2M \omega_{\ell n}$  & \multicolumn{1}{c} {$ {\cal B}_{\ell n}$ }& \multicolumn{1}{c} {$|{\cal B}_{\ell n}|$}&  \\

&   &   &  $  \mathrm{for} \, {\tilde \alpha}\rightarrow0   $ &\multicolumn{1}{c}{ $  \mathrm{for} \, {\tilde \alpha}\rightarrow0   $ }& $  \mathrm{for} \, {\tilde \alpha}\rightarrow0   $

& $ \mathrm{for} \, {\tilde \alpha}={\tilde \alpha}_{\ell n}   $  &  $ \mathrm{for} \, {\tilde \alpha}={\tilde \alpha}_{\ell n}   $&$ \mathrm{for} \, {\tilde \alpha}={\tilde \alpha}_{\ell n}   $\\

\hline

& $(0,0)$ & $0.47912$  & $0.220910\, -0.209791 i$  & $-0.2123\, +0.0593 i$ & $0.2205$  &  $0.426995\, -0.033401 i$    &\hphantom{0} $-0.7147\, -0.1121 i$ & $0.7234$   \\

& $(0,1)$ & $0.26048$  & $0.172234\, -0.696105 i$ & $0.0635\, +0.0661 i$ & $0.0917$  &$0.159227\, -0.682980 i$    & \hphantom{000}$0.0612\, +0.0696 i$ & $0.0926$  \\

\hline
\hline

\end{tabular}%
}
\end{table}

\begin{table}

\caption{\label{tab:table4} Fierz-Pauli field (odd-parity $\ell=1$ QNMs). A sample of the first quasinormal frequencies $\omega_{\ell n}$ and excitation factors ${\cal B}_{\ell n}$ for ${\tilde \alpha} \to 0$ and for ${\tilde \alpha}={\tilde \alpha}_{\ell n} $ (Fierz-Pauli field with the mass parameter corresponding to the maximum of the excitation factor). For a given angular momentum index $\ell$, only the excitation factors of the lowest overtones present a strong resonant behavior.}

\smallskip

\centering

\resizebox{\textwidth}{!}{%

\begin{tabular}{ccccrcccrcc}

\hline
\hline
& $(\ell,n)$  &  \multicolumn{1}{c}{$ {\tilde \alpha}_{\ell n}$} &

$2M \omega_{\ell n}$ &

\multicolumn{1}{c}{${\cal B}_{\ell n}$} & $|{\cal B}_{\ell n}|$ &  $2M \omega_{\ell n}$  & \multicolumn{1}{c} {$ {\cal B}_{\ell n}$ }& \multicolumn{1}{c} {$|{\cal B}_{\ell n}|$}&  \\

&   &   &  $  \mathrm{for} \, {\tilde \alpha}\rightarrow0   $ &\multicolumn{1}{c}{ $  \mathrm{for} \, {\tilde \alpha}\rightarrow0   $ }& $  \mathrm{for} \, {\tilde \alpha}\rightarrow0   $

& $ \mathrm{for} \, {\tilde \alpha}={\tilde \alpha}_{\ell n}   $  &  $ \mathrm{for} \, {\tilde \alpha}={\tilde \alpha}_{\ell n}   $&$ \mathrm{for} \, {\tilde \alpha}={\tilde \alpha}_{\ell n}   $\\

\hline

& $(1,0)$ & $0.89757$  & $0.496527\, -0.184975 i$  & $0.1614\, -0.0119 i$ & $0.1618$  &  $0.859691\, -0.038782 i$    &\hphantom{0} $3.2524\, +19.2819 i$ & $19.5543$   \\

& $(1,1)$ & $0.82081$  & $0.429031\, -0.587335 i$ & $-0.0118\, -0.1809 i$ & $0.1813$  &  $0.397013\, -0.261246 i$    & \hphantom{0}$0.1969\, -0.4443 i$ & $0.4860$  \\

& $(1,2)$ & $1.08214$  & $0.349547\, -1.050375 i$  & $-0.0816\, +0.0721 i$ & $0.1089$   &  $0.223343\, -0.920334 i$    &  $-0.1215\, +0.0669 i$ & $0.1387$  \\

\hline
\hline

\end{tabular}%
}
\end{table}

\end{widetext}

In Figs.~\ref{fig:QNM30_INT} and \ref{fig:QNM31_INT}, we respectively consider the $(\ell=3,n=0)$ QNM and the $(\ell=3,n=1)$ QNM of the massive scalar field. {\it Mutatis mutandis}, all the effects already noted for the $(\ell=2,n=0)$ QNM are still present and we shall not discuss them again. We can, however, note that the maximum of the quasinormal excitation factor seems to increase rapidly with the angular momentum index $\ell$, a rather surprising behavior, and to decrease with the overtone index $n$. In Table~\ref{tab:table1}, where we consider a rather large sample of QMNs, we confirm these behaviors. As a consequence, for a given value of $\ell$, the giant BH ringings corresponding to the fundamental tone are the most interesting ones and, moreover, as $\ell$ increases, their amplitudes rapidly increase. We are quite disturbed by this last result which we are not able to explain simply.

In Fig.~\ref{fig:QNM20_s1odd_INT} and in Table~\ref{tab:table2}, we consider the odd-parity QNMs of the Proca field. We can observe that their behavior (or, more precisely, the behavior of the complex quasinormal frequencies, of the quasinormal excitation factors and of the ringings) is rather similar to that of the QNMs of the massive scalar field.

In Table~\ref{tab:table3} , we consider the even-parity $\ell=0$ QNMs of the Proca field. We can observe that the resonant behavior of the quasinormal excitation factors is little pronounced, even for the fundamental tone.

In Fig.~\ref{fig:QNM10_s2odd_INT} and in Table~\ref{tab:table4}, we consider the odd-parity $\ell=1$ QNMs of the Fierz-Pauli field and we observe that they presents lot of similarities with the $\ell=2$ QNMs of the massive scalar field and the odd-parity $\ell=2$ QNMs of the Proca field. We can note, in particular, the strong resonant behavior of the quasinormal excitation factor of the fundamental tone and the associated giant ringings. As already discussed in Ref.~\cite{Decanini:2014kha}, such ringings could have fascinating observational consequences.

\section{Resonant behavior of the quasinormal excitation factors : A semiclassical analysis.}

\begin{figure}
\centering
\vspace{0.2cm}
\includegraphics[height=4cm,width=8.5cm ]{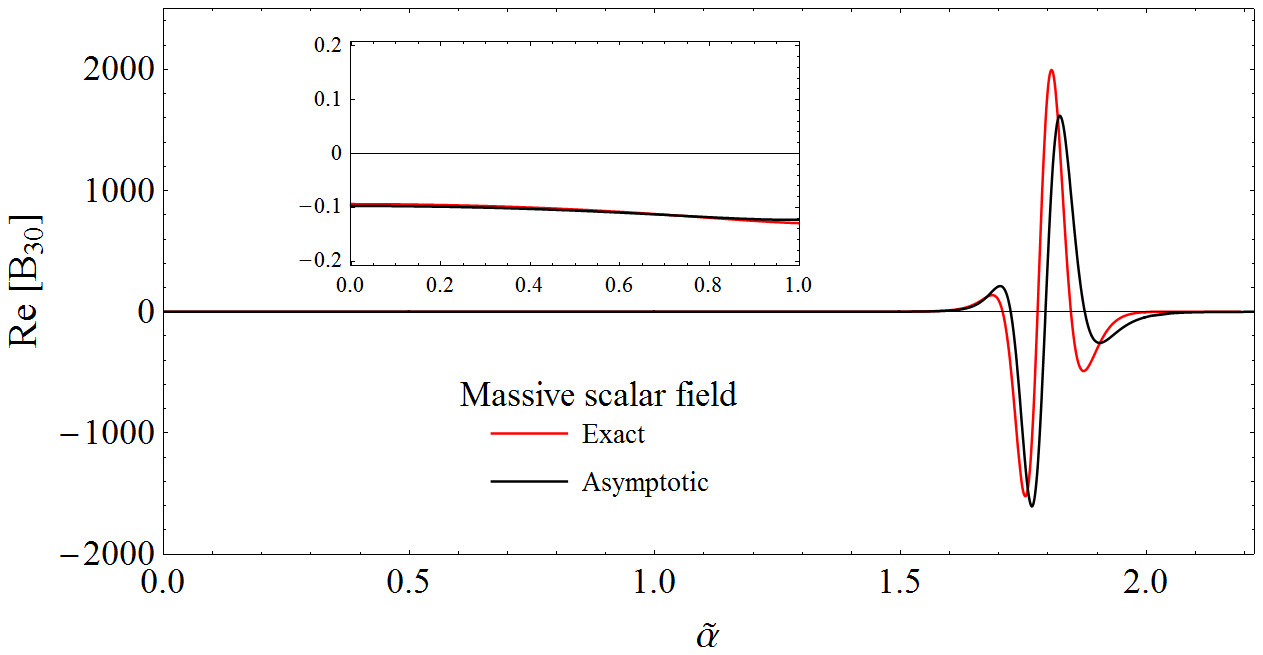}
\centering
\vspace{0.2cm}
\includegraphics[height=4cm,width=8.5cm ]{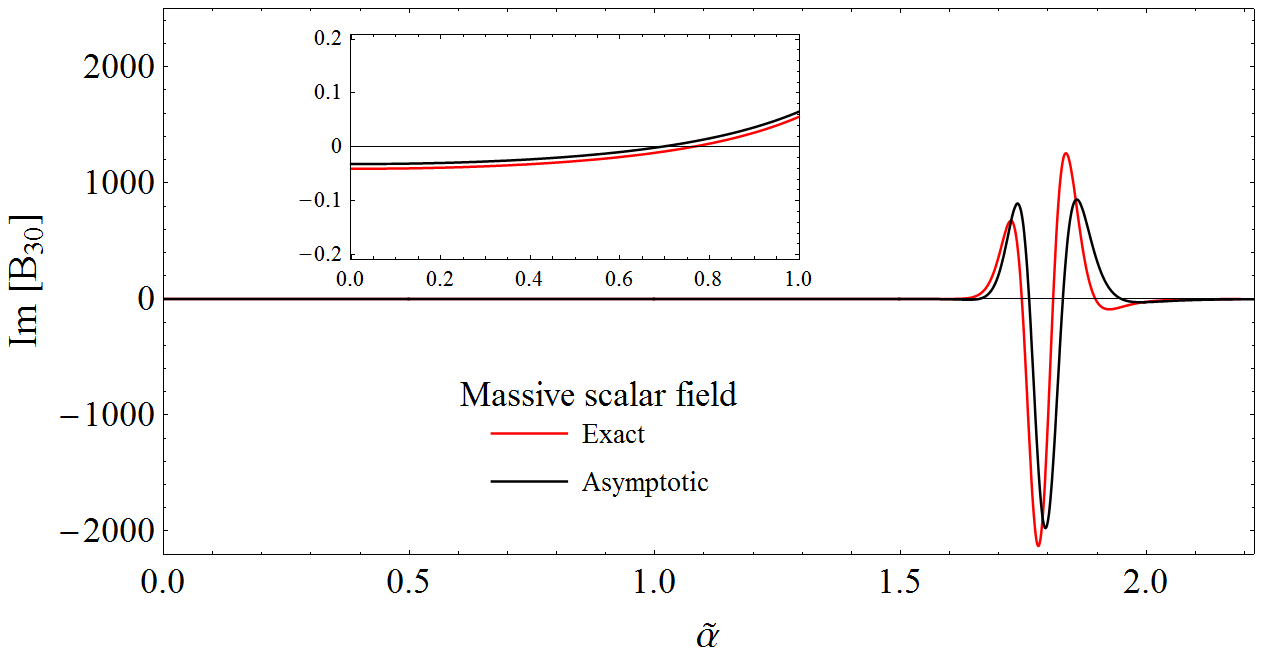}
\vspace{-0.3cm}
\caption{\label{fig:BLN30} The quasinormal excitation factor of the $(\ell = 3, n = 0)$ QNM of the scalar field. Exact and asymptotic behaviors.}
\end{figure}

\begin{figure}
\centering
\vspace{0.2cm}
\includegraphics[height=4cm,width=8.5cm ]{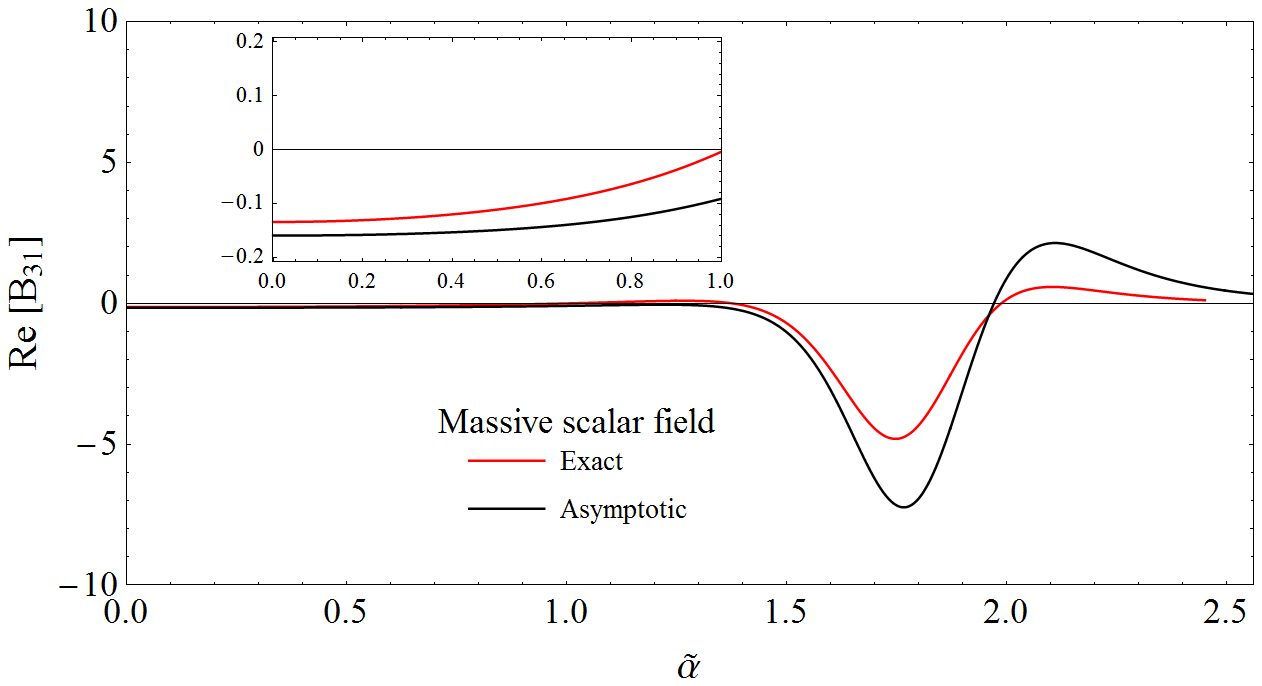}
\centering
\vspace{0.2cm}
\includegraphics[height=4cm,width=8.5cm ]{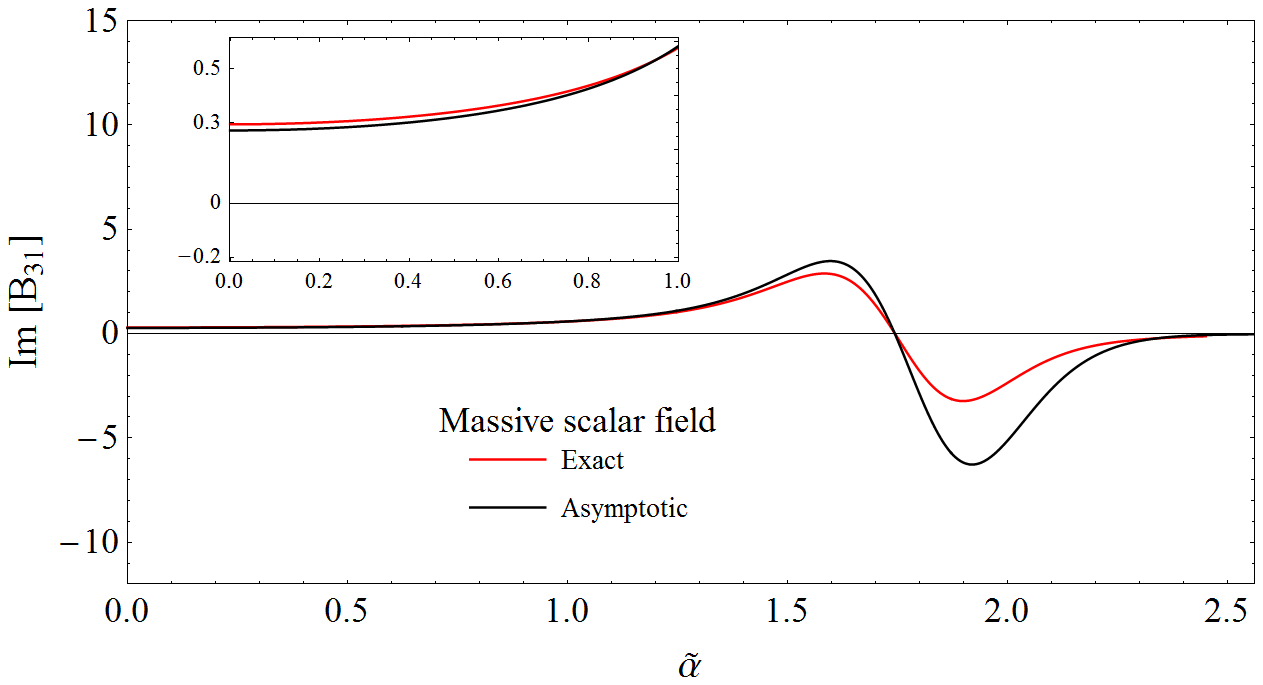}
\vspace{-0.3cm}
\caption{\label{fig:BLN31} The quasinormal excitation factor of the $(\ell = 3, n = 1)$ QNM of the scalar field. Exact and asymptotic behaviors.}
\end{figure}

\begin{figure}
\centering
\vspace{0.2cm}
\includegraphics[height=4cm,width=8.5cm ]{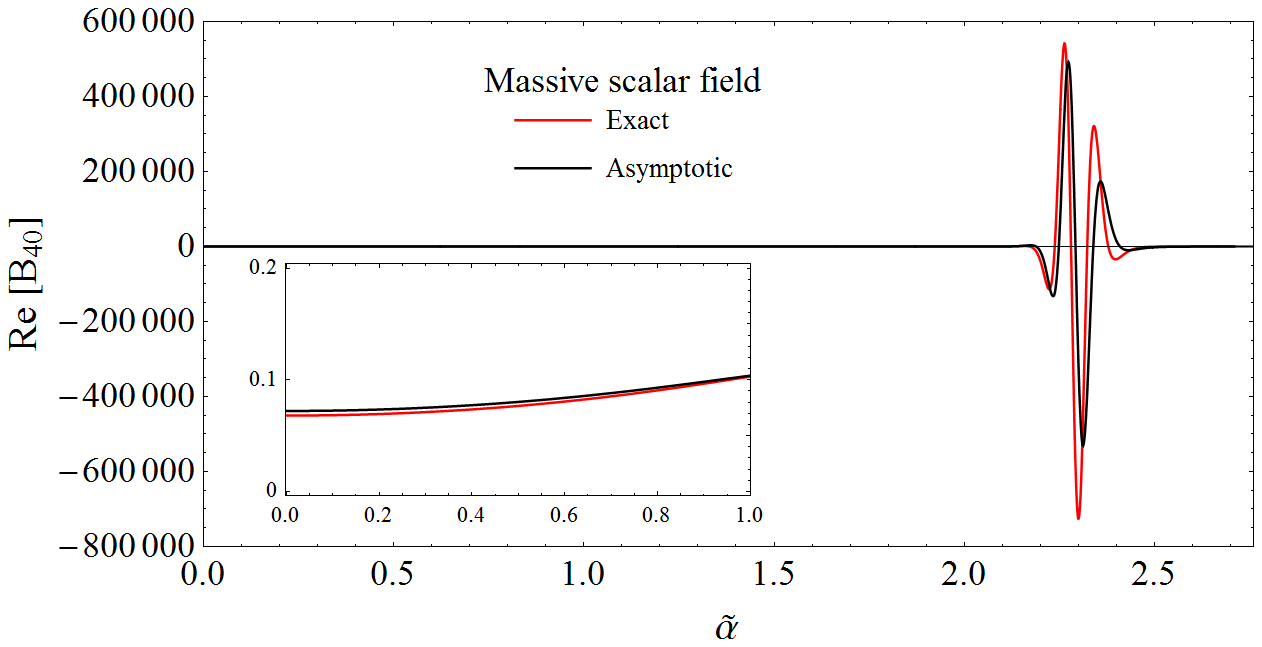}
\centering
\vspace{0.2cm}
\includegraphics[height=4cm,width=8.5cm ]{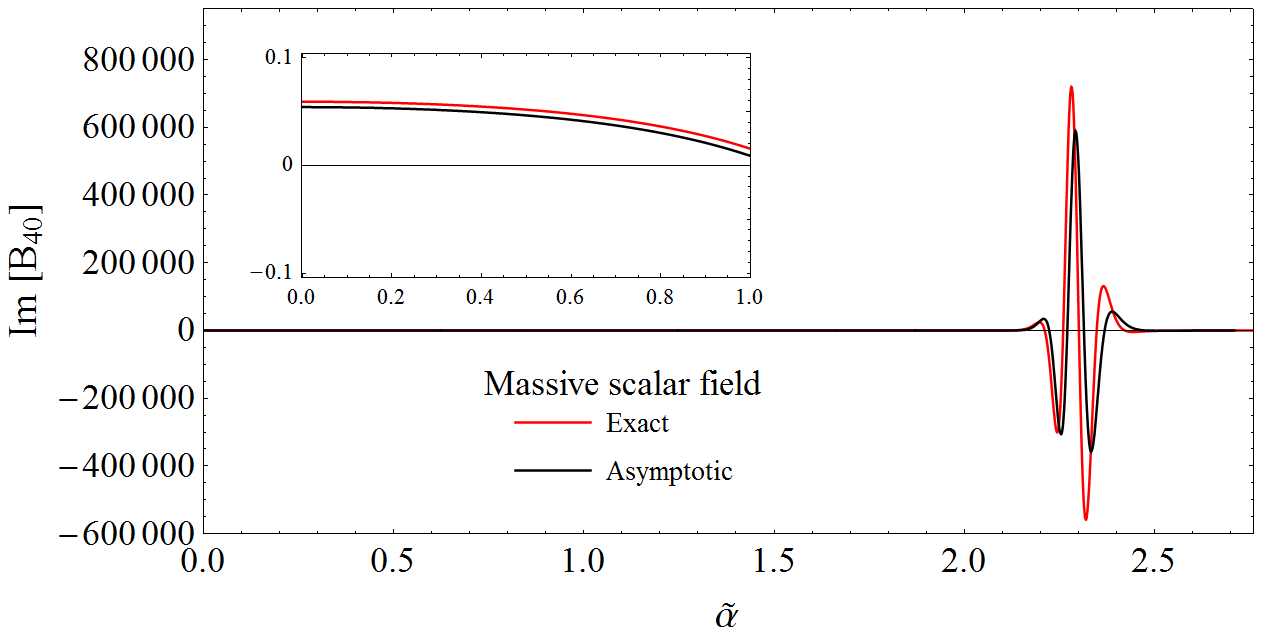}
\vspace{-0.3cm}
\caption{\label{fig:BLN40} The quasinormal excitation factor of the $(\ell = 4, n = 0)$ QNM of the scalar field. Exact and asymptotic behaviors.}
\end{figure}

It is well known that the weakly damped QNMs of BHs which are associated with massless fields can be interpreted in terms of waves trapped close to the so-called photon sphere, i.e., the hypersurface on which a massless particle can orbit the BH on unstable circular null geodesics. This appealing interpretation has been suggested a long time ago by Goebel \cite{1972ApJ...172L..95G}, has been implemented in various articles (see, e.g., Refs.~\cite{Ferrari:1984zz,Vanzo:2004fy,Cardoso:2008bp} for an approach based on eikonal considerations and Refs.~\cite{Decanini:2002ha,Decanini:2009mu,Decanini:2010fz,Decanini:2011eh} for an approach based on Regge pole techniques) and allows us to provide analytical approximations for the complex frequencies of the QNMs. In a recent work \cite{Dolan:2009nk}, Dolan and Ottewill introduced a novel and powerful ansatz for the QNMs of spherically symmetric BHs \cite{Dolan:2009nk}. This ansatz, which agrees with Goebel's interpretation, permitted them not only to determine the quasinormal frequencies but also to obtain an analytical expression for the quasinormal excitation factors ${\cal B}_{\ell n}$ of the Schwarzschild BH formally valid for large values of the angular momentum index $\ell$ \cite{Dolan:2011fh}. In this section, we briefly explain how to extend the Dolan-Ottewill approach to massive fields and we then give an approximation for the quasinormal excitation factors discussed in Sec.~II. It should be noted that we do not intend to enter into the technical aspects of its derivation because, {\it mutatis mutandis}, we extend ``trivially" the calculations presented in Appendix A of Ref.~\cite{Dolan:2011fh}. We finally show that the semiclassical formula obtained describes very correctly the resonant behavior of the quasinormal excitation factors. It is important to note that, in this section, we only focus on the modes for which the large $\ell$ limit can be taken, i.e., on those of the scalar field and those in the odd-parity sector of the Proca field.

It is crucial to remark that two geometrical parameters are involved in the Dolan-Ottewill ansatz [see Eqs.~(18) and (A13) of Ref.~\cite{Dolan:2011fh} as well as the discussion following Eq.~(18) and, for more precisions and physical motivations, Sec.~5 of Ref.~\cite{Dolan:2009nk}] : the ``radius" $r_c=3M$ of the photon sphere and the corresponding impact parameter $b_c=3 \sqrt{3} M$. For massive fields, the situation is much more complicated (see also Ref.~\cite{Decanini:2011eh} where the Dolan-Ottewill method has been considered in the context of Regge pole techniques). Indeed, the corresponding geometrical parameters depend not only on the BH mass $M$ but also on the rest mass $\mu$ and energy $\omega$ of the particle associated with the massive field considered (see, e.g., Ref.~\cite{Unruh:1976fm}). The sphere on which the massive particle can orbit the BH on unstable
circular timelike geodesics is located at $r=r_c(\omega) \in ]3M,4M[$ given by
\begin{equation} \label{crit_massive1}
r_c(\omega)=2M  \left( \frac{3+\left(1+8v^2(\omega)\right)^{1/2}}
{1+\left(1+8v^2(\omega)\right)^{1/2}} \right).
\end{equation}
Here
\begin{equation} \label{vitesse}
v(\omega)=\sqrt{1-\frac{\mu^2}{\omega^2}}
\end{equation}
denotes the particle speed at large distances from the BH which can be expressed in term of the particle momentum $p(\omega)=\sqrt{\omega^2-\mu^2}$ by $v(\omega)=p(\omega)/\omega$. The critical radius $r_c(\omega)$ defines an associated critical impact parameter
\begin{eqnarray}\label{crit_massive2}
&& b_c(\omega)=\frac{M}{\sqrt{2} \, v^2(\omega)}\left[
8v^4(\omega)+20v^2(\omega)-1  \phantom{{\left(v^2(\omega)\right)^{3/2}}}  \right. \nonumber \\
&&  \qquad \qquad \qquad \qquad \qquad \left.
+\left(1+8v^2(\omega)\right)^{3/2} \right]^{1/2}.
\end{eqnarray}
We recall that any massive particle sent toward the Schwarzschild BH with an impact
parameter $b<b_c(\omega)$ is captured while particles with impact parameter
$b>b_c(\omega)$ are scattered. It should be also noted that, for $\mu = 0$, $v(\omega)=1$ and from
Eqs.~(\ref{crit_massive1}) and (\ref{crit_massive2}) we can then recover the
parameters used by Dolan and Ottewill.

These preliminary considerations permit us to understand that, in order to describe the QNMs of massive fields and to derive their excitation factors, the Dolan-Ottewill ansatz (A13) of Ref.~\cite{Dolan:2011fh} must be replaced by
\begin{widetext}
\begin{equation} \label{DO-Ansatz1a}
\phi^{\pm}_{\omega  \ell}(r)=  \exp\left[{\pm}i p(\omega)
\int^{r_\ast} \left(1+\frac{2M
b_c(\omega)^2/r_c(\omega)^2}{r'}\right)^{1/2}\left(1-\frac{r_c(\omega)}{r'}\right)dr'_\ast\right] v^{\pm}_{\omega  \ell}(r)
\end{equation}
\end{widetext}
where the functions $v^{\pm}_{\omega  \ell}(r)$ are assumed to be regular for $r \to 2M$, i.e., at the event horizon, and for $r \to +\infty $, i.e., at spatial infinity. It is then easy to show that the usual boundary conditions for the solutions of (\ref{RW}) with (\ref{pot_RW_Schw}) are automatically satisfied, i.e., that $\phi^{\pm}_{\omega  \ell}(r) \sim \exp[{\pm} i\omega r_\ast]$ for $r_\ast \to -\infty$ and $\phi^{\pm}_{\omega  \ell}(r) \sim \exp[{\pm} i\lbrace{p(\omega) r_\ast+[M\mu^2/p(\omega)] \ln(r/M)\rbrace}]$ for $r_\ast \to +\infty$.

{\it Mutatis mutandis}, the derivation of the quasinormal excitation factors ${\cal B}_{\ell n}$ can be realized by following the different steps of Appendix A of Ref.~\cite{Dolan:2011fh} and, more precisely, by using the standard WKB techniques as well as the usual matching procedures. After a tedious calculation, we obtain
\begin{widetext}
\begin{eqnarray}\label{asym_BLN}
&& {\cal B}_{\ell n}=\frac{i (\ell+1/2)^{-1}}{\sqrt{8\pi}  n!}\left( \frac{-216i(\ell +1/2)}{\xi}\right)^{n+1/2} \exp[2i p(\omega_{\ell n})\zeta_c(\omega_{\ell n})]
\end{eqnarray}
where $\xi= 7+4\sqrt{3}$ and with $\zeta_c(\omega)$ given by
\begin{eqnarray}\label{zetaC}
&& \frac{\zeta_c(\omega)}{2M}= \frac{b^2_c(\omega)}{2r^2_c(\omega)}-\frac{r_c(\omega)}{2M} \sqrt{1+\frac{2M b^2_c(\omega)}{r^3_c(\omega)}}-\left( \frac{b^2_c(\omega)}{2r^2_c(\omega)}-\frac{r_c(\omega)}{2M} +1 \right)\ln\left[\frac{b^2_c(\omega)}{2r^2_c(\omega)}+\frac{r_c(\omega)}{2M}+ \frac{r_c(\omega)}{2M} \sqrt{1+\frac{2M b^2_c(\omega)}{r^3_c(\omega)}} \right] \nonumber \\
&& \qquad + \left(\frac{r_c(\omega)}{2M}-1 \right)\sqrt{1+\frac{b^2_c(\omega)}{r^2_c(\omega)}}\ln\left[\left(\frac{r_c(\omega)}{2M}-1 \right) \left( \frac{b^2_c(\omega)}{2r^2_c(\omega)}+1+ \sqrt{1+\frac{b^2_c(\omega)}{r^2_c(\omega)}} \right) \right]\nonumber \\
&& \qquad - \left(\frac{r_c(\omega)}{2M}-1 \right)\sqrt{1+\frac{b^2_c(\omega)}{r^2_c(\omega)}}\ln \left[\frac{b^2_c(\omega)}{2r^2_c(\omega)}\left(\frac{r_c(\omega)}{2M}+1 \right)+\frac{r_c(\omega)}{2M}+\frac{r_c(\omega)}{2M}\sqrt{\left(1+\frac{2Mb^2_c(\omega)}{r^3_c(\omega)}
\right)\left( 1+\frac{b^2_c(\omega)}{r^2_c(\omega)} \right)}  \right] + \ln 2.\nonumber \\
&&
\end{eqnarray}
\end{widetext}
It is straightforward to show that, for $\mu \to 0$, $\zeta_c(\omega)=(3-3\sqrt{3}+4\ln 2 -3 \ln \xi)M$ and Eq.~(\ref{asym_BLN}) then reduces to Eq.~(31) of Ref~\cite{Dolan:2011fh}. It is also important (i) to recall that the Dolan-Ottewill approach we have extended for massive fields only provides, for the quasinormal excitation factors, a leading-order expansion in $\ell+1/2$ and (ii) to note that the spin dependence, or more precisely the coefficient $\beta$ in (\ref{pot_RW_Schw}), does not play any role.

\begin{figure}
\centering
\vspace{0.2cm}
\includegraphics[height=4cm,width=8.5cm ]{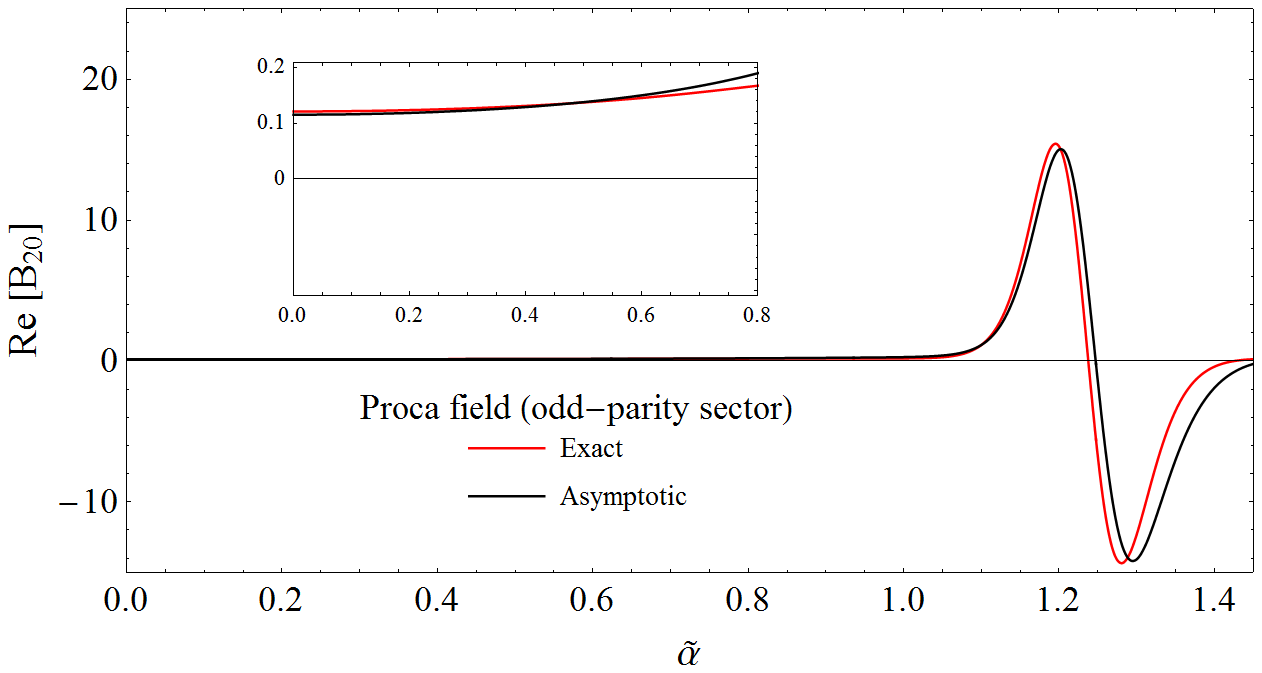}
\centering
\vspace{0.2cm}
\includegraphics[height=4cm,width=8.5cm ]{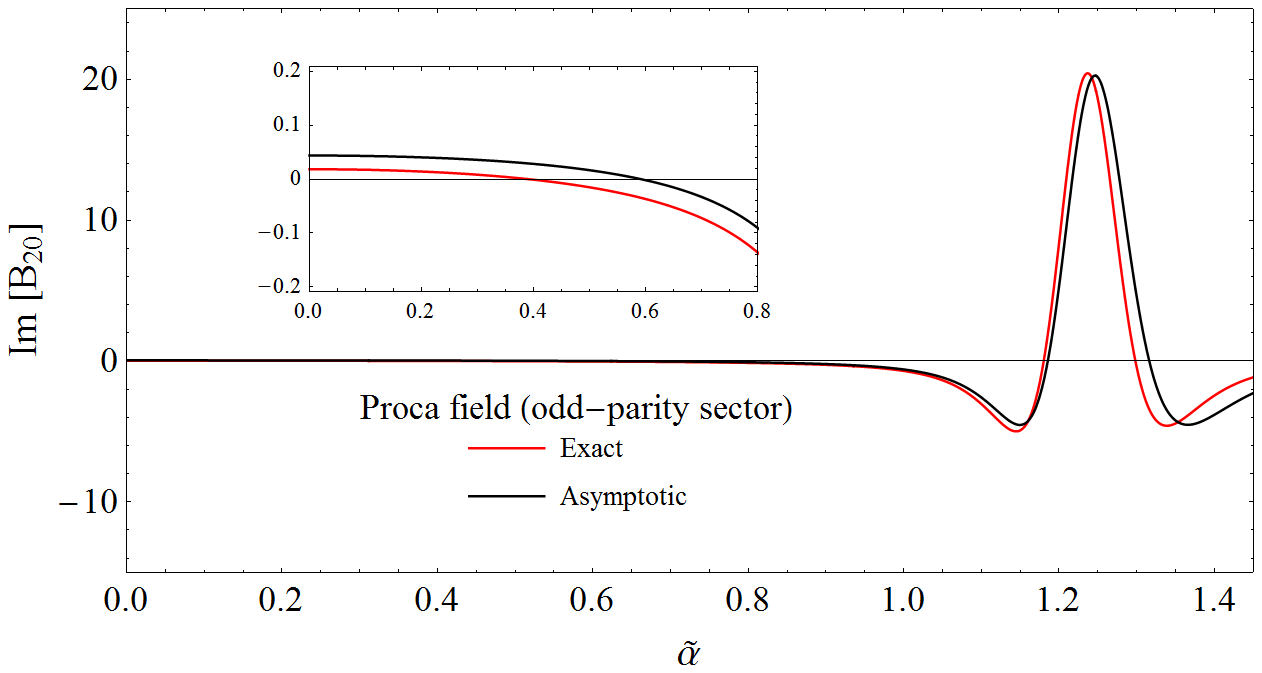}
\vspace{-0.3cm}
\caption{\label{fig:BLN20_Proca} The quasinormal excitation factor of the odd-parity $(\ell = 2, n = 0)$ QNM of the Proca field. Exact and asymptotic behaviors.}
\end{figure}

In Figs.~(\ref{fig:BLN30}), (\ref{fig:BLN31}) and (\ref{fig:BLN40}), we plot the behaviors of some quasinormal excitation factors. We consider the massive scalar field and the $(\ell = 3, n = 0)$, $(\ell = 3, n = 1)$ and $(\ell = 4, n = 0)$ QNMs. We compare the numerical results obtained in Sec.~II (``exact" results) with the asymptotic results provided by formula (\ref{asym_BLN}). It should be noted that we have put into this formula the ``exact" behavior of the complex quasinormal frequencies $\omega_{\ell n}$ we numerically obtained in Sec.~II. For the two fundamental QNMs considered [see Figs.~(\ref{fig:BLN30}) and (\ref{fig:BLN40})], the agreement is impressive. In particular, the strong resonant behavior of the quasinormal excitation factors is very well predicted. For the overtone QNM [see Fig.~(\ref{fig:BLN31})], the agreement remains satisfactory and is even very good for low values of the mass. This is also the case for the $(\ell=2,n=0)$ fundamental QNM while the quasinormal excitation factors of the $\ell=0$ and $1$ QNMs cannot be correctly described by formula (\ref{asym_BLN}). Similar considerations apply for the QNMs in the odd-parity sector of the Proca field and formula (\ref{asym_BLN}) can be also used efficiently in this context. It gives results that are even more accurate because the coefficient $\beta$, which has been neglected in the derivation of (\ref{asym_BLN}), vanishes for these QNMs. In Fig.~(\ref{fig:BLN20_Proca}), we focus on the quasinormal excitation factor of the odd-parity $(\ell = 2, n = 0)$ QNM of the Proca field. The agreement between the exact and asymptotic results is again impressive.

\section{Extrinsic giant ringings.}

The quasinormal frequencies $\omega_{\ell n}$ and the associated excitation factors ${\cal B}_{\ell n}$ discussed in the two previous sections are intrinsic properties of the BH interacting with a massive bosonic field. As a consequence, the extraordinary ringings we have exhibited in Sec.~II and which are constructed from the quasinormal retarded Green function (\ref{Gret_ell_QNM}) are not directly relevant to physics or astrophysics because they do not take into account the external mechanism which generates the BH distortion. In fact, it is necessary to check that if we consider a realistic perturbation of the BH, there still exists a giant ringing into the response obtained by convolution of the source of the perturbation with the retarded Green function. Of course, with astrophysical considerations in mind, it would be very interesting to consider as a source a ``particle" falling radially or plunging into the BH but, in the framework of massive field theories, this is a very difficult problem which, to our knowledge, has never been addressed. To simplify our purpose, we shall consider a more simple problem, but despite that, it will provide us with interesting information.

We consider that the BH perturbation is generated by an initial value problem with (nonlocalized) Gaussian initial data \cite{Leaver:1986gd,Andersson:1996cm,Berti:2006wq}. More precisely, we assume that the partial amplitude $\phi_\ell (t,r)$ solution of (\ref{Phi_ell1}) is given, at $t=0$, by
\begin{equation}\label{Cauchy_data}
\phi_\ell (t=0,r)= \phi_0 \exp \left[-\frac{a^2}{(2M)^2} [r_\ast (r) - r_\ast(r_0)]^2 \right]
\end{equation}
and, moreover, satisfies $\partial_t\phi_\ell (t=0,r)=0$. By Green's theorem and using (\ref{Phi_ell1}) and (\ref{Gret}), we can show that
\begin{equation}\label{TimeEvolution}
\phi_\ell (t,r)=\int \partial_t G_\ell^\mathrm{ret}(t;r,r') \phi_\ell (t=0,r')   dr'_\ast
\end{equation}
describes the time evolution of $\phi_\ell (t,r)$ for $t>0$. We can now insert (\ref{Gret_om}) into (\ref{TimeEvolution}) and deform again the contour of integration in the complex $\omega$ plane in order to capture the contribution of the QNMs. This allows us to isolate the BH ringing generated by the initial data. We have
\begin{equation}\label{TimeEvolution_QNM_sum}
\phi^\mathrm{QNM}_\ell (t,r)= \sum_n \phi^\mathrm{QNM}_{\ell n} (t,r)
\end{equation}
with
\begin{eqnarray}\label{TimeEvolution_QNM}
&& \phi^\mathrm{QNM}_{\ell n} (t,r)= 2 \, \mathrm{Re} \left[ i\omega_{\ell n} {\cal C}_{\ell n} \vphantom{e^{i[M\mu^2/p(\omega_{\ell n})] \ln(r/M)}} \right. \nonumber \\
&& \left. \qquad \times e^{-i[\omega_{\ell n} t - p(\omega_{\ell n})r_\ast - [M\mu^2/p(\omega_{\ell n})] \ln(r/M)]} \right].
\end{eqnarray}
Here ${\cal C}_{\ell n}$ denotes the excitation coefficient of the $(\ell,n)$ QNM. It is defined from the corresponding excitation factor ${\cal B}_{\ell n}$ but, in addition, it takes explicitly into account the role of the source of the BH perturbation. We have
 \begin{equation}\label{EC}
{\cal C}_{\ell n}={\cal B}_{\ell n} \int \frac{\phi_\ell (t=0,r') \phi^\mathrm{in}_{\omega_{\ell n} \ell}(r')}{\sqrt{\omega_{\ell n}/p(\omega_{\ell n})}A_\ell^{(+)}(\omega_{\ell n})}   dr'_\ast.
\end{equation}

\begin{figure}
\centering
\vspace{0.2cm}
\includegraphics[height=4cm,width=8.5cm ]{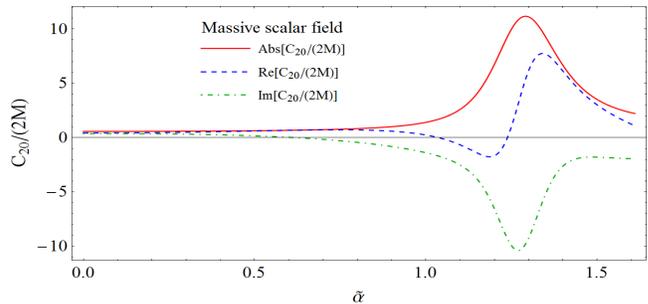}
\setlength\abovecaptionskip{0.25ex}
\vspace{-0.3cm}
\caption{\label{fig:C20_s0} Resonant behavior of the excitation coefficient ${\cal C}_{20}$ of the $(\ell = 2, n = 0)$ QNM of the massive scalar field. It is obtained from (\ref{EC}) by using (\ref{Cauchy_data}) with $\phi_0=1$, $a=1$ and $r_0=10M$.}
\end{figure}

\begin{figure}
\centering
\vspace{0.2cm}
\includegraphics[height=4cm,width=8.5cm ]{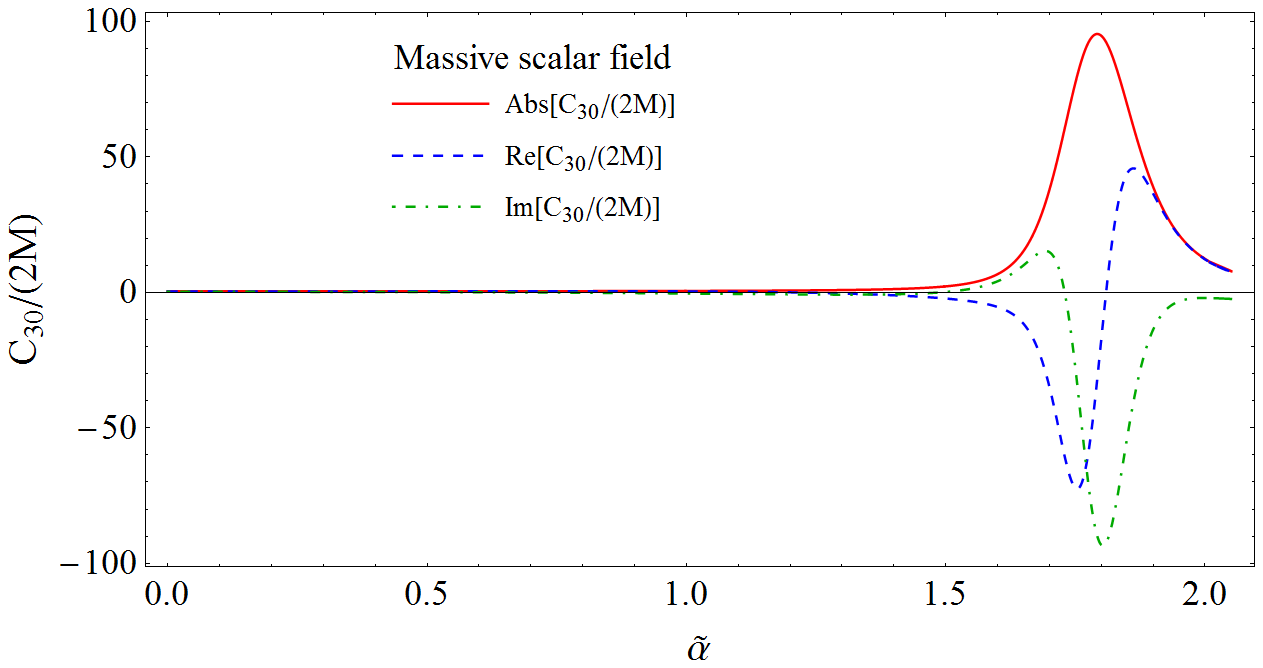}
\setlength\abovecaptionskip{0.25ex}
\vspace{-0.3cm}
\caption{\label{fig:C30_s0} Resonant behavior of the excitation coefficient ${\cal C}_{30}$ of the $(\ell = 3, n = 0)$ QNM of the massive scalar field. It is obtained from (\ref{EC}) by using (\ref{Cauchy_data}) with $\phi_0=1$, $a=1$ and $r_0=10M$.}
\end{figure}

The excitation coefficients ${\cal C}_{\ell n}$ can be ``easily" calculated. Indeed, in order to obtain the excitation factors ${\cal B}_{\ell n}$ we have already constructed the functions $\phi^\mathrm{in}_{\omega_{\ell n} \ell}(r)$ and determined the coefficients $A_\ell^{(+)}(\omega_{\ell n})$ by solving numerically the Regge-Wheeler equation (\ref{RW}).  The evaluation of the integral in Eq.~(\ref{EC}) is then elementary. It should be, however, noted that the numerical instabilities we discussed in Sec.~II and that occur near the values of the mass parameter for which the ${\cal B}_{\ell n}$ vanish are still present. They forbid us to investigate the behavior of the BH ringing (\ref{TimeEvolution_QNM}) for the corresponding masses. It should be also noted that by describing the BH ringing by formula (\ref{TimeEvolution_QNM}) we are again confronted with the "time-shift problem". Here, we shall consider that this formula can be used beyond the starting time $t_\mathrm{start} \approx [r_\ast + r_\ast(r_0)]\mathrm{Re}[\omega_{\ell n}]/\mathrm{Re}[p(\omega_{\ell n})]$. This seems rather reasonable if the width of the Gaussian function (\ref{Cauchy_data}) is not too large, i.e., if $a$ is not too small.

In Fig.~\ref{fig:C20_s0} where we consider the $(\ell=2,n=0)$ QNM of the massive scalar field, we  exhibit the strong resonant behavior of ${\cal C}_{20}$ for particular values of the parameters $a$ and $r_0$ defining the initial data (\ref{Cauchy_data}). It should be noted that, in fact, it depends little on these parameters. By comparing Fig.~\ref{fig:C20_s0} and Fig.~\ref{fig:QNM20_INT}(b), we can observe that the resonant behaviors of ${\cal C}_{20}$ and ${\cal B}_{20}$ are rather similar, both occurring for masses in a range where the QNM is a long-lived mode, but that the first one is more attenuated than the second one. It should be also noted that the positions of the maximums of $|{\cal C}_{20}|$ and $|{\cal B}_{20}|$ are slightly shifted. Similar considerations seem to apply for the other excitation coefficients ${\cal C}_{\ell n}$ of the massive scalar field [see, e.g., Fig.~\ref{fig:C30_s0} and Fig.~\ref{fig:QNM30_INT}(b) where we consider respectively the excitation coefficient ${\cal C}_{30}$ and the excitation factor ${\cal B}_{30}$ of the $(\ell=3,n=0)$ QNM] and for all the other massive bosonic fields. However, it is important to note that, for overtones, the resonance phenomenon is more and more attenuated as the overtone index $n$ increases. As a consequence, we can predict that the extrinsic ringings generated by the fundamental QNMs are certainly the most interesting and that, like the corresponding intrinsic ringings, they have huge and slowly decaying amplitudes.

\begin{figure}
\centering
\vspace{0.2cm}
\includegraphics[height=4cm,width=8.5cm ]{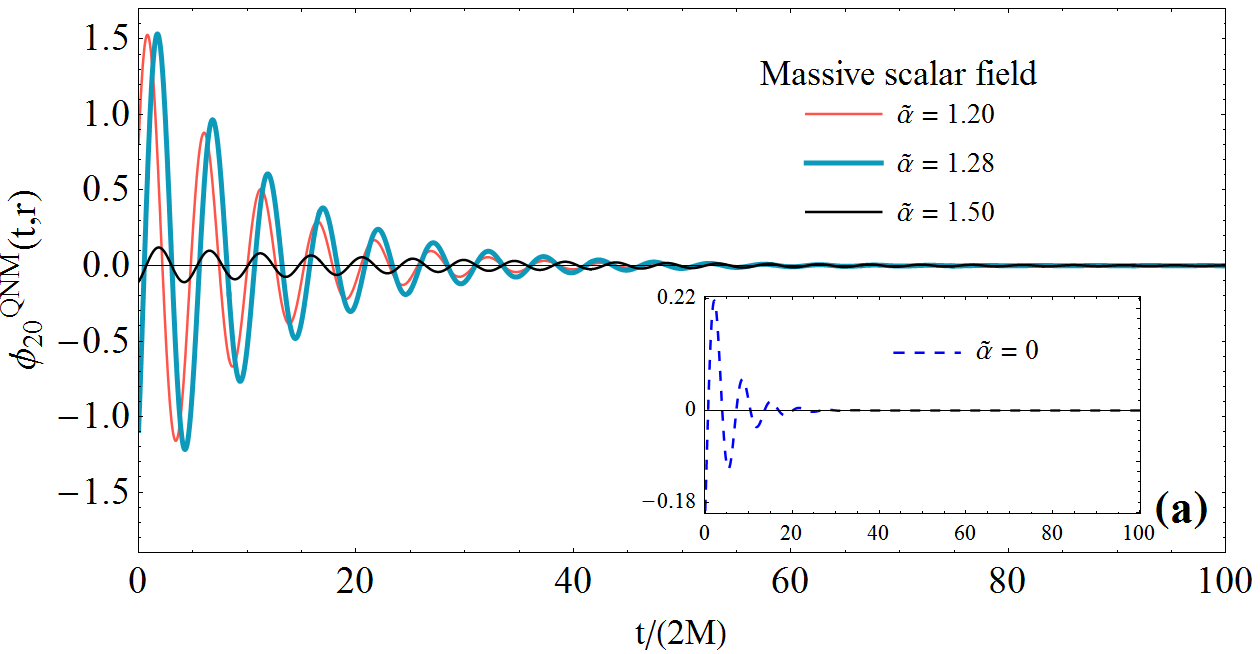}
\centering
\vspace{0.2cm}
\includegraphics[height=4cm,width=8.5cm ]{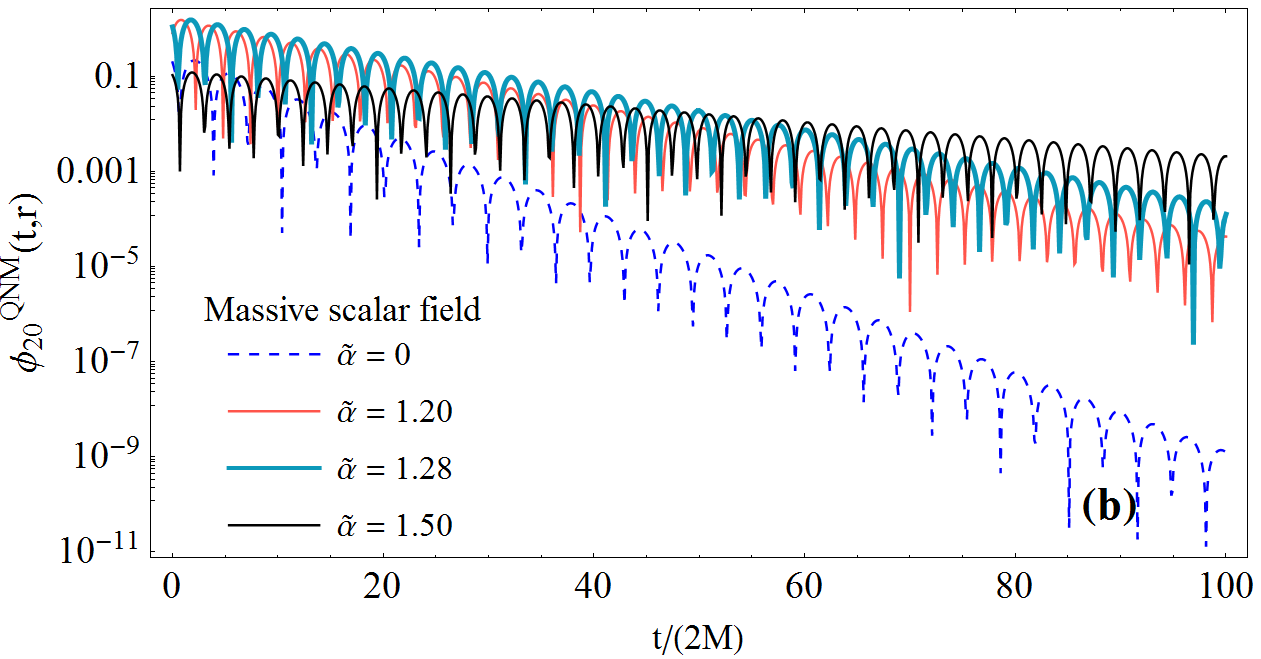}
\setlength\abovecaptionskip{0.25ex}
\vspace{-0.3cm}
\caption{\label{fig:QNM20_EXT} The $(\ell = 2, n = 0)$ QNM of the massive scalar field. (a) and (b) Some extrinsic ringings corresponding to values of the mass near and above the critical value  $\tilde\alpha_{20}$ and comparison with the ringing associated with the massless scalar field. Here the results are obtained from $(\ref{TimeEvolution_QNM})$ with $r = 50M$ and by using $(\ref{Cauchy_data})$ with $\phi_0=1$, $a=1$ and $r_0 = 10M$.}
\end{figure}

\begin{figure}
\centering
\vspace{0.2cm}
\includegraphics[height=4cm,width=8.5cm ]{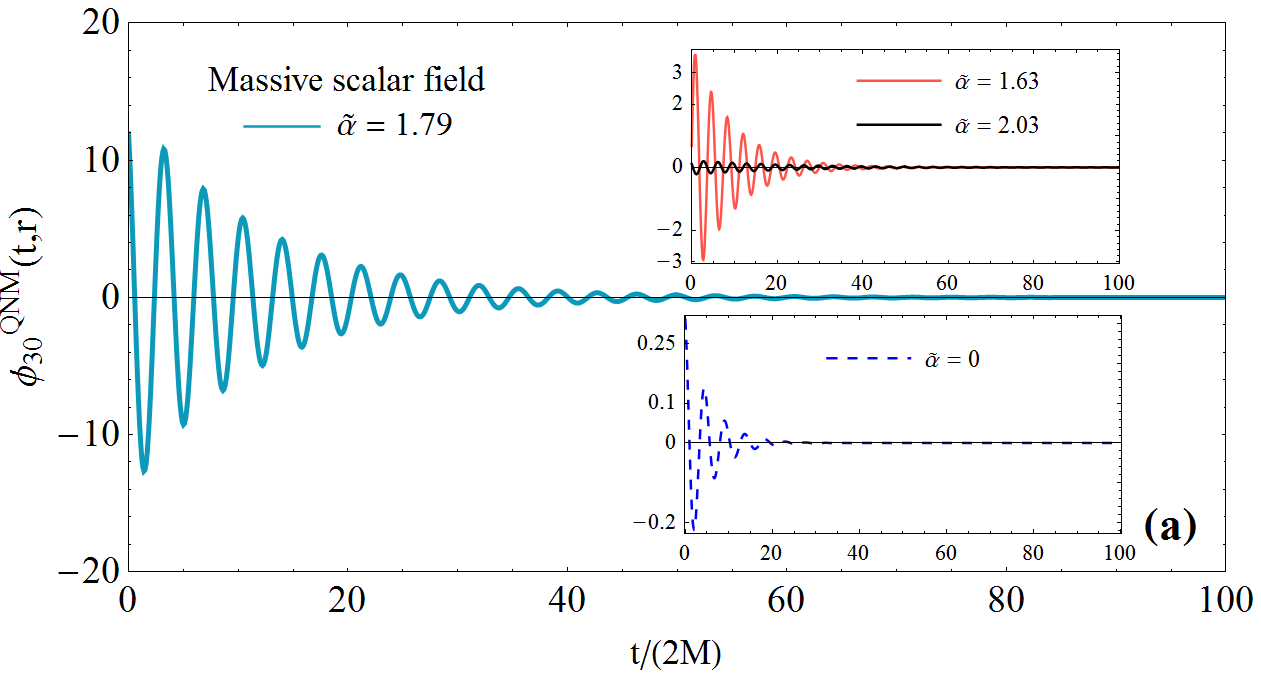}
\centering
\vspace{0.2cm}
\includegraphics[height=4cm,width=8.5cm ]{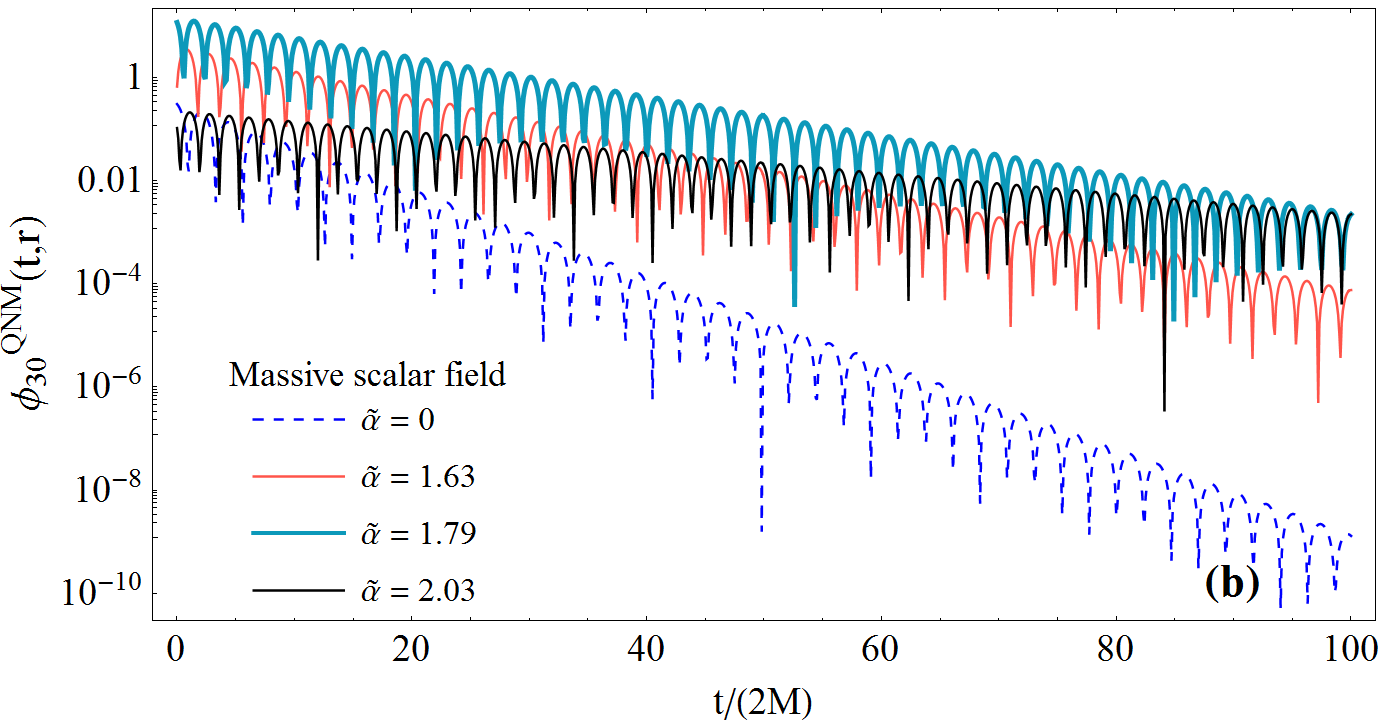}
\setlength\abovecaptionskip{0.25ex}
\vspace{-0.3cm}
\caption{\label{fig:QNM30_EXT} The $(\ell = 3, n = 0)$ QNM of the massive scalar field. (a) and (b) Some extrinsic ringings corresponding to values of the mass near and above the critical value  $\tilde\alpha_{30}$ and comparison with the ringing associated with the massless scalar field. Here the results are obtained from $(\ref{TimeEvolution_QNM})$ with $r = 50M$ and by using $(\ref{Cauchy_data})$ with $\phi_0=1$, $a=1$ and $r_0 = 10M$.}
\end{figure}

\begin{figure}
\centering
\vspace{0.2cm}
\includegraphics[height=4cm,width=8.5cm ]{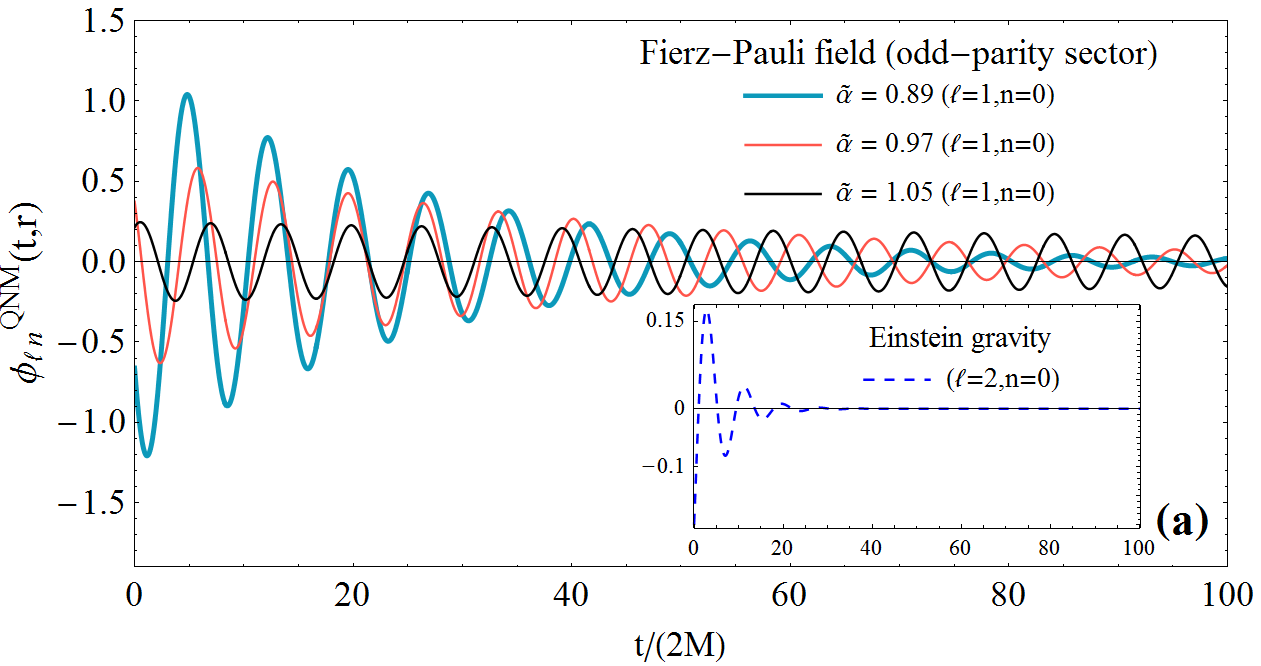}
\centering
\vspace{0.2cm}
\includegraphics[height=4cm,width=8.5cm ]{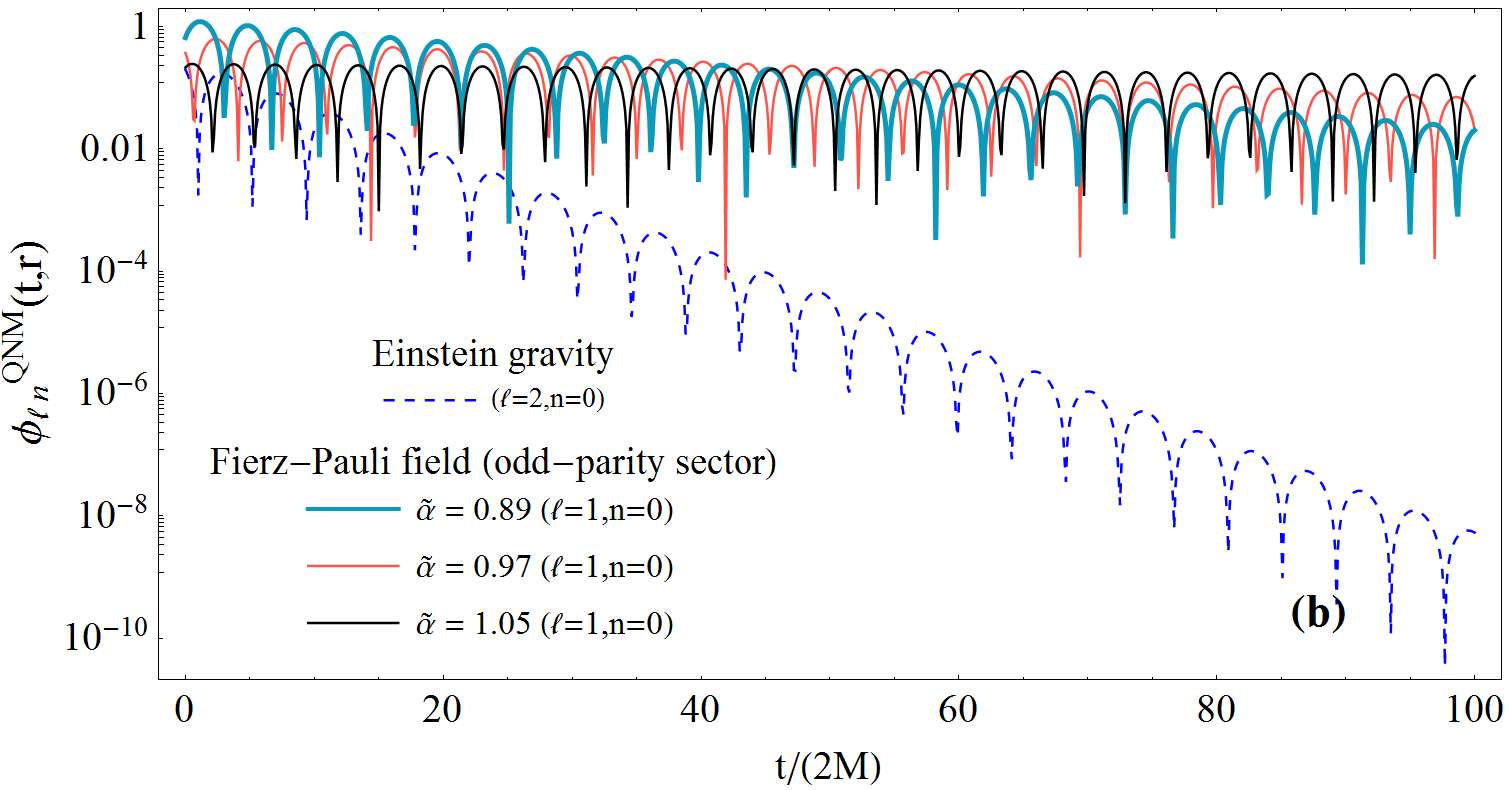}
\setlength\abovecaptionskip{0.25ex}
\vspace{-0.3cm}
\caption{\label{fig:QNM10_s2odd_EXT10} The odd-parity $(\ell = 1, n = 0)$ QNM of the Fierz-Pauli field. (a) and (b) Some extrinsic ringings corresponding to values of the mass near and above the critical value  $\tilde\alpha_{10}$ and comparison with the ringing associated with the odd-parity $(\ell = 2, n = 0)$ QNM of the massless spin-2 field. Here the results are obtained from $(\ref{TimeEvolution_QNM})$ with $r = 50M$ and by using $(\ref{Cauchy_data})$ with $\phi_0=1$, $a=1$ and $r_0 = 10M$.}
\end{figure}

\begin{figure}
\centering
\vspace{0.2cm}
\includegraphics[height=4cm,width=8.5cm ]{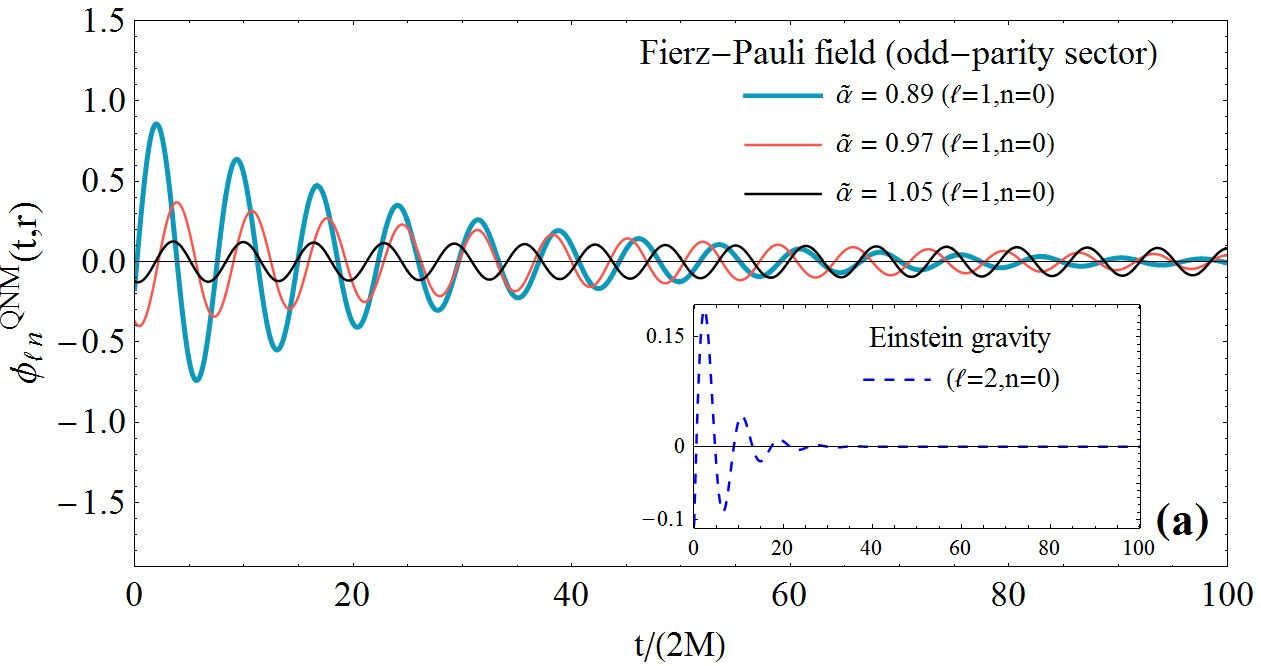}
\centering
\vspace{0.2cm}
\includegraphics[height=4cm,width=8.5cm ]{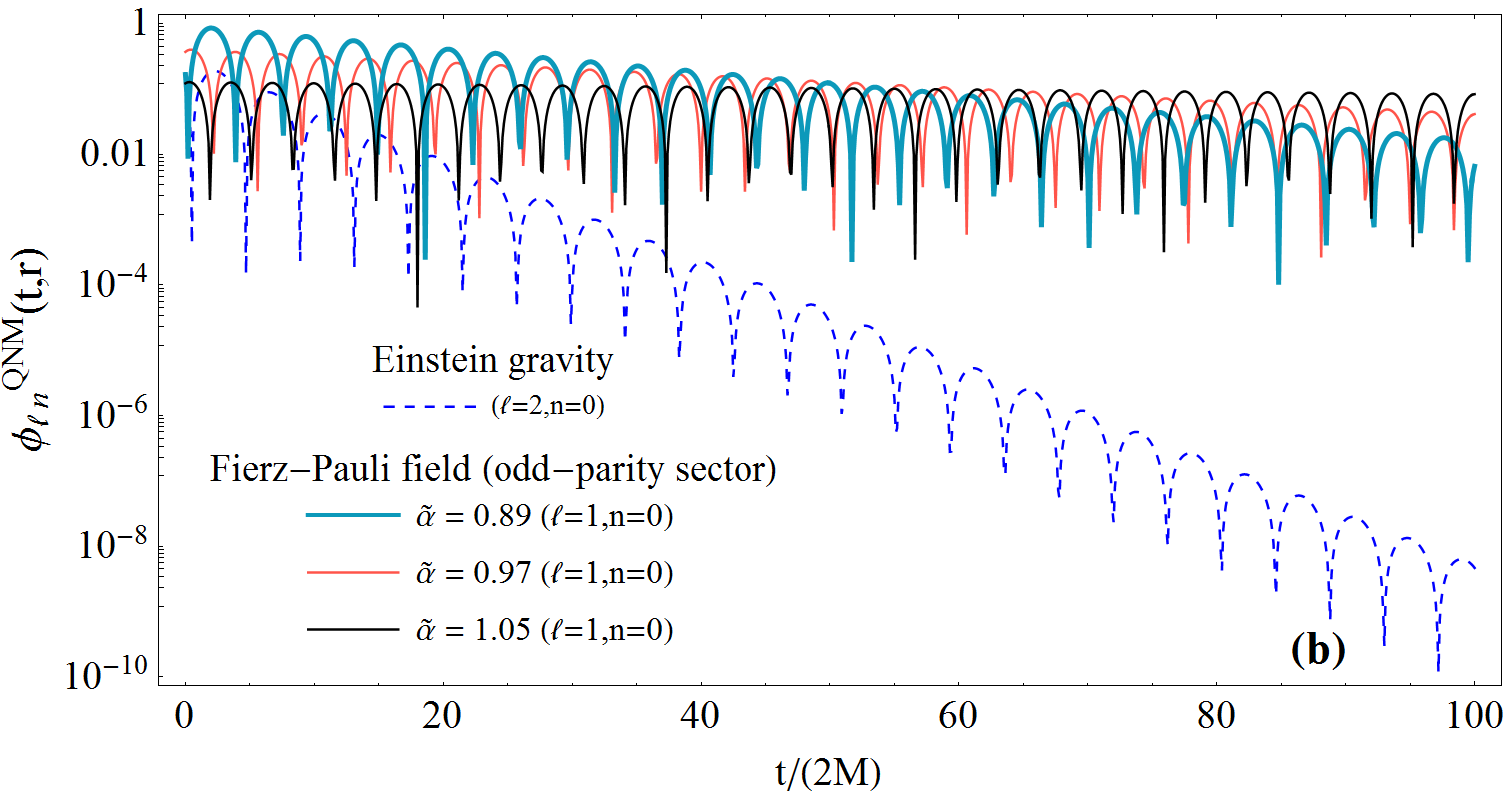}
\setlength\abovecaptionskip{0.25ex}
\vspace{-0.3cm}
\caption{\label{fig:QNM10_s2odd_EXT20} The odd-parity $(\ell = 1, n = 0)$ QNM of the Fierz-Pauli field. (a) and (b) Some extrinsic ringings corresponding to values of the mass near and above the critical value  $\tilde\alpha_{10}$ and comparison with the ringing associated with the odd-parity $(\ell = 2, n = 0)$ QNM of the massless spin-2 field. Here the results are obtained from $(\ref{TimeEvolution_QNM})$ with $r = 50M$ and by using $(\ref{Cauchy_data})$ with $\phi_0=1$, $a=1$ and $r_0 = 20M$.}
\end{figure}

In Figs.~\ref{fig:QNM20_EXT} and \ref{fig:QNM30_EXT}, we focus again on the $(\ell=2,n=0)$ and $(\ell=3,n=0)$ QNMs of the massive scalar field. We plot, for the various masses already considered in Figs.~\ref{fig:QNM20_INT} and \ref{fig:QNM30_INT}, the BH ``extrinsic" ringings defined by (\ref{TimeEvolution_QNM}) and we compare them with the extrinsic ringings generated by the massless scalar field. We can immediately observe that giant and slowly decaying ringings also exist when we take into account the role of the source of the BH perturbation.

It should be noted that, in Figs.~\ref{fig:QNM20_EXT} and \ref{fig:QNM30_EXT}, we have plotted the ringings for $r = 50M$ and $r_0 = 10M$ but similar results can be obtained for various locations of the observer and for various values of the parameters $a$ and $r_0$ defining the initial data (\ref{Cauchy_data}). In particular, in Figs.~\ref{fig:QNM10_s2odd_EXT10} and \ref{fig:QNM10_s2odd_EXT20} where we highlight the extraordinary extrinsic ringings generated in the context of massive gravity by the odd-parity $(\ell = 1, n = 0)$ QNM of the Fierz-Pauli field, we point out the influence of $r_0$.

\section{Conclusion and perspectives}

In this article, by considering three important massive bosonic field theories in the Schwarzschild spacetime, we have shown that the excitation factors of their long-lived QNMs have a strong resonant behavior with, as a consequence, the existence of giant and slowly decaying ringings in rather large domains of the mass parameter. We have more particularly studied the role of the angular momentum index $\ell $ and of the overtone index $n$ of the QNMs and we have also checked that the resonant effect and the associated giant ringing phenomenon still exist when the source of the BH perturbation is described by an initial value problem with Gaussian initial data. Of course, we should now consider more realistic BH excitations such as the excitation by a ``particle" falling radially or plunging and, with in mind astrophysical implications, to extend our study to the Kerr BH. Furthermore, it would be necessary to examine the full response of the BH to an external perturbation and not only the part associated with QNMs. Indeed, for massless fields, the QNM contribution can be easily identified into the full response of the BH but, for massive fields, the situation might be different due to their dispersive character. But even if this is the case, i.e., if the QNM contribution is blurred, it is nevertheless clear that the resonant behavior of the excitation factors of the long-lived QNMs must induce a huge amplification of the BH response.

It should be noted that, in the present study, we have only focussed on the QNMs which are governed by a single differential equation of Regge-Wheeler-type so our task has been greatly simplified. But what happens for the even-parity $\ell \ge 1$ QNMs of the Proca field as well as for the even-parity $\ell \ge 1$ QNMs and the odd-parity $\ell \ge 2$ QNMs of the Fierz-Pauli field? Do they generate giant ringings? Of course, this seems quite natural (we have just shown that, for the QNMs of the massive scalar field and for the odd-parity QNMs of the Proca field, giant ringings exist for all values of the angular momentum $\ell$) but remains to be proved which is far from obvious : indeed, depending on the parity sector and the angular momentum index, these QNMs are governed by two or three coupled differential equations \cite{Konoplya:2005hr,Rosa:2011my,Brito:2013wya} and, as a consequence, the corresponding excitation factors and ringings are more challenging to compute while, at first sight, a semiclassical approach of the problem seems out of reach.

It should be also noted that we have numerically and semiclassically shown that the maximum of the amplitude of the quasinormal excitation factors increases rapidly with the angular momentum index $\ell$. We are quite disturbed by this puzzling result which we are not able to explain simply but we believe that it is harmless. Indeed :

\qquad (i) if it was limited to the spin-$2$ field, we might associate it with a new kind of BH instability but, because it also exists for the other two bosonic fields, such an explanation is not satisfactory,

\qquad  (ii) it does not seem to lead to a divergent behavior of sums like (\ref{Gret_ell_QNMsum}) or (\ref{TimeEvolution_QNM_sum}) because, even if we consider a value of the mass parameter for which one of the QNM generates a huge contribution, the contributions of all the other QNMs vanish or are neglectable.

We have motivated our work by the possible implications for ultralight bosonic fields but, of course, our results could have also consequences in the context of ordinary massive bosonic fields interacting with primordial BHs. In this context, it would be interesting to consider, in addition, massive fermionic fields and to check that the resonant effects studied in this article also exist for such fields.

Let us finally note that the strong resonant behavior of the excitation factors of the long-lived QNMs is a new effect in BH physics so lot of work remains to be done to understand all its physical consequences. This article and our previous one \cite{Decanini:2014kha} are a first step in this direction, limited to astrophysical implications, but we think that this effect could also have interesting implications in the context of quantum field theory in curved spacetime (Hawking effect and absorption cross section, renormalized stress-energy tensors and associated correlators, etc.) and perhaps in string theory and, more particularly, in the context of the Kerr/CFT correspondence. Some of these directions will be explored shortly.

\begin{acknowledgments}

We wish to thank Andrei Belokogne for various discussions and the ``Collectivit\'e Territoriale de Corse" for its support through the COMPA project. We are grateful to Vitor Cardoso for comments on our recent work \cite{Decanini:2014kha} and for pointing us to an important reference.

\end{acknowledgments}

\bibliography{Resonant_excitation_of_BH}

\begin{thebibliography}{46}%
\makeatletter
\providecommand \@ifxundefined [1]{%
 \@ifx{#1\undefined}
}%
\providecommand \@ifnum [1]{%
 \ifnum #1\expandafter \@firstoftwo
 \else \expandafter \@secondoftwo
 \fi
}%
\providecommand \@ifx [1]{%
 \ifx #1\expandafter \@firstoftwo
 \else \expandafter \@secondoftwo
 \fi
}%
\providecommand \natexlab [1]{#1}%
\providecommand \enquote  [1]{``#1''}%
\providecommand \bibnamefont  [1]{#1}%
\providecommand \bibfnamefont [1]{#1}%
\providecommand \citenamefont [1]{#1}%
\providecommand \href@noop [0]{\@secondoftwo}%
\providecommand \href [0]{\begingroup \@sanitize@url \@href}%
\providecommand \@href[1]{\@@startlink{#1}\@@href}%
\providecommand \@@href[1]{\endgroup#1\@@endlink}%
\providecommand \@sanitize@url [0]{\catcode `\\12\catcode `\$12\catcode
  `\&12\catcode `\#12\catcode `\^12\catcode `\_12\catcode `\%12\relax}%
\providecommand \@@startlink[1]{}%
\providecommand \@@endlink[0]{}%
\providecommand \url  [0]{\begingroup\@sanitize@url \@url }%
\providecommand \@url [1]{\endgroup\@href {#1}{\urlprefix }}%
\providecommand \urlprefix  [0]{URL }%
\providecommand \Eprint [0]{\href }%
\providecommand \doibase [0]{http://dx.doi.org/}%
\providecommand \selectlanguage [0]{\@gobble}%
\providecommand \bibinfo  [0]{\@secondoftwo}%
\providecommand \bibfield  [0]{\@secondoftwo}%
\providecommand \translation [1]{[#1]}%
\providecommand \BibitemOpen [0]{}%
\providecommand \bibitemStop [0]{}%
\providecommand \bibitemNoStop [0]{.\EOS\space}%
\providecommand \EOS [0]{\spacefactor3000\relax}%
\providecommand \BibitemShut  [1]{\csname bibitem#1\endcsname}%
\let\auto@bib@innerbib\@empty
%</preamble>
\bibitem [{\citenamefont {Arvanitaki}\ \emph {et~al.}(2010)\citenamefont
  {Arvanitaki}, \citenamefont {Dimopoulos}, \citenamefont {Dubovsky},
  \citenamefont {Kaloper},\ and\ \citenamefont
  {March-Russell}}]{Arvanitaki:2009fg}%
  \BibitemOpen
  \bibfield  {author} {\bibinfo {author} {\bibfnamefont {A.}~\bibnamefont
  {Arvanitaki}}, \bibinfo {author} {\bibfnamefont {S.}~\bibnamefont
  {Dimopoulos}}, \bibinfo {author} {\bibfnamefont {S.}~\bibnamefont
  {Dubovsky}}, \bibinfo {author} {\bibfnamefont {N.}~\bibnamefont {Kaloper}}, \
  and\ \bibinfo {author} {\bibfnamefont {J.}~\bibnamefont {March-Russell}},\
  }\href {\doibase 10.1103/PhysRevD.81.123530} {\bibfield  {journal} {\bibinfo
  {journal} {Phys.\ Rev.\ D}\ }\textbf {\bibinfo {volume} {81}},\ \bibinfo
  {pages} {123530} (\bibinfo {year} {2010})},\ \Eprint
  {http://arxiv.org/abs/0905.4720} {arXiv:0905.4720 [hep-th]} \BibitemShut
  {NoStop}%
%%CITATION = ARXIV:0905.4720;%%
\bibitem [{\citenamefont {Rosa}(2010)}]{Rosa:2009ei}%
  \BibitemOpen
  \bibfield  {author} {\bibinfo {author} {\bibfnamefont {J.}~\bibnamefont
  {Rosa}},\ }\href {\doibase 10.1007/JHEP06(2010)015} {\bibfield  {journal}
  {\bibinfo  {journal} {JHEP}\ }\textbf {\bibinfo {volume} {1006}},\ \bibinfo
  {pages} {015} (\bibinfo {year} {2010})},\ \Eprint
  {http://arxiv.org/abs/0912.1780} {arXiv:0912.1780 [hep-th]} \BibitemShut
  {NoStop}%
%%CITATION = ARXIV:0912.1780;%%
\bibitem [{\citenamefont {Arvanitaki}\ and\ \citenamefont
  {Dubovsky}(2011)}]{Arvanitaki:2010sy}%
  \BibitemOpen
  \bibfield  {author} {\bibinfo {author} {\bibfnamefont {A.}~\bibnamefont
  {Arvanitaki}}\ and\ \bibinfo {author} {\bibfnamefont {S.}~\bibnamefont
  {Dubovsky}},\ }\href {\doibase 10.1103/PhysRevD.83.044026} {\bibfield
  {journal} {\bibinfo  {journal} {Phys.\ Rev.\ D}\ }\textbf {\bibinfo {volume}
  {83}},\ \bibinfo {pages} {044026} (\bibinfo {year} {2011})},\ \Eprint
  {http://arxiv.org/abs/1004.3558} {arXiv:1004.3558 [hep-th]} \BibitemShut
  {NoStop}%
%%CITATION = ARXIV:1004.3558;%%
\bibitem [{\citenamefont {Barranco}\ \emph {et~al.}(2011)\citenamefont
  {Barranco}, \citenamefont {Bernal}, \citenamefont {Degollado}, \citenamefont
  {Diez-Tejedor}, \citenamefont {Megevand}, \citenamefont {Alcubierre},
  \citenamefont {N{\'u}{\~n}ez},\ and\ \citenamefont {Sarbach}}]{Burt:2011pv}%
  \BibitemOpen
  \bibfield  {author} {\bibinfo {author} {\bibfnamefont {J.}~\bibnamefont
  {Barranco}}, \bibinfo {author} {\bibfnamefont {A.}~\bibnamefont {Bernal}},
  \bibinfo {author} {\bibfnamefont {J.~C.}\ \bibnamefont {Degollado}}, \bibinfo
  {author} {\bibfnamefont {A.}~\bibnamefont {Diez-Tejedor}}, \bibinfo {author}
  {\bibfnamefont {M.}~\bibnamefont {Megevand}}, \bibinfo {author}
  {\bibfnamefont {M.}~\bibnamefont {Alcubierre}}, \bibinfo {author}
  {\bibfnamefont {D.}~\bibnamefont {N{\'u}{\~n}ez}}, \ and\ \bibinfo {author}
  {\bibfnamefont {O.}~\bibnamefont {Sarbach}},\ }\href {\doibase
  10.1103/PhysRevD.84.083008} {\bibfield  {journal} {\bibinfo  {journal}
  {Phys.\ Rev.\ D}\ }\textbf {\bibinfo {volume} {84}},\ \bibinfo {pages}
  {083008} (\bibinfo {year} {2011})},\ \Eprint {http://arxiv.org/abs/1108.0931}
  {arXiv:1108.0931 [gr-qc]} \BibitemShut {NoStop}%
%%CITATION = ARXIV:1108.0931;%%
\bibitem [{\citenamefont {Cardoso}\ \emph {et~al.}(2011)\citenamefont
  {Cardoso}, \citenamefont {Chakrabarti}, \citenamefont {Pani}, \citenamefont
  {Berti},\ and\ \citenamefont {Gualtieri}}]{Cardoso:2011xi}%
  \BibitemOpen
  \bibfield  {author} {\bibinfo {author} {\bibfnamefont {V.}~\bibnamefont
  {Cardoso}}, \bibinfo {author} {\bibfnamefont {S.}~\bibnamefont
  {Chakrabarti}}, \bibinfo {author} {\bibfnamefont {P.}~\bibnamefont {Pani}},
  \bibinfo {author} {\bibfnamefont {E.}~\bibnamefont {Berti}}, \ and\ \bibinfo
  {author} {\bibfnamefont {L.}~\bibnamefont {Gualtieri}},\ }\href {\doibase
  10.1103/PhysRevLett.107.241101} {\bibfield  {journal} {\bibinfo  {journal}
  {Phys.\ Rev.\ Lett.}\ }\textbf {\bibinfo {volume} {107}},\ \bibinfo {pages}
  {241101} (\bibinfo {year} {2011})},\ \Eprint {http://arxiv.org/abs/1109.6021}
  {arXiv:1109.6021 [gr-qc]} \BibitemShut {NoStop}%
%%CITATION = ARXIV:1109.6021;%%
\bibitem [{\citenamefont {Rosa}\ and\ \citenamefont
  {Dolan}(2012)}]{Rosa:2011my}%
  \BibitemOpen
  \bibfield  {author} {\bibinfo {author} {\bibfnamefont {J.~G.}\ \bibnamefont
  {Rosa}}\ and\ \bibinfo {author} {\bibfnamefont {S.~R.}\ \bibnamefont
  {Dolan}},\ }\href {\doibase 10.1103/PhysRevD.85.044043} {\bibfield  {journal}
  {\bibinfo  {journal} {Phys.\ Rev.\ D}\ }\textbf {\bibinfo {volume} {85}},\
  \bibinfo {pages} {044043} (\bibinfo {year} {2012})},\ \Eprint
  {http://arxiv.org/abs/1110.4494} {arXiv:1110.4494 [hep-th]} \BibitemShut
  {NoStop}%
%%CITATION = ARXIV:1110.4494;%%
\bibitem [{\citenamefont {Barranco}\ \emph {et~al.}(2012)\citenamefont
  {Barranco}, \citenamefont {Bernal}, \citenamefont {Degollado}, \citenamefont
  {Diez-Tejedor}, \citenamefont {Megevand}, \citenamefont {Alcubierre},
  \citenamefont {N{\'u}{\~n}ez},\ and\ \citenamefont
  {Sarbach}}]{Barranco:2012qs}%
  \BibitemOpen
  \bibfield  {author} {\bibinfo {author} {\bibfnamefont {J.}~\bibnamefont
  {Barranco}}, \bibinfo {author} {\bibfnamefont {A.}~\bibnamefont {Bernal}},
  \bibinfo {author} {\bibfnamefont {J.~C.}\ \bibnamefont {Degollado}}, \bibinfo
  {author} {\bibfnamefont {A.}~\bibnamefont {Diez-Tejedor}}, \bibinfo {author}
  {\bibfnamefont {M.}~\bibnamefont {Megevand}}, \bibinfo {author}
  {\bibfnamefont {M.}~\bibnamefont {Alcubierre}}, \bibinfo {author}
  {\bibfnamefont {D.}~\bibnamefont {N{\'u}{\~n}ez}}, \ and\ \bibinfo {author}
  {\bibfnamefont {O.}~\bibnamefont {Sarbach}},\ }\href {\doibase
  10.1103/PhysRevLett.109.081102} {\bibfield  {journal} {\bibinfo  {journal}
  {Phys.\ Rev.\ Lett.}\ }\textbf {\bibinfo {volume} {109}},\ \bibinfo {pages}
  {081102} (\bibinfo {year} {2012})},\ \Eprint {http://arxiv.org/abs/1207.2153}
  {arXiv:1207.2153 [gr-qc]} \BibitemShut {NoStop}%
%%CITATION = ARXIV:1207.2153;%%
\bibitem [{\citenamefont {Pani}\ \emph {et~al.}(2012)\citenamefont {Pani},
  \citenamefont {Cardoso}, \citenamefont {Gualtieri}, \citenamefont {Berti},\
  and\ \citenamefont {Ishibashi}}]{Pani:2012vp}%
  \BibitemOpen
  \bibfield  {author} {\bibinfo {author} {\bibfnamefont {P.}~\bibnamefont
  {Pani}}, \bibinfo {author} {\bibfnamefont {V.}~\bibnamefont {Cardoso}},
  \bibinfo {author} {\bibfnamefont {L.}~\bibnamefont {Gualtieri}}, \bibinfo
  {author} {\bibfnamefont {E.}~\bibnamefont {Berti}}, \ and\ \bibinfo {author}
  {\bibfnamefont {A.}~\bibnamefont {Ishibashi}},\ }\href {\doibase
  10.1103/PhysRevLett.109.131102} {\bibfield  {journal} {\bibinfo  {journal}
  {Phys.\ Rev.\ Lett.}\ }\textbf {\bibinfo {volume} {109}},\ \bibinfo {pages}
  {131102} (\bibinfo {year} {2012})},\ \Eprint {http://arxiv.org/abs/1209.0465}
  {arXiv:1209.0465 [gr-qc]} \BibitemShut {NoStop}%
%%CITATION = ARXIV:1209.0465;%%
\bibitem [{\citenamefont {Dolan}(2013)}]{Dolan:2012yt}%
  \BibitemOpen
  \bibfield  {author} {\bibinfo {author} {\bibfnamefont {S.~R.}\ \bibnamefont
  {Dolan}},\ }\href {\doibase 10.1103/PhysRevD.87.124026} {\bibfield  {journal}
  {\bibinfo  {journal} {Phys.\ Rev.\ D}\ }\textbf {\bibinfo {volume} {87}},\
  \bibinfo {pages} {124026} (\bibinfo {year} {2013})},\ \Eprint
  {http://arxiv.org/abs/1212.1477} {arXiv:1212.1477 [gr-qc]} \BibitemShut
  {NoStop}%
%%CITATION = ARXIV:1212.1477;%%
\bibitem [{\citenamefont {Brito}\ \emph
  {et~al.}(2013{\natexlab{a}})\citenamefont {Brito}, \citenamefont {Cardoso},\
  and\ \citenamefont {Pani}}]{Brito:2013wya}%
  \BibitemOpen
  \bibfield  {author} {\bibinfo {author} {\bibfnamefont {R.}~\bibnamefont
  {Brito}}, \bibinfo {author} {\bibfnamefont {V.}~\bibnamefont {Cardoso}}, \
  and\ \bibinfo {author} {\bibfnamefont {P.}~\bibnamefont {Pani}},\ }\href
  {\doibase 10.1103/PhysRevD.88.023514} {\bibfield  {journal} {\bibinfo
  {journal} {Phys.\ Rev.\ D}\ }\textbf {\bibinfo {volume} {88}},\ \bibinfo
  {pages} {023514} (\bibinfo {year} {2013}{\natexlab{a}})},\ \Eprint
  {http://arxiv.org/abs/1304.6725} {arXiv:1304.6725 [gr-qc]} \BibitemShut
  {NoStop}%
%%CITATION = ARXIV:1304.6725;%%
\bibitem [{\citenamefont {Brito}\ \emph
  {et~al.}(2013{\natexlab{b}})\citenamefont {Brito}, \citenamefont {Cardoso},\
  and\ \citenamefont {Pani}}]{Brito:2013yxa}%
  \BibitemOpen
  \bibfield  {author} {\bibinfo {author} {\bibfnamefont {R.}~\bibnamefont
  {Brito}}, \bibinfo {author} {\bibfnamefont {V.}~\bibnamefont {Cardoso}}, \
  and\ \bibinfo {author} {\bibfnamefont {P.}~\bibnamefont {Pani}},\ }\href
  {\doibase 10.1103/PhysRevD.87.124024} {\bibfield  {journal} {\bibinfo
  {journal} {Phys.\ Rev.\ D}\ }\textbf {\bibinfo {volume} {87}},\ \bibinfo
  {pages} {124024} (\bibinfo {year} {2013}{\natexlab{b}})},\ \Eprint
  {http://arxiv.org/abs/1306.0908} {arXiv:1306.0908 [gr-qc]} \BibitemShut
  {NoStop}%
%%CITATION = ARXIV:1306.0908;%%
\bibitem [{\citenamefont {Babichev}\ and\ \citenamefont
  {Fabbri}(2013)}]{Babichev:2013una}%
  \BibitemOpen
  \bibfield  {author} {\bibinfo {author} {\bibfnamefont {E.}~\bibnamefont
  {Babichev}}\ and\ \bibinfo {author} {\bibfnamefont {A.}~\bibnamefont
  {Fabbri}},\ }\href {\doibase 10.1088/0264-9381/30/15/152001} {\bibfield
  {journal} {\bibinfo  {journal} {Classical Quantum Gravity}\ }\textbf
  {\bibinfo {volume} {30}},\ \bibinfo {pages} {152001} (\bibinfo {year}
  {2013})},\ \Eprint {http://arxiv.org/abs/1304.5992} {arXiv:1304.5992 [gr-qc]}
  \BibitemShut {NoStop}%
%%CITATION = ARXIV:1304.5992;%%
\bibitem [{\citenamefont {Brito}\ \emph
  {et~al.}(2013{\natexlab{c}})\citenamefont {Brito}, \citenamefont {Cardoso},\
  and\ \citenamefont {Pani}}]{Brito:2013xaa}%
  \BibitemOpen
  \bibfield  {author} {\bibinfo {author} {\bibfnamefont {R.}~\bibnamefont
  {Brito}}, \bibinfo {author} {\bibfnamefont {V.}~\bibnamefont {Cardoso}}, \
  and\ \bibinfo {author} {\bibfnamefont {P.}~\bibnamefont {Pani}},\ }\href
  {\doibase 10.1103/PhysRevD.88.064006} {\bibfield  {journal} {\bibinfo
  {journal} {Phys.\ Rev.\ D}\ }\textbf {\bibinfo {volume} {88}},\ \bibinfo
  {pages} {064006} (\bibinfo {year} {2013}{\natexlab{c}})},\ \Eprint
  {http://arxiv.org/abs/1309.0818} {arXiv:1309.0818 [gr-qc]} \BibitemShut
  {NoStop}%
%%CITATION = ARXIV:1309.0818;%%
\bibitem [{\citenamefont {Hod}(2013)}]{Hod:2013dka}%
  \BibitemOpen
  \bibfield  {author} {\bibinfo {author} {\bibfnamefont {S.}~\bibnamefont
  {Hod}},\ }\href {\doibase 10.1088/0264-9381/30/23/237002} {\bibfield
  {journal} {\bibinfo  {journal} {Classical Quantum Gravity}\ }\textbf
  {\bibinfo {volume} {30}},\ \bibinfo {pages} {237002} (\bibinfo {year}
  {2013})}\BibitemShut {NoStop}%
%%CITATION = CQGRD,30,237002;%%
\bibitem [{\citenamefont {Barranco}\ \emph {et~al.}(2013)\citenamefont
  {Barranco}, \citenamefont {Bernal}, \citenamefont {Degollado}, \citenamefont
  {Diez-Tejedor}, \citenamefont {Megevand} \emph {et~al.}}]{Barranco:2013rua}%
  \BibitemOpen
  \bibfield  {author} {\bibinfo {author} {\bibfnamefont {J.}~\bibnamefont
  {Barranco}}, \bibinfo {author} {\bibfnamefont {A.}~\bibnamefont {Bernal}},
  \bibinfo {author} {\bibfnamefont {J.~C.}\ \bibnamefont {Degollado}}, \bibinfo
  {author} {\bibfnamefont {A.}~\bibnamefont {Diez-Tejedor}}, \bibinfo {author}
  {\bibfnamefont {M.}~\bibnamefont {Megevand}},  \emph {et~al.},\ }\href@noop
  {} {\  (\bibinfo {year} {2013})},\ \Eprint {http://arxiv.org/abs/1312.5808}
  {arXiv:1312.5808 [gr-qc]} \BibitemShut {NoStop}%
%%CITATION = ARXIV:1312.5808;%%
\bibitem [{\citenamefont {Decanini}\ \emph {et~al.}(2014)\citenamefont
  {Decanini}, \citenamefont {Folacci},\ and\ \citenamefont {Ould
  El~Hadj}}]{Decanini:2014kha}%
  \BibitemOpen
  \bibfield  {author} {\bibinfo {author} {\bibfnamefont {Y.}~\bibnamefont
  {Decanini}}, \bibinfo {author} {\bibfnamefont {A.}~\bibnamefont {Folacci}}, \
  and\ \bibinfo {author} {\bibfnamefont {M.}~\bibnamefont {Ould El~Hadj}},\
  }\href@noop {} {\  (\bibinfo {year} {2014})},\ \Eprint
  {http://arxiv.org/abs/1401.0321} {arXiv:1401.0321 [gr-qc]} \BibitemShut
  {NoStop}%
%%CITATION = ARXIV:1401.0321;%%
\bibitem [{\citenamefont {Fierz}(1939)}]{Fierz:1939zz}%
  \BibitemOpen
  \bibfield  {author} {\bibinfo {author} {\bibfnamefont {M.}~\bibnamefont
  {Fierz}},\ }\href {\doibase 10.5169/seals-110930} {\bibfield  {journal}
  {\bibinfo  {journal} {Helv.\ Phys.\ Acta}\ }\textbf {\bibinfo {volume}
  {12}},\ \bibinfo {pages} {3} (\bibinfo {year} {1939})}\BibitemShut {NoStop}%
%%CITATION = HPACA,12,3;%%
\bibitem [{\citenamefont {Fierz}\ and\ \citenamefont
  {Pauli}(1939)}]{Fierz:1939ix}%
  \BibitemOpen
  \bibfield  {author} {\bibinfo {author} {\bibfnamefont {M.}~\bibnamefont
  {Fierz}}\ and\ \bibinfo {author} {\bibfnamefont {W.}~\bibnamefont {Pauli}},\
  }\href {\doibase 10.1098/rspa.1939.0140} {\bibfield  {journal} {\bibinfo
  {journal} {Proc.\ Roy.\ Soc.\ Lond.\ A}\ }\textbf {\bibinfo {volume} {173}},\
  \bibinfo {pages} {211} (\bibinfo {year} {1939})}\BibitemShut {NoStop}%
%%CITATION = PRSLA,A173,211;%%
\bibitem [{\citenamefont {Hassan}\ \emph {et~al.}(2013)\citenamefont {Hassan},
  \citenamefont {Schmidt-May},\ and\ \citenamefont {von
  Strauss}}]{Hassan:2012wr}%
  \BibitemOpen
  \bibfield  {author} {\bibinfo {author} {\bibfnamefont {S.}~\bibnamefont
  {Hassan}}, \bibinfo {author} {\bibfnamefont {A.}~\bibnamefont {Schmidt-May}},
  \ and\ \bibinfo {author} {\bibfnamefont {M.}~\bibnamefont {von Strauss}},\
  }\href {\doibase 10.1007/JHEP05(2013)086} {\bibfield  {journal} {\bibinfo
  {journal} {JHEP}\ }\textbf {\bibinfo {volume} {1305}},\ \bibinfo {pages}
  {086} (\bibinfo {year} {2013})},\ \Eprint {http://arxiv.org/abs/1208.1515}
  {arXiv:1208.1515 [hep-th]} \BibitemShut {NoStop}%
%%CITATION = ARXIV:1208.1515;%%
\bibitem [{\citenamefont {de~Rham}\ and\ \citenamefont
  {Gabadadze}(2010)}]{deRham:2010ik}%
  \BibitemOpen
  \bibfield  {author} {\bibinfo {author} {\bibfnamefont {C.}~\bibnamefont
  {de~Rham}}\ and\ \bibinfo {author} {\bibfnamefont {G.}~\bibnamefont
  {Gabadadze}},\ }\href {\doibase 10.1103/PhysRevD.82.044020} {\bibfield
  {journal} {\bibinfo  {journal} {Phys.\ Rev.\ D}\ }\textbf {\bibinfo {volume}
  {82}},\ \bibinfo {pages} {044020} (\bibinfo {year} {2010})},\ \Eprint
  {http://arxiv.org/abs/1007.0443} {arXiv:1007.0443 [hep-th]} \BibitemShut
  {NoStop}%
%%CITATION = ARXIV:1007.0443;%%
\bibitem [{\citenamefont {de~Rham}\ \emph {et~al.}(2011)\citenamefont
  {de~Rham}, \citenamefont {Gabadadze},\ and\ \citenamefont
  {Tolley}}]{deRham:2010kj}%
  \BibitemOpen
  \bibfield  {author} {\bibinfo {author} {\bibfnamefont {C.}~\bibnamefont
  {de~Rham}}, \bibinfo {author} {\bibfnamefont {G.}~\bibnamefont {Gabadadze}},
  \ and\ \bibinfo {author} {\bibfnamefont {A.~J.}\ \bibnamefont {Tolley}},\
  }\href {\doibase 10.1103/PhysRevLett.106.231101} {\bibfield  {journal}
  {\bibinfo  {journal} {Phys.\ Rev.\ Lett.}\ }\textbf {\bibinfo {volume}
  {106}},\ \bibinfo {pages} {231101} (\bibinfo {year} {2011})},\ \Eprint
  {http://arxiv.org/abs/1011.1232} {arXiv:1011.1232 [hep-th]} \BibitemShut
  {NoStop}%
%%CITATION = ARXIV:1011.1232;%%
\bibitem [{\citenamefont {Konoplya}(2006)}]{Konoplya:2005hr}%
  \BibitemOpen
  \bibfield  {author} {\bibinfo {author} {\bibfnamefont {R.}~\bibnamefont
  {Konoplya}},\ }\href {\doibase 10.1103/PhysRevD.73.024009} {\bibfield
  {journal} {\bibinfo  {journal} {Phys.\ Rev.\ D}\ }\textbf {\bibinfo {volume}
  {73}},\ \bibinfo {pages} {024009} (\bibinfo {year} {2006})},\ \Eprint
  {http://arxiv.org/abs/gr-qc/0509026} {arXiv:gr-qc/0509026} \BibitemShut
  {NoStop}%
%%CITATION = GR-QC/0509026;%%
\bibitem [{\citenamefont {Deruelle}\ and\ \citenamefont
  {Ruffini}(1974)}]{Deruelle:1974zy}%
  \BibitemOpen
  \bibfield  {author} {\bibinfo {author} {\bibfnamefont {N.}~\bibnamefont
  {Deruelle}}\ and\ \bibinfo {author} {\bibfnamefont {R.}~\bibnamefont
  {Ruffini}},\ }\href {\doibase 10.1016/0370-2693(74)90119-1} {\bibfield
  {journal} {\bibinfo  {journal} {Phys.\ Lett.\ B}\ }\textbf {\bibinfo {volume}
  {52}},\ \bibinfo {pages} {437} (\bibinfo {year} {1974})}\BibitemShut
  {NoStop}%
%%CITATION = PHLTA,B52,437;%%
\bibitem [{\citenamefont {Damour}\ \emph {et~al.}(1976)\citenamefont {Damour},
  \citenamefont {Deruelle},\ and\ \citenamefont {Ruffini}}]{Damour:1976kh}%
  \BibitemOpen
  \bibfield  {author} {\bibinfo {author} {\bibfnamefont {T.}~\bibnamefont
  {Damour}}, \bibinfo {author} {\bibfnamefont {N.}~\bibnamefont {Deruelle}}, \
  and\ \bibinfo {author} {\bibfnamefont {R.}~\bibnamefont {Ruffini}},\ }\href
  {\doibase 10.1007/BF02725534} {\bibfield  {journal} {\bibinfo  {journal}
  {Lett.\ Nuovo Cim.}\ }\textbf {\bibinfo {volume} {15}},\ \bibinfo {pages}
  {257} (\bibinfo {year} {1976})}\BibitemShut {NoStop}%
%%CITATION = NCLTA,15,257;%%
\bibitem [{\citenamefont {Zouros}\ and\ \citenamefont
  {Eardley}(1979)}]{Zouros:1979iw}%
  \BibitemOpen
  \bibfield  {author} {\bibinfo {author} {\bibfnamefont {T.}~\bibnamefont
  {Zouros}}\ and\ \bibinfo {author} {\bibfnamefont {D.}~\bibnamefont
  {Eardley}},\ }\href {\doibase 10.1016/0003-4916(79)90237-9} {\bibfield
  {journal} {\bibinfo  {journal} {Ann.\ Phys.\ (N.Y.)}\ }\textbf {\bibinfo
  {volume} {118}},\ \bibinfo {pages} {139} (\bibinfo {year}
  {1979})}\BibitemShut {NoStop}%
%%CITATION = APNYA,118,139;%%
\bibitem [{\citenamefont {Detweiler}(1980{\natexlab{a}})}]{Detweiler:1980uk}%
  \BibitemOpen
  \bibfield  {author} {\bibinfo {author} {\bibfnamefont {S.~L.}\ \bibnamefont
  {Detweiler}},\ }\href {\doibase 10.1103/PhysRevD.22.2323} {\bibfield
  {journal} {\bibinfo  {journal} {Phys.\ Rev.\ D}\ }\textbf {\bibinfo {volume}
  {22}},\ \bibinfo {pages} {2323} (\bibinfo {year}
  {1980}{\natexlab{a}})}\BibitemShut {NoStop}%
%%CITATION = PHRVA,D22,2323;%%
\bibitem [{\citenamefont {Detweiler}(1980{\natexlab{b}})}]{Detweiler:1980gk}%
  \BibitemOpen
  \bibfield  {author} {\bibinfo {author} {\bibfnamefont {S.~L.}\ \bibnamefont
  {Detweiler}},\ }\href {\doibase 10.1086/158109} {\bibfield  {journal}
  {\bibinfo  {journal} {Astrophys.\ J.}\ }\textbf {\bibinfo {volume} {239}},\
  \bibinfo {pages} {292} (\bibinfo {year} {1980}{\natexlab{b}})}\BibitemShut
  {NoStop}%
%%CITATION = ASJOA,239,292;%%
\bibitem [{\citenamefont {Simone}\ and\ \citenamefont
  {Will}(1992)}]{Simone:1991wn}%
  \BibitemOpen
  \bibfield  {author} {\bibinfo {author} {\bibfnamefont {L.~E.}\ \bibnamefont
  {Simone}}\ and\ \bibinfo {author} {\bibfnamefont {C.~M.}\ \bibnamefont
  {Will}},\ }\href {\doibase 10.1088/0264-9381/9/4/012} {\bibfield  {journal}
  {\bibinfo  {journal} {Classical Quantum Gravity}\ }\textbf {\bibinfo {volume}
  {9}},\ \bibinfo {pages} {963} (\bibinfo {year} {1992})}\BibitemShut {NoStop}%
%%CITATION = CQGRD,9,963;%%
\bibitem [{\citenamefont {Ferrari}\ and\ \citenamefont
  {Mashhoon}(1984)}]{Ferrari:1984zz}%
  \BibitemOpen
  \bibfield  {author} {\bibinfo {author} {\bibfnamefont {V.}~\bibnamefont
  {Ferrari}}\ and\ \bibinfo {author} {\bibfnamefont {B.}~\bibnamefont
  {Mashhoon}},\ }\href {\doibase 10.1103/PhysRevD.30.295} {\bibfield  {journal}
  {\bibinfo  {journal} {Phys.\ Rev.\ D}\ }\textbf {\bibinfo {volume} {30}},\
  \bibinfo {pages} {295} (\bibinfo {year} {1984})}\BibitemShut {NoStop}%
%%CITATION = PHRVA,D30,295;%%
\bibitem [{\citenamefont {Berti}\ and\ \citenamefont
  {Cardoso}(2006)}]{Berti:2006wq}%
  \BibitemOpen
  \bibfield  {author} {\bibinfo {author} {\bibfnamefont {E.}~\bibnamefont
  {Berti}}\ and\ \bibinfo {author} {\bibfnamefont {V.}~\bibnamefont
  {Cardoso}},\ }\href {\doibase 10.1103/PhysRevD.74.104020} {\bibfield
  {journal} {\bibinfo  {journal} {Phys.\ Rev.\ D}\ }\textbf {\bibinfo {volume}
  {74}},\ \bibinfo {pages} {104020} (\bibinfo {year} {2006})},\ \Eprint
  {http://arxiv.org/abs/gr-qc/0605118} {arXiv:gr-qc/0605118} \BibitemShut
  {NoStop}%
%%CITATION = GR-QC/0605118;%%
\bibitem [{\citenamefont {Leaver}(1986)}]{Leaver:1986gd}%
  \BibitemOpen
  \bibfield  {author} {\bibinfo {author} {\bibfnamefont {E.~W.}\ \bibnamefont
  {Leaver}},\ }\href {\doibase 10.1103/PhysRevD.34.384} {\bibfield  {journal}
  {\bibinfo  {journal} {Phys.\ Rev.\ D}\ }\textbf {\bibinfo {volume} {34}},\
  \bibinfo {pages} {384} (\bibinfo {year} {1986})}\BibitemShut {NoStop}%
%%CITATION = PHRVA,D34,384;%%
\bibitem [{\citenamefont {Andersson}(1997)}]{Andersson:1996cm}%
  \BibitemOpen
  \bibfield  {author} {\bibinfo {author} {\bibfnamefont {N.}~\bibnamefont
  {Andersson}},\ }\href {\doibase 10.1103/PhysRevD.55.468} {\bibfield
  {journal} {\bibinfo  {journal} {Phys.\ Rev.\ D}\ }\textbf {\bibinfo {volume}
  {55}},\ \bibinfo {pages} {468} (\bibinfo {year} {1997})},\ \Eprint
  {http://arxiv.org/abs/gr-qc/9607064} {arXiv:gr-qc/9607064} \BibitemShut
  {NoStop}%
%%CITATION = GR-QC/9607064;%%
\bibitem [{\citenamefont {Leaver}(1985)}]{Leaver:1985ax}%
  \BibitemOpen
  \bibfield  {author} {\bibinfo {author} {\bibfnamefont {E.}~\bibnamefont
  {Leaver}},\ }\href {\doibase 10.1098/rspa.1985.0119} {\bibfield  {journal}
  {\bibinfo  {journal} {Proc.\ Roy.\ Soc.\ Lond.\ A}\ }\textbf {\bibinfo
  {volume} {402}},\ \bibinfo {pages} {285} (\bibinfo {year}
  {1985})}\BibitemShut {NoStop}%
%%CITATION = PRSLA,A402,285;%%
\bibitem [{\citenamefont {Konoplya}\ and\ \citenamefont
  {Zhidenko}(2005)}]{Konoplya:2004wg}%
  \BibitemOpen
  \bibfield  {author} {\bibinfo {author} {\bibfnamefont {R.}~\bibnamefont
  {Konoplya}}\ and\ \bibinfo {author} {\bibfnamefont {A.}~\bibnamefont
  {Zhidenko}},\ }\href {\doibase 10.1016/j.physletb.2005.01.078} {\bibfield
  {journal} {\bibinfo  {journal} {Phys.\ Lett.\ B}\ }\textbf {\bibinfo {volume}
  {609}},\ \bibinfo {pages} {377} (\bibinfo {year} {2005})},\ \Eprint
  {http://arxiv.org/abs/gr-qc/0411059} {arXiv:gr-qc/0411059} \BibitemShut
  {NoStop}%
%%CITATION = GR-QC/0411059;%%
\bibitem [{\citenamefont {Majumdar}\ and\ \citenamefont
  {Panchapakesan}(1989)}]{mp}%
  \BibitemOpen
  \bibfield  {author} {\bibinfo {author} {\bibfnamefont {B.}~\bibnamefont
  {Majumdar}}\ and\ \bibinfo {author} {\bibfnamefont {N.}~\bibnamefont
  {Panchapakesan}},\ }\href {\doibase 10.1103/PhysRevD.40.2568} {\bibfield
  {journal} {\bibinfo  {journal} {Phys.\ Rev.\ D}\ }\textbf {\bibinfo {volume}
  {40}},\ \bibinfo {pages} {2568} (\bibinfo {year} {1989})}\BibitemShut
  {NoStop}%
\bibitem [{\citenamefont {Bender}\ and\ \citenamefont
  {Orszag}(1978)}]{BenderOrszag1978}%
  \BibitemOpen
  \bibfield  {author} {\bibinfo {author} {\bibfnamefont {C.~M.}\ \bibnamefont
  {Bender}}\ and\ \bibinfo {author} {\bibfnamefont {S.~A.}\ \bibnamefont
  {Orszag}},\ }\href@noop {} {\emph {\bibinfo {title} {Advanced Mathematical
  Methods for Scientists and Engineers}}}\ (\bibinfo  {publisher} {McGraw-Hill
  Book Co, Singapore},\ \bibinfo {year} {1978})\BibitemShut {NoStop}%
\bibitem [{\citenamefont {{Goebel}}(1972)}]{1972ApJ...172L..95G}%
  \BibitemOpen
  \bibfield  {author} {\bibinfo {author} {\bibfnamefont {C.~J.}\ \bibnamefont
  {{Goebel}}},\ }\href {\doibase 10.1086/180898} {\bibfield  {journal}
  {\bibinfo  {journal} {Astrophys.\ J.}\ }\textbf {\bibinfo {volume} {172}},\
  \bibinfo {pages} {L95} (\bibinfo {year} {1972})}\BibitemShut {NoStop}%
\bibitem [{\citenamefont {Vanzo}\ and\ \citenamefont
  {Zerbini}(2004)}]{Vanzo:2004fy}%
  \BibitemOpen
  \bibfield  {author} {\bibinfo {author} {\bibfnamefont {L.}~\bibnamefont
  {Vanzo}}\ and\ \bibinfo {author} {\bibfnamefont {S.}~\bibnamefont
  {Zerbini}},\ }\href {\doibase 10.1103/PhysRevD.70.044030} {\bibfield
  {journal} {\bibinfo  {journal} {Phys.\ Rev.\ D}\ }\textbf {\bibinfo {volume}
  {70}},\ \bibinfo {pages} {044030} (\bibinfo {year} {2004})},\ \Eprint
  {http://arxiv.org/abs/hep-th/0402103} {arXiv:hep-th/0402103} \BibitemShut
  {NoStop}%
\bibitem [{\citenamefont {Cardoso}\ \emph {et~al.}(2009)\citenamefont
  {Cardoso}, \citenamefont {Miranda}, \citenamefont {Berti}, \citenamefont
  {Witek},\ and\ \citenamefont {Zanchin}}]{Cardoso:2008bp}%
  \BibitemOpen
  \bibfield  {author} {\bibinfo {author} {\bibfnamefont {V.}~\bibnamefont
  {Cardoso}}, \bibinfo {author} {\bibfnamefont {A.~S.}\ \bibnamefont
  {Miranda}}, \bibinfo {author} {\bibfnamefont {E.}~\bibnamefont {Berti}},
  \bibinfo {author} {\bibfnamefont {H.}~\bibnamefont {Witek}}, \ and\ \bibinfo
  {author} {\bibfnamefont {V.~T.}\ \bibnamefont {Zanchin}},\ }\href {\doibase
  10.1103/PhysRevD.79.064016} {\bibfield  {journal} {\bibinfo  {journal}
  {Phys.\ Rev.\ D}\ }\textbf {\bibinfo {volume} {79}},\ \bibinfo {pages}
  {064016} (\bibinfo {year} {2009})},\ \Eprint {http://arxiv.org/abs/0812.1806}
  {arXiv:0812.1806 [hep-th]} \BibitemShut {NoStop}%
%%CITATION = ARXIV:0812.1806;%%
\bibitem [{\citenamefont {Decanini}\ \emph {et~al.}(2003)\citenamefont
  {Decanini}, \citenamefont {Folacci},\ and\ \citenamefont
  {Jensen}}]{Decanini:2002ha}%
  \BibitemOpen
  \bibfield  {author} {\bibinfo {author} {\bibfnamefont {Y.}~\bibnamefont
  {Decanini}}, \bibinfo {author} {\bibfnamefont {A.}~\bibnamefont {Folacci}}, \
  and\ \bibinfo {author} {\bibfnamefont {B.}~\bibnamefont {Jensen}},\ }\href
  {\doibase 10.1103/PhysRevD.67.124017} {\bibfield  {journal} {\bibinfo
  {journal} {Phys.\ Rev.\ D}\ }\textbf {\bibinfo {volume} {67}},\ \bibinfo
  {pages} {124017} (\bibinfo {year} {2003})},\ \Eprint
  {http://arxiv.org/abs/gr-qc/0212093} {arXiv:gr-qc/0212093} \BibitemShut
  {NoStop}%
\bibitem [{\citenamefont {Decanini}\ and\ \citenamefont
  {Folacci}(2010)}]{Decanini:2009mu}%
  \BibitemOpen
  \bibfield  {author} {\bibinfo {author} {\bibfnamefont {Y.}~\bibnamefont
  {Decanini}}\ and\ \bibinfo {author} {\bibfnamefont {A.}~\bibnamefont
  {Folacci}},\ }\href {\doibase 10.1103/PhysRevD.81.024031} {\bibfield
  {journal} {\bibinfo  {journal} {Phys.\ Rev.\ D}\ }\textbf {\bibinfo {volume}
  {81}},\ \bibinfo {pages} {024031} (\bibinfo {year} {2010})},\ \Eprint
  {http://arxiv.org/abs/0906.2601} {arXiv:0906.2601 [gr-qc]} \BibitemShut
  {NoStop}%
%%CITATION = ARXIV:0906.2601;%%
\bibitem [{\citenamefont {Decanini}\ \emph {et~al.}(2010)\citenamefont
  {Decanini}, \citenamefont {Folacci},\ and\ \citenamefont
  {Raffaelli}}]{Decanini:2010fz}%
  \BibitemOpen
  \bibfield  {author} {\bibinfo {author} {\bibfnamefont {Y.}~\bibnamefont
  {Decanini}}, \bibinfo {author} {\bibfnamefont {A.}~\bibnamefont {Folacci}}, \
  and\ \bibinfo {author} {\bibfnamefont {B.}~\bibnamefont {Raffaelli}},\ }\href
  {\doibase 10.1103/PhysRevD.81.104039} {\bibfield  {journal} {\bibinfo
  {journal} {Phys.\ Rev.\ D}\ }\textbf {\bibinfo {volume} {81}},\ \bibinfo
  {pages} {104039} (\bibinfo {year} {2010})},\ \Eprint
  {http://arxiv.org/abs/1002.0121} {arXiv:1002.0121 [gr-qc]} \BibitemShut
  {NoStop}%
%%CITATION = ARXIV:1002.0121;%%
\bibitem [{\citenamefont {Decanini}\ \emph {et~al.}(2011)\citenamefont
  {Decanini}, \citenamefont {Folacci},\ and\ \citenamefont
  {Raffaelli}}]{Decanini:2011eh}%
  \BibitemOpen
  \bibfield  {author} {\bibinfo {author} {\bibfnamefont {Y.}~\bibnamefont
  {Decanini}}, \bibinfo {author} {\bibfnamefont {A.}~\bibnamefont {Folacci}}, \
  and\ \bibinfo {author} {\bibfnamefont {B.}~\bibnamefont {Raffaelli}},\ }\href
  {\doibase 10.1103/PhysRevD.84.084035} {\bibfield  {journal} {\bibinfo
  {journal} {Phys.\ Rev.\ D}\ }\textbf {\bibinfo {volume} {84}},\ \bibinfo
  {pages} {084035} (\bibinfo {year} {2011})},\ \Eprint
  {http://arxiv.org/abs/1108.5076} {arXiv:1108.5076 [gr-qc]} \BibitemShut
  {NoStop}%
%%CITATION = ARXIV:1108.5076;%%
\bibitem [{\citenamefont {Dolan}\ and\ \citenamefont
  {Ottewill}(2009)}]{Dolan:2009nk}%
  \BibitemOpen
  \bibfield  {author} {\bibinfo {author} {\bibfnamefont {S.~R.}\ \bibnamefont
  {Dolan}}\ and\ \bibinfo {author} {\bibfnamefont {A.~C.}\ \bibnamefont
  {Ottewill}},\ }\href {\doibase 10.1088/0264-9381/26/22/225003} {\bibfield
  {journal} {\bibinfo  {journal} {Classical Quantum Gravity}\ }\textbf
  {\bibinfo {volume} {26}},\ \bibinfo {pages} {225003} (\bibinfo {year}
  {2009})},\ \Eprint {http://arxiv.org/abs/0908.0329} {arXiv:0908.0329 [gr-qc]}
  \BibitemShut {NoStop}%
%%CITATION = ARXIV:0908.0329;%%
\bibitem [{\citenamefont {Dolan}\ and\ \citenamefont
  {Ottewill}(2011)}]{Dolan:2011fh}%
  \BibitemOpen
  \bibfield  {author} {\bibinfo {author} {\bibfnamefont {S.~R.}\ \bibnamefont
  {Dolan}}\ and\ \bibinfo {author} {\bibfnamefont {A.~C.}\ \bibnamefont
  {Ottewill}},\ }\href {\doibase 10.1103/PhysRevD.84.104002} {\bibfield
  {journal} {\bibinfo  {journal} {Phys.\ Rev.\ D}\ }\textbf {\bibinfo {volume}
  {84}},\ \bibinfo {pages} {104002} (\bibinfo {year} {2011})},\ \Eprint
  {http://arxiv.org/abs/1106.4318} {arXiv:1106.4318 [gr-qc]} \BibitemShut
  {NoStop}%
%%CITATION = ARXIV:1106.4318;%%
\bibitem [{\citenamefont {Unruh}(1976)}]{Unruh:1976fm}%
  \BibitemOpen
  \bibfield  {author} {\bibinfo {author} {\bibfnamefont {W.}~\bibnamefont
  {Unruh}},\ }\href {\doibase 10.1103/PhysRevD.14.3251} {\bibfield  {journal}
  {\bibinfo  {journal} {Phys.\ Rev.\ D}\ }\textbf {\bibinfo {volume} {14}},\
  \bibinfo {pages} {3251} (\bibinfo {year} {1976})}\BibitemShut {NoStop}%
%%CITATION = PHRVA,D14,3251;%%
\end{thebibliography}%

\end{document}